%% file: Perceived_ABC_Trustworthiness.tex
\documentclass[sigconf,nonacm,balance=false]{acmart}\usepackage[]{graphicx}\usepackage[]{xcolor}
\makeatletter
\def\maxwidth{ %
  \ifdim\Gin@nat@width>\linewidth
    \linewidth
  \else
    \Gin@nat@width
  \fi
}
\makeatother

\definecolor{fgcolor}{rgb}{0.345, 0.345, 0.345}

\usepackage{framed}
\makeatletter
\newenvironment{kframe}{%
 \def\at@end@of@kframe{}%
 \ifinner\ifhmode%
  \def\at@end@of@kframe{\end{minipage}}%
  \begin{minipage}{\columnwidth}%
 \fi\fi%
 \def\FrameCommand##1{\hskip\@totalleftmargin \hskip-\fboxsep
 \colorbox{shadecolor}{##1}\hskip-\fboxsep
     \hskip-\linewidth \hskip-\@totalleftmargin \hskip\columnwidth}%
 \MakeFramed {\advance\hsize-\width
   \@totalleftmargin\z@ \linewidth\hsize
   \@setminipage}}%
 {\par\unskip\endMakeFramed%
 \at@end@of@kframe}
\makeatother

\definecolor{shadecolor}{rgb}{.97, .97, .97}
\definecolor{messagecolor}{rgb}{0, 0, 0}
\definecolor{warningcolor}{rgb}{1, 0, 1}
\definecolor{errorcolor}{rgb}{1, 0, 0}

\usepackage{alltt}
\setcopyright{none}

\usepackage[utf8]{inputenc}

\usepackage{epsfig,endnotes,url}

\usepackage{xspace}

\usepackage{xcolor}

\usepackage{amsmath, amsthm, latexsym} 
\usepackage{ulem}

\usepackage{graphicx}

\usepackage{multirow}
\usepackage{booktabs}
\usepackage{longtable}
\usepackage{rotating}

\usepackage{pdfpages}

\usepackage{paralist}

\usepackage{balance}

\usepackage{versions}
\includeversion{DocumentVersionConference}
\excludeversion{DocumentVersionTR}
\includeversion{DocumentVersionLNCS}
\excludeversion{RedundantContent}
\excludeversion{DocumentVersionBootstrapping}
\excludeversion{DocumentVersionAppendix}

\usepackage[caption=false]{subfig}

\newtheorem{researchquestion}{RQ}
\theoremstyle{definition} 

\newcommand{\vari}[1]{\ensuremath{\mathit{#1}\xspace}}

\newcommand{\CASCAde}{ERC Starting Grant CASCAde (GA n\textsuperscript{o}716980)}
\IfFileExists{upquote.sty}{\usepackage{upquote}}{}
\begin{document}
















%
%

\newcommand{\conditionTable}{
\begin{table*}[ht]
\centering
\caption{Experimental conditions and descriptives of trust and behavioral intention.} 
\label{tab:conditions}
\begingroup\footnotesize
\begin{tabular}{lllllllrrrrr}
  \toprule
Survey & \textsf{provider} & \textsf{benefits} & \textsf{usage} & \textsf{simplicity} & \textsf{people} & \textsf{support} & $n$ & $\mathsf{avg}(\mathsf{trust})$ & $\mathsf{sd}(\mathsf{trust})$ & $\mathsf{avg}(\mathsf{bi})$ & $\mathsf{sd}(\mathsf{bi})$ \\ 
  \midrule
Intrinsic & Company & Privacy & Everyday & simple & photo & nosupport &  24 & 4.28 & 1.25 & 4.49 & 1.53 \\ 
  Intrinsic & Company & Privacy & Tech & simple & photo & nosupport &  29 & 4.48 & 0.88 & 4.44 & 1.04 \\ 
  Intrinsic & Company & User & Everyday & simple & photo & nosupport &  34 & 4.24 & 1.17 & 4.06 & 1.54 \\ 
  Intrinsic & Company & User & Tech & simple & photo & nosupport &  30 & 4.17 & 1.25 & 4.41 & 1.76 \\ 
  Intrinsic & Government & Privacy & Everyday & simple & photo & nosupport &  26 & 4.40 & 1.16 & 4.37 & 1.44 \\ 
  Intrinsic & Government & Privacy & Tech & simple & photo & nosupport &  35 & 4.14 & 1.05 & 4.58 & 1.34 \\ 
  Intrinsic & Government & User & Everyday & simple & photo & nosupport &  41 & 4.24 & 1.08 & 4.54 & 1.34 \\ 
  Intrinsic & Government & User & Tech & simple & photo & nosupport &  37 & 4.23 & 1.18 & 4.24 & 1.63 \\ 
  Intrinsic & University & Privacy & Everyday & simple & photo & nosupport &  26 & 4.60 & 0.86 & 4.51 & 1.32 \\ 
  Intrinsic & University & Privacy & Tech & simple & photo & nosupport &  39 & 4.64 & 0.90 & 4.36 & 1.40 \\ 
  Intrinsic & University & User & Everyday & simple & photo & nosupport &  34 & 4.07 & 0.94 & 4.51 & 1.21 \\ 
  Intrinsic & University & User & Tech & simple & photo & nosupport &  24 & 4.75 & 0.96 & 5.11 & 1.14 \\ 
  Presentation & none & User & Everyday & complex & nophoto & contact &  27 & 4.44 & 1.06 & 4.64 & 1.69 \\ 
  Presentation & none & User & Everyday & complex & nophoto & fullsupport &  40 & 4.35 & 0.87 & 4.33 & 1.38 \\ 
  Presentation & none & User & Everyday & complex & nophoto & nosupport &  35 & 4.14 & 0.94 & 3.90 & 1.21 \\ 
  Presentation & none & User & Everyday & complex & photo & contact &  32 & 4.38 & 1.10 & 4.28 & 1.46 \\ 
  Presentation & none & User & Everyday & complex & photo & fullsupport &  40 & 4.36 & 0.88 & 4.28 & 1.45 \\ 
  Presentation & none & User & Everyday & complex & photo & nosupport &  35 & 4.00 & 0.90 & 4.16 & 1.31 \\ 
  Presentation & none & User & Everyday & simple & nophoto & contact &  30 & 4.32 & 1.25 & 4.97 & 1.22 \\ 
  Presentation & none & User & Everyday & simple & nophoto & fullsupport &  43 & 4.28 & 1.27 & 4.58 & 1.37 \\ 
  Presentation & none & User & Everyday & simple & nophoto & nosupport &  46 & 4.05 & 0.97 & 4.46 & 1.25 \\ 
  Presentation & none & User & Everyday & simple & photo & contact &  33 & 4.60 & 0.87 & 4.65 & 1.22 \\ 
  Presentation & none & User & Everyday & simple & photo & fullsupport &  31 & 4.20 & 0.75 & 4.45 & 1.33 \\ 
  Presentation & none & User & Everyday & simple & photo & nosupport &  41 & 4.35 & 0.92 & 4.76 & 1.26 \\ 
   \bottomrule
\end{tabular}
\endgroup
\end{table*}
}




%
%

%
%

\newcommand{\semPlotModel}{
\begin{figure*}[p]
\centering
\vspace{-2cm}

\includegraphics[width=\maxwidth]{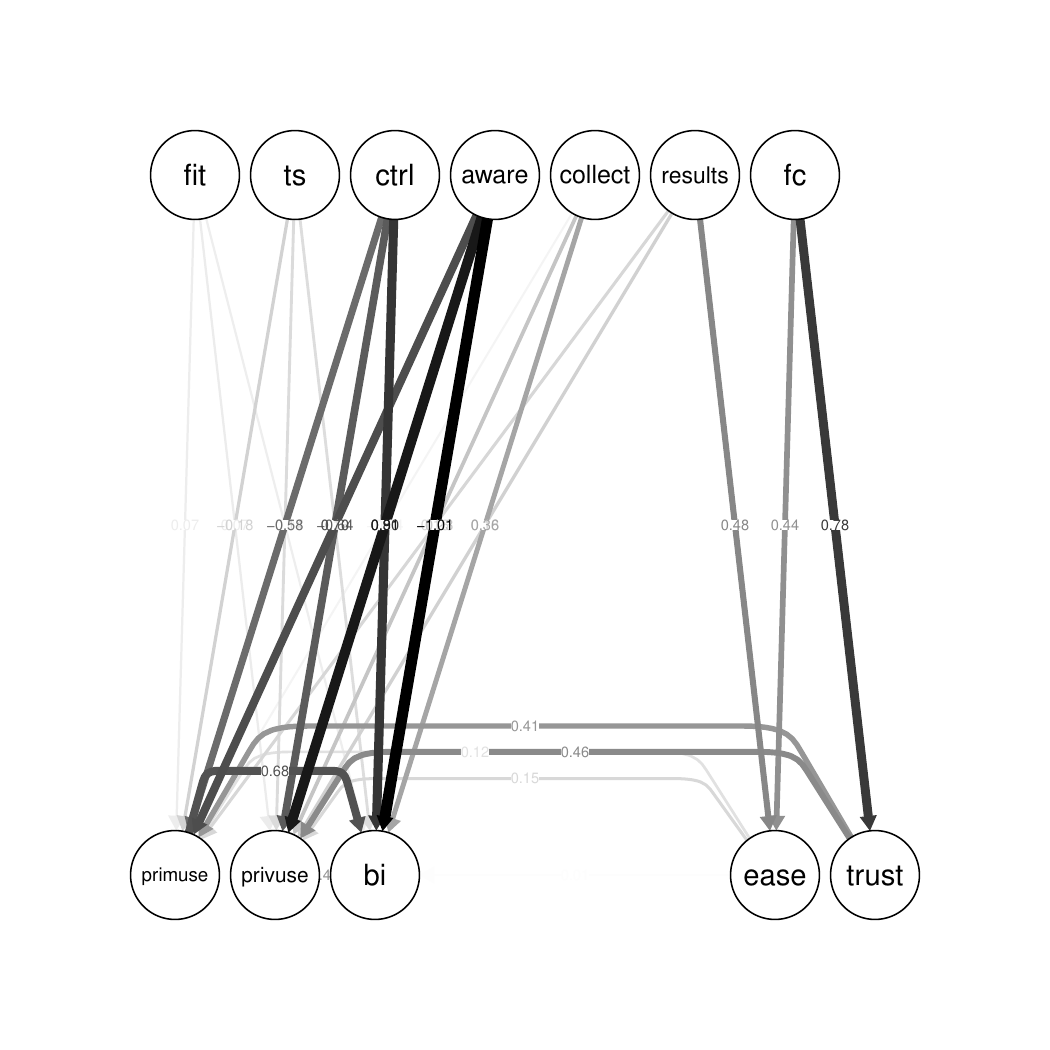} 
\vspace{-2.5cm}
\caption{Correlational SEM model restricted to latent variables (standardized coefficients).}
\label{fig:semPlotModel}
\end{figure*}
}
\newcommand{\semPlotModelCausal}{
\begin{figure*}[p]
\centering
\vspace{-2cm}
\begin{kframe}

{\ttfamily\noindent\bfseries\color{errorcolor}{\#\# Error in dimnames(x) <- dn: length of 'dimnames' [2] not equal to array extent}}\end{kframe}\vspace{-2.5cm}
\caption{Causal SEM model restricted to latent variables (standardized coefficients).}
\label{fig:semPlotModelCausal}
\end{figure*}
}

\newcommand{\fitCFAMeasurement}{
\begin{table*}[ht]
\centering
\caption{Fit measures for the WLSMV-estimated CFA of the measurement model.} 
\label{tab:fitCFAMeasurement}
\begingroup\footnotesize
\begin{tabular}{rrrrrrrrrr}
  \toprule
$\chi^2$ & $\mathit{df}$ & $p_{\chi^2}$ & \textsf{CFI} & \textsf{TFI} & \textsf{RMSEA} & \textsf{LL} & \textsf{UL} & $p_{\mathsf{rmsea}}$ & \textsf{SRMR} \\ 
  \midrule
3501.88 & 879.00 & $<.001$ & 0.97 & 0.97 & 0.06 & 0.06 & 0.06 & $<.001$ & 0.05 \\ 
   \bottomrule
\end{tabular}
\endgroup
\end{table*}
}
\newcommand{\residualsMM}{
\begin{table*}[tbp]
\centering\caption{Residuals of fitted model}
\label{tab:residualsMM}
\captionsetup{position=top}
\subfloat[Correlation residuals]{
\label{tab:residualsMMcor}
\centering
\begingroup\footnotesize
\begin{tabular}{rlllllllllllllllllllllllllllllllllllllllllllll}
  \toprule
 & 1 & 2 & 3 & 4 & 5 & 6 & 7 & 8 & 9 & 10 & 11 & 12 & 13 & 14 & 15 & 16 & 17 & 18 & 19 & 20 & 21 & 22 & 23 & 24 & 25 & 26 & 27 & 28 & 29 & 30 & 31 & 32 & 33 & 34 & 35 & 36 & 37 & 38 & 39 & 40 & 41 & 42 & 43 & 44 & 45 \\ 
  \midrule
1. MKF1 & --- &  &  &  &  &  &  &  &  &  &  &  &  &  &  &  &  &  &  &  &  &  &  &  &  &  &  &  &  &  &  &  &  &  &  &  &  &  &  &  &  &  &  &  &  \\ 
  2. MKF2 & 0.024 & --- &  &  &  &  &  &  &  &  &  &  &  &  &  &  &  &  &  &  &  &  &  &  &  &  &  &  &  &  &  &  &  &  &  &  &  &  &  &  &  &  &  &  &  \\ 
  3. MKF3 & 0.018 & 0.023 & --- &  &  &  &  &  &  &  &  &  &  &  &  &  &  &  &  &  &  &  &  &  &  &  &  &  &  &  &  &  &  &  &  &  &  &  &  &  &  &  &  &  &  \\ 
  4. MKF4 & -0.019 & -0.046 & -0.037 & --- &  &  &  &  &  &  &  &  &  &  &  &  &  &  &  &  &  &  &  &  &  &  &  &  &  &  &  &  &  &  &  &  &  &  &  &  &  &  &  &  &  \\ 
  5. MKTS1 & -0.022 & -0.031 & -0.056 & 0.049 & --- &  &  &  &  &  &  &  &  &  &  &  &  &  &  &  &  &  &  &  &  &  &  &  &  &  &  &  &  &  &  &  &  &  &  &  &  &  &  &  &  \\ 
  6. MKTS2 & -0.007 & -0.033 & -0.051 & 0.056 & 0.012 & --- &  &  &  &  &  &  &  &  &  &  &  &  &  &  &  &  &  &  &  &  &  &  &  &  &  &  &  &  &  &  &  &  &  &  &  &  &  &  &  \\ 
  7. MKTS3 & 0.017 & -0.007 & -0.003 & \textbf{0.104} & -0.034 & -0.037 & --- &  &  &  &  &  &  &  &  &  &  &  &  &  &  &  &  &  &  &  &  &  &  &  &  &  &  &  &  &  &  &  &  &  &  &  &  &  &  \\ 
  8. ctrl1 & 0.046 & 0.038 & 0.034 & 0.04 & 0.058 & 0.065 & 0.051 & --- &  &  &  &  &  &  &  &  &  &  &  &  &  &  &  &  &  &  &  &  &  &  &  &  &  &  &  &  &  &  &  &  &  &  &  &  &  \\ 
  9. ctrl2 & -0.059 & -0.017 & 0.004 & 0.044 & -0.006 & 0.021 & 0.022 & 0.083 & --- &  &  &  &  &  &  &  &  &  &  &  &  &  &  &  &  &  &  &  &  &  &  &  &  &  &  &  &  &  &  &  &  &  &  &  &  \\ 
  10. ctrl3 & -0.045 & -0.032 & -0.04 & -0.081 & \textbf{-0.132} & \textbf{-0.105} & -0.02 & \textbf{-0.131} & \textbf{-0.121} & --- &  &  &  &  &  &  &  &  &  &  &  &  &  &  &  &  &  &  &  &  &  &  &  &  &  &  &  &  &  &  &  &  &  &  &  \\ 
  11. awa1 & -0.066 & 0.005 & -0.016 & 0.019 & -0.066 & -0.02 & 0.024 & -0.022 & -0.015 & \textbf{0.12} & --- &  &  &  &  &  &  &  &  &  &  &  &  &  &  &  &  &  &  &  &  &  &  &  &  &  &  &  &  &  &  &  &  &  &  \\ 
  12. awa2 & -0.042 & -0.018 & -0.025 & 0.057 & -0.021 & 0.017 & 0.049 & -0.017 & 0.001 & \textbf{0.107} & 0.076 & --- &  &  &  &  &  &  &  &  &  &  &  &  &  &  &  &  &  &  &  &  &  &  &  &  &  &  &  &  &  &  &  &  &  \\ 
  13. awa3 & 0.008 & 0.019 & -0.006 & 0.072 & -0.013 & 0.01 & 0.026 & -0.098 & -0.078 & 0.026 & \textbf{-0.132} & \textbf{-0.116} & --- &  &  &  &  &  &  &  &  &  &  &  &  &  &  &  &  &  &  &  &  &  &  &  &  &  &  &  &  &  &  &  &  \\ 
  14. coll1 & -0.033 & -0.039 & -0.031 & -0.049 & -0.027 & -0.016 & -0.014 & \textbf{-0.162} & \textbf{-0.138} & \textbf{0.126} & \textbf{-0.175} & \textbf{-0.12} & 0.072 & --- &  &  &  &  &  &  &  &  &  &  &  &  &  &  &  &  &  &  &  &  &  &  &  &  &  &  &  &  &  &  &  \\ 
  15. coll2 & 0.022 & 0.018 & 0.005 & 0.021 & -0.01 & 0 & 0.035 & -0.042 & -0.046 & \textbf{0.188} & -0.061 & 0.003 & \textbf{0.109} & 0.027 & --- &  &  &  &  &  &  &  &  &  &  &  &  &  &  &  &  &  &  &  &  &  &  &  &  &  &  &  &  &  &  \\ 
  16. coll3 & 0.032 & 0.021 & 0.025 & 0.005 & -0.003 & -0.007 & 0.025 & -0.078 & \textbf{-0.101} & \textbf{0.157} & \textbf{-0.127} & -0.082 & \textbf{0.103} & 0.011 & -0.006 & --- &  &  &  &  &  &  &  &  &  &  &  &  &  &  &  &  &  &  &  &  &  &  &  &  &  &  &  &  &  \\ 
  17. coll4 & -0.005 & 0.008 & -0.017 & 0.008 & -0.012 & 0.027 & 0.015 & -0.063 & -0.052 & \textbf{0.184} & -0.024 & -0.017 & \textbf{0.143} & -0.012 & -0.034 & 0.003 & --- &  &  &  &  &  &  &  &  &  &  &  &  &  &  &  &  &  &  &  &  &  &  &  &  &  &  &  &  \\ 
  18. STPE1 & 0.015 & 0.012 & 0.041 & 0.04 & -0.021 & -0.013 & 0.064 & 0.066 & 0.028 & \textbf{0.101} & \textbf{0.113} & 0.076 & 0.065 & 0.02 & 0.059 & 0.058 & 0.1 & --- &  &  &  &  &  &  &  &  &  &  &  &  &  &  &  &  &  &  &  &  &  &  &  &  &  &  &  \\ 
  19. STPE2 & -0.003 & -0.013 & 0.029 & -0.032 & -0.005 & -0.001 & 0.026 & 0.023 & -0.022 & 0.035 & -0.012 & -0.059 & 0.077 & -0.005 & 0.01 & 0.014 & 0.064 & -0.03 & --- &  &  &  &  &  &  &  &  &  &  &  &  &  &  &  &  &  &  &  &  &  &  &  &  &  &  \\ 
  20. STPE3 & 0.005 & 0.004 & -0.036 & -0.004 & -0.022 & -0.011 & 0.036 & -0.029 & -0.066 & 0.028 & -0.041 & \textbf{-0.108} & -0.012 & -0.067 & -0.037 & -0.061 & -0.024 & \textbf{-0.195} & -0.024 & --- &  &  &  &  &  &  &  &  &  &  &  &  &  &  &  &  &  &  &  &  &  &  &  &  &  \\ 
  21. STPE4 & -0.005 & -0.012 & -0.015 & -0.026 & -0.025 & -0.019 & 0.019 & -0.053 & -0.075 & -0.002 & -0.067 & \textbf{-0.11} & 0.013 & -0.057 & -0.051 & -0.055 & -0.007 & \textbf{-0.196} & -0.035 & 0.087 & --- &  &  &  &  &  &  &  &  &  &  &  &  &  &  &  &  &  &  &  &  &  &  &  &  \\ 
  22. TAU1 & -0.016 & 0.005 & 0.006 & -0.009 & -0.011 & -0.015 & 0.03 & 0.045 & 0.03 & 0.051 & 0.054 & 0.031 & -0.003 & -0.031 & 0.073 & 0.031 & 0.07 & \textbf{0.111} & 0.025 & -0.029 & -0.05 & --- &  &  &  &  &  &  &  &  &  &  &  &  &  &  &  &  &  &  &  &  &  &  &  \\ 
  23. TAU2 & -0.024 & 0.014 & -0.002 & 0.001 & -0.022 & -0.03 & 0.011 & 0.007 & 0.002 & -0.018 & 0.025 & -0.011 & -0.067 & -0.088 & 0.014 & 0.007 & 0.025 & 0.069 & -0.027 & -0.06 & -0.072 & -0.009 & --- &  &  &  &  &  &  &  &  &  &  &  &  &  &  &  &  &  &  &  &  &  &  \\ 
  24. TAU3 & -0.02 & -0.008 & -0.007 & 0.027 & -0.006 & -0.026 & 0.03 & -0.015 & -0.024 & -0.03 & 0.029 & 0.03 & -0.056 & \textbf{-0.106} & -0.008 & -0.03 & 0.005 & 0.039 & -0.058 & \textbf{-0.105} & \textbf{-0.116} & -0.009 & 0.022 & --- &  &  &  &  &  &  &  &  &  &  &  &  &  &  &  &  &  &  &  &  &  \\ 
  25. TAU4 & 0.01 & 0.01 & 0.019 & 0.002 & 0.024 & 0.018 & 0.015 & -0.021 & -0.03 & -0.015 & 0.021 & 0.026 & -0.044 & -0.057 & 0.015 & 0.017 & 0.045 & 0.095 & 0.042 & 0.006 & -0.013 & -0.011 & -0.052 & -0.03 & --- &  &  &  &  &  &  &  &  &  &  &  &  &  &  &  &  &  &  &  &  \\ 
  26. TAE1 & -0.012 & 0.009 & -0.015 & -0.01 & -0.008 & -0.022 & 0.042 & 0.005 & 0.017 & 0.003 & -0.059 & -0.05 & 0.078 & -0.015 & 0.035 & 0.002 & 0.009 & 0.048 & 0.051 & 0.051 & 0.031 & 0.037 & -0.008 & 0.002 & 0.087 & --- &  &  &  &  &  &  &  &  &  &  &  &  &  &  &  &  &  &  &  \\ 
  27. TAE2 & -0.017 & 0.012 & -0.01 & 0.051 & -0.058 & -0.037 & -0.023 & -0.008 & 0.022 & 0.062 & 0.047 & 0.019 & 0.017 & -0.031 & 0.014 & -0.021 & 0.015 & 0 & -0.036 & 0.007 & -0.018 & -0.024 & -0.052 & -0.041 & 0.041 & -0.014 & --- &  &  &  &  &  &  &  &  &  &  &  &  &  &  &  &  &  &  \\ 
  28. TAE3 & -0.003 & -0.002 & 0.023 & 0.018 & -0.008 & 0.003 & 0.06 & -0.015 & -0.003 & -0.034 & -0.035 & -0.054 & 0.011 & -0.044 & -0.015 & -0.013 & -0.004 & -0.03 & -0.023 & 0.017 & 0.007 & -0.023 & -0.011 & -0.016 & 0.053 & -0.044 & 0.037 & --- &  &  &  &  &  &  &  &  &  &  &  &  &  &  &  &  &  \\ 
  29. TAE4 & -0.02 & -0.004 & 0.002 & -0.01 & 0.014 & 0.026 & 0.023 & -0.009 & -0.013 & -0.025 & -0.029 & -0.04 & 0.068 & -0.006 & 0.009 & 0.035 & 0.03 & -0.064 & -0.032 & -0.04 & -0.048 & -0.029 & -0.081 & -0.076 & 0.012 & -0.076 & 0.045 & 0.04 & --- &  &  &  &  &  &  &  &  &  &  &  &  &  &  &  &  \\ 
  30. STBI1 & -0.028 & -0.002 & 0.005 & -0.018 & -0.022 & -0.016 & 0.037 & -0.005 & -0.019 & 0.01 & 0.008 & -0.049 & 0.051 & -0.008 & -0.005 & 0.008 & 0.036 & 0.043 & 0.013 & -0.095 & -0.098 & 0.018 & -0.049 & -0.066 & 0.034 & 0.014 & 0.003 & -0.032 & -0.027 & --- &  &  &  &  &  &  &  &  &  &  &  &  &  &  &  \\ 
  31. STBI2 & -0.022 & 0.054 & 0.029 & -0.022 & 0.01 & 0.008 & 0.053 & 0.019 & -0.016 & -0.009 & -0.036 & -0.079 & 0.07 & 0.007 & -0.037 & 0.022 & 0.068 & 0.042 & 0.083 & -0.034 & -0.033 & 0.017 & -0.037 & -0.07 & 0.02 & 0.056 & -0.004 & -0.009 & 0.001 & 0 & --- &  &  &  &  &  &  &  &  &  &  &  &  &  &  \\ 
  32. STBI4 & 0.002 & -0.002 & -0.001 & 0.004 & -0.037 & -0.043 & 0.019 & 0.019 & -0.008 & 0.014 & 0.011 & -0.032 & 0.029 & -0.062 & -0.018 & -0.019 & 0.002 & 0.083 & 0.004 & -0.093 & \textbf{-0.111} & 0.056 & -0.006 & -0.015 & 0.07 & 0.014 & 0.018 & -0.029 & -0.046 & 0.016 & -0.051 & --- &  &  &  &  &  &  &  &  &  &  &  &  &  \\ 
  33. TAR1 & -0.015 & -0.03 & 0.003 & 0.014 & -0.007 & 0.005 & 0.059 & -0.02 & -0.021 & -0.046 & -0.045 & -0.011 & 0.049 & -0.01 & 0.021 & 0.04 & 0.055 & 0.045 & 0.035 & -0.019 & -0.017 & 0.002 & -0.019 & -0.029 & 0.023 & -0.039 & 0.003 & 0.021 & 0.076 & 0.008 & 0.034 & -0.001 & --- &  &  &  &  &  &  &  &  &  &  &  &  \\ 
  34. TAR2 & 0.005 & 0.011 & 0.002 & 0.023 & -0.05 & -0.034 & 0.015 & 0.044 & 0.047 & 0.013 & 0.024 & 0.017 & 0.022 & -0.046 & 0.012 & 0.002 & 0.023 & 0.032 & 0.007 & -0.035 & -0.067 & 0.01 & -0.031 & -0.039 & 0.006 & -0.088 & 0.007 & -0.053 & 0.022 & 0.006 & 0.012 & 0.006 & 0.036 & --- &  &  &  &  &  &  &  &  &  &  &  \\ 
  35. TAR3 & -0.017 & 0.018 & -0.01 & 0.007 & -0.029 & 0.005 & 0.038 & 0.005 & -0.001 & -0.023 & -0.001 & -0.027 & -0.031 & -0.075 & 0.009 & -0.019 & 0.002 & 0.028 & 0.008 & -0.005 & -0.033 & 0.031 & -0.001 & -0.013 & 0.035 & 0.011 & 0.006 & -0.035 & -0.002 & -0.035 & -0.003 & -0.017 & \textbf{-0.124} & 0.051 & --- &  &  &  &  &  &  &  &  &  &  \\ 
  36. STFC1 & -0.063 & -0.067 & -0.022 & 0.032 & -0.08 & -0.051 & 0.022 & 0 & 0.027 & 0.02 & 0.068 & -0.003 & 0.004 & -0.002 & 0.059 & 0.063 & 0.038 & -0.013 & -0.019 & -0.02 & -0.045 & 0.003 & -0.019 & -0.027 & -0.012 & -0.021 & 0.046 & -0.011 & 0.039 & -0.041 & -0.046 & -0.06 & -0.03 & 0.048 & -0.012 & --- &  &  &  &  &  &  &  &  &  \\ 
  37. STFC2 & -0.042 & -0.026 & -0.022 & -0.036 & -0.039 & -0.055 & 0.036 & -0.018 & 0.013 & 0.002 & -0.035 & -0.024 & 0.013 & -0.043 & 0.068 & 0.026 & 0.017 & -0.024 & -0.007 & -0.025 & -0.022 & -0.024 & -0.035 & -0.072 & -0.008 & 0.019 & 0.087 & 0.033 & 0.071 & -0.031 & -0.042 & -0.065 & -0.015 & 0.08 & 0.021 & \textbf{0.149} & --- &  &  &  &  &  &  &  &  \\ 
  38. STFC3 & 0.031 & 0.044 & 0.021 & 0.07 & -0.006 & 0.006 & 0.039 & 0.021 & 0.028 & -0.062 & -0.025 & -0.018 & -0.015 & -0.055 & 0.002 & 0.015 & -0.019 & 0.039 & 0.045 & -0.024 & -0.011 & 0.039 & 0.045 & 0.038 & 0.017 & -0.062 & -0.063 & -0.045 & -0.039 & 0.059 & 0.037 & 0.028 & -0.026 & -0.013 & -0.058 & -0.028 & -0.081 & --- &  &  &  &  &  &  &  \\ 
  39. STFC4 & 0.041 & -0.001 & 0.026 & 0.059 & 0.055 & 0.046 & 0.069 & -0.002 & 0.039 & -0.098 & -0.021 & 0.013 & 0.043 & -0.082 & -0.034 & -0.034 & -0.003 & 0.029 & 0.051 & 0.005 & 0.029 & 0.036 & -0.011 & -0.003 & 0.041 & -0.029 & -0.073 & -0.042 & -0.006 & 0.058 & 0.036 & 0.027 & 0.018 & 0.01 & -0.024 & \textbf{-0.126} & \textbf{-0.162} & -0.022 & --- &  &  &  &  &  &  \\ 
  40. STT1 & -0.008 & -0.037 & -0.009 & -0.065 & -0.1 & -0.087 & -0.044 & 0.037 & 0.031 & -0.058 & 0.055 & 0.01 & 0.046 & -0.026 & -0.01 & 0.06 & 0.035 & 0.064 & 0.056 & -0.018 & 0.027 & 0.021 & -0.001 & 0.008 & 0.04 & 0.029 & -0.049 & -0.018 & -0.062 & 0.011 & 0.035 & 0.037 & -0.022 & -0.098 & -0.006 & -0.083 & -0.074 & 0.029 & 0.021 & --- &  &  &  &  &  \\ 
  41. STT2 & -0.011 & -0.038 & -0.036 & -0.051 & -0.056 & -0.07 & -0.009 & 0.043 & 0.045 & -0.082 & 0.021 & -0.015 & 0.007 & -0.039 & -0.007 & 0.026 & 0.007 & 0.02 & 0.02 & -0.055 & -0.019 & -0.018 & -0.04 & -0.019 & 0.023 & -0.002 & -0.07 & -0.028 & -0.073 & -0.011 & 0.016 & 0.014 & -0.034 & \textbf{-0.102} & -0.011 & -0.061 & -0.061 & 0.042 & 0.029 & 0.013 & --- &  &  &  &  \\ 
  42. STT3 & 0.015 & 0.028 & 0.036 & -0.028 & 0.055 & 0.055 & 0.076 & -0.011 & 0.005 & \textbf{-0.115} & -0.036 & -0.031 & 0.016 & -0.038 & -0.044 & 0.024 & 0.002 & -0.05 & -0.023 & -0.056 & -0.041 & -0.034 & -0.045 & -0.025 & -0.003 & 0.02 & -0.089 & -0.027 & -0.026 & -0.051 & -0.027 & -0.052 & -0.011 & -0.059 & 0.039 & -0.089 & -0.078 & 0.019 & 0.061 & -0.052 & 0.023 & --- &  &  &  \\ 
  43. STT4 & 0.063 & 0.034 & 0.021 & 0.017 & 0.024 & 0.03 & 0.055 & 0.046 & 0.048 & -0.093 & -0.011 & -0.022 & \textbf{0.103} & -0.019 & -0.041 & 0.017 & -0.001 & -0.022 & 0.006 & -0.035 & 0.016 & -0.01 & -0.029 & -0.031 & -0.016 & 0.08 & -0.038 & 0.013 & 0.006 & -0.037 & 0.01 & -0.051 & 0.027 & -0.052 & 0.059 & -0.062 & -0.04 & 0.056 & 0.083 & -0.078 & -0.05 & \textbf{0.11} & --- &  &  \\ 
  44. STT5 & 0.032 & 0.033 & 0.006 & -0.028 & 0.081 & 0.086 & \textbf{0.114} & -0.031 & -0.037 & \textbf{-0.119} & \textbf{-0.174} & \textbf{-0.175} & -0.066 & -0.021 & \textbf{-0.105} & -0.011 & -0.014 & -0.092 & 0.038 & -0.044 & 0 & -0.097 & -0.09 & -0.093 & -0.049 & 0.049 & -0.042 & 0.014 & 0.013 & -0.024 & 0.045 & -0.07 & -0.003 & -0.07 & -0.02 & -0.095 & -0.045 & 0.003 & 0.035 & -0.043 & -0.024 & \textbf{0.109} & \textbf{0.112} & --- &  \\ 
  45. STT6 & -0.007 & 0.022 & 0.045 & -0.001 & -0.058 & -0.067 & 0.012 & 0.083 & \textbf{0.103} & 0.004 & 0.082 & 0.077 & -0.017 & -0.034 & 0.067 & 0.04 & 0.039 & 0.074 & 0.031 & -0.009 & -0.016 & 0.087 & 0.069 & 0.049 & 0.078 & 0.069 & 0.059 & 0.066 & 0.038 & 0.006 & 0.006 & 0.043 & 0.082 & 0.05 & \textbf{0.121} & 0.021 & 0.017 & \textbf{0.107} & 0.096 & \textbf{-0.13} & \textbf{-0.117} & -0.073 & -0.077 & -0.044 & --- \\ 
   \bottomrule
\multicolumn{46}{c}{\emph{Note:} Correlation residuals in absolute $> 0.1$ are marked}\\
\end{tabular}
\endgroup
}

\subfloat[Standardized residuals]{%
\label{tab:residualsMMstandardized}
\centering
\begingroup\footnotesize
\begin{tabular}{rlllllllllllllllllllllllllllllllllllllllllllll}
  \toprule
 & 1 & 2 & 3 & 4 & 5 & 6 & 7 & 8 & 9 & 10 & 11 & 12 & 13 & 14 & 15 & 16 & 17 & 18 & 19 & 20 & 21 & 22 & 23 & 24 & 25 & 26 & 27 & 28 & 29 & 30 & 31 & 32 & 33 & 34 & 35 & 36 & 37 & 38 & 39 & 40 & 41 & 42 & 43 & 44 & 45 \\ 
  \midrule
1. MKF1 & --- &  &  &  &  &  &  &  &  &  &  &  &  &  &  &  &  &  &  &  &  &  &  &  &  &  &  &  &  &  &  &  &  &  &  &  &  &  &  &  &  &  &  &  &  \\ 
  2. MKF2 & 1.339 & --- &  &  &  &  &  &  &  &  &  &  &  &  &  &  &  &  &  &  &  &  &  &  &  &  &  &  &  &  &  &  &  &  &  &  &  &  &  &  &  &  &  &  &  \\ 
  3. MKF3 & 1.028 & 1.158 & --- &  &  &  &  &  &  &  &  &  &  &  &  &  &  &  &  &  &  &  &  &  &  &  &  &  &  &  &  &  &  &  &  &  &  &  &  &  &  &  &  &  &  \\ 
  4. MKF4 & -0.876 & \textbf{-2.178} & -1.498 & --- &  &  &  &  &  &  &  &  &  &  &  &  &  &  &  &  &  &  &  &  &  &  &  &  &  &  &  &  &  &  &  &  &  &  &  &  &  &  &  &  &  \\ 
  5. MKTS1 & -1.338 & -1.85 & \textbf{-3.102} & \textbf{2.418} & --- &  &  &  &  &  &  &  &  &  &  &  &  &  &  &  &  &  &  &  &  &  &  &  &  &  &  &  &  &  &  &  &  &  &  &  &  &  &  &  &  \\ 
  6. MKTS2 & -0.43 & -1.862 & \textbf{-2.718} & \textbf{2.683} & \textbf{4.009} & --- &  &  &  &  &  &  &  &  &  &  &  &  &  &  &  &  &  &  &  &  &  &  &  &  &  &  &  &  &  &  &  &  &  &  &  &  &  &  &  \\ 
  7. MKTS3 & 0.89 & -0.362 & -0.15 & \textbf{4.793} & \textbf{-2.782} & \textbf{-3.039} & --- &  &  &  &  &  &  &  &  &  &  &  &  &  &  &  &  &  &  &  &  &  &  &  &  &  &  &  &  &  &  &  &  &  &  &  &  &  &  \\ 
  8. ctrl1 & \textbf{1.993} & 1.64 & 1.424 & 1.547 & \textbf{2.685} & \textbf{2.99} & 1.888 & --- &  &  &  &  &  &  &  &  &  &  &  &  &  &  &  &  &  &  &  &  &  &  &  &  &  &  &  &  &  &  &  &  &  &  &  &  &  \\ 
  9. ctrl2 & \textbf{-2.506} & -0.661 & 0.182 & 1.546 & -0.275 & 1.058 & 0.824 & \textbf{6.548} & --- &  &  &  &  &  &  &  &  &  &  &  &  &  &  &  &  &  &  &  &  &  &  &  &  &  &  &  &  &  &  &  &  &  &  &  &  \\ 
  10. ctrl3 & -1.413 & -1.044 & -1.285 & \textbf{-2.382} & \textbf{-4.498} & \textbf{-3.51} & -0.625 & \textbf{-5.764} & \textbf{-4.944} & --- &  &  &  &  &  &  &  &  &  &  &  &  &  &  &  &  &  &  &  &  &  &  &  &  &  &  &  &  &  &  &  &  &  &  &  \\ 
  11. awa1 & \textbf{-2.17} & 0.153 & -0.486 & 0.557 & \textbf{-1.996} & -0.648 & 0.642 & -0.84 & -0.55 & \textbf{3.951} & --- &  &  &  &  &  &  &  &  &  &  &  &  &  &  &  &  &  &  &  &  &  &  &  &  &  &  &  &  &  &  &  &  &  &  \\ 
  12. awa2 & -1.293 & -0.613 & -0.765 & 1.713 & -0.684 & 0.581 & 1.373 & -0.704 & 0.035 & \textbf{4.017} & \textbf{5.646} & --- &  &  &  &  &  &  &  &  &  &  &  &  &  &  &  &  &  &  &  &  &  &  &  &  &  &  &  &  &  &  &  &  &  \\ 
  13. awa3 & 0.262 & 0.629 & -0.208 & \textbf{2.112} & -0.446 & 0.372 & 0.791 & \textbf{-3.802} & \textbf{-2.627} & 1.031 & \textbf{-4.917} & \textbf{-4.306} & --- &  &  &  &  &  &  &  &  &  &  &  &  &  &  &  &  &  &  &  &  &  &  &  &  &  &  &  &  &  &  &  &  \\ 
  14. coll1 & -1.301 & -1.503 & -1.174 & -1.641 & -1.206 & -0.736 & -0.535 & \textbf{-6.84} & \textbf{-5.545} & \textbf{4.93} & \textbf{-5.594} & \textbf{-3.975} & \textbf{2.716} & --- &  &  &  &  &  &  &  &  &  &  &  &  &  &  &  &  &  &  &  &  &  &  &  &  &  &  &  &  &  &  &  \\ 
  15. coll2 & 0.897 & 0.667 & 0.199 & 0.674 & -0.443 & 0.016 & 1.265 & -1.796 & -1.8 & \textbf{7.432} & -1.878 & 0.107 & \textbf{3.909} & \textbf{3.695} & --- &  &  &  &  &  &  &  &  &  &  &  &  &  &  &  &  &  &  &  &  &  &  &  &  &  &  &  &  &  &  \\ 
  16. coll3 & 1.46 & 0.899 & 1.046 & 0.183 & -0.136 & -0.369 & 0.992 & \textbf{-3.966} & \textbf{-4.661} & \textbf{6.836} & \textbf{-5.226} & \textbf{-3.231} & \textbf{4.414} & \textbf{2.115} & -1.181 & --- &  &  &  &  &  &  &  &  &  &  &  &  &  &  &  &  &  &  &  &  &  &  &  &  &  &  &  &  &  \\ 
  17. coll4 & -0.203 & 0.33 & -0.718 & 0.281 & -0.575 & 1.336 & 0.586 & \textbf{-2.924} & \textbf{-2.251} & \textbf{8.485} & -0.853 & -0.659 & \textbf{6.328} & \textbf{-1.976} & \textbf{-5.762} & 0.906 & --- &  &  &  &  &  &  &  &  &  &  &  &  &  &  &  &  &  &  &  &  &  &  &  &  &  &  &  &  \\ 
  18. STPE1 & 0.597 & 0.46 & 1.684 & 1.467 & -1.027 & -0.595 & \textbf{2.675} & \textbf{2.461} & 1.122 & \textbf{3.663} & \textbf{4.221} & \textbf{2.828} & \textbf{2.326} & 0.837 & \textbf{2.567} & \textbf{2.713} & \textbf{4.72} & --- &  &  &  &  &  &  &  &  &  &  &  &  &  &  &  &  &  &  &  &  &  &  &  &  &  &  &  \\ 
  19. STPE2 & -0.149 & -0.604 & 1.402 & -1.225 & -0.28 & -0.059 & 1.216 & 1.014 & -0.985 & 1.221 & -0.433 & \textbf{-2.201} & \textbf{2.907} & -0.244 & 0.447 & 0.777 & \textbf{3.062} & \textbf{-2.851} & --- &  &  &  &  &  &  &  &  &  &  &  &  &  &  &  &  &  &  &  &  &  &  &  &  &  &  \\ 
  20. STPE3 & 0.217 & 0.21 & -1.699 & -0.161 & -1.165 & -0.593 & 1.713 & -1.307 & \textbf{-3.056} & 0.966 & -1.595 & \textbf{-3.93} & -0.444 & \textbf{-2.852} & -1.62 & \textbf{-3.202} & -1.172 & \textbf{-14.01} & \textbf{-2.709} & --- &  &  &  &  &  &  &  &  &  &  &  &  &  &  &  &  &  &  &  &  &  &  &  &  &  \\ 
  21. STPE4 & -0.264 & -0.551 & -0.719 & -0.989 & -1.448 & -1.056 & 0.952 & \textbf{-2.331} & \textbf{-3.56} & -0.076 & \textbf{-2.431} & \textbf{-3.645} & 0.459 & \textbf{-2.47} & \textbf{-2.333} & \textbf{-2.768} & -0.343 & \textbf{-12.454} & \textbf{-3.684} & \textbf{12.114} & --- &  &  &  &  &  &  &  &  &  &  &  &  &  &  &  &  &  &  &  &  &  &  &  &  \\ 
  22. TAU1 & -0.83 & 0.257 & 0.345 & -0.411 & -0.636 & -0.936 & 1.383 & \textbf{2.187} & 1.56 & \textbf{1.986} & \textbf{2.276} & 1.232 & -0.13 & -1.33 & \textbf{3.135} & 1.695 & \textbf{3.465} & \textbf{8.555} & 1.853 & \textbf{-2.107} & \textbf{-3.245} & --- &  &  &  &  &  &  &  &  &  &  &  &  &  &  &  &  &  &  &  &  &  &  &  \\ 
  23. TAU2 & -1.38 & 0.819 & -0.14 & 0.034 & -1.498 & \textbf{-2.139} & 0.607 & 0.339 & 0.09 & -0.65 & 0.922 & -0.412 & \textbf{-2.738} & \textbf{-3.848} & 0.605 & 0.364 & 1.195 & \textbf{5.187} & \textbf{-2.022} & \textbf{-4.756} & \textbf{-5.073} & \textbf{-2.204} & --- &  &  &  &  &  &  &  &  &  &  &  &  &  &  &  &  &  &  &  &  &  &  \\ 
  24. TAU3 & -1.102 & -0.422 & -0.377 & 1.168 & -0.359 & -1.721 & 1.393 & -0.671 & -1.198 & -1.078 & 1.108 & 1.147 & \textbf{-2.265} & \textbf{-4.734} & -0.35 & -1.689 & 0.232 & \textbf{2.985} & \textbf{-4.476} & \textbf{-7.794} & \textbf{-7.891} & -1.876 & \textbf{9.063} & --- &  &  &  &  &  &  &  &  &  &  &  &  &  &  &  &  &  &  &  &  &  \\ 
  25. TAU4 & 0.511 & 0.463 & 0.969 & 0.069 & 1.389 & 1.11 & 0.667 & -0.869 & -1.356 & -0.526 & 0.683 & 0.891 & -1.599 & \textbf{-2.365} & 0.645 & 0.855 & \textbf{2.215} & \textbf{6.587} & \textbf{3.233} & 0.443 & -0.883 & -1.723 & \textbf{-8.955} & \textbf{-5.57} & --- &  &  &  &  &  &  &  &  &  &  &  &  &  &  &  &  &  &  &  &  \\ 
  26. TAE1 & -0.544 & 0.489 & -0.683 & -0.415 & -0.411 & -1.209 & \textbf{2.105} & 0.225 & 0.73 & 0.083 & -1.908 & -1.696 & \textbf{2.621} & -0.627 & 1.465 & 0.106 & 0.411 & \textbf{3.353} & \textbf{4.257} & \textbf{3.906} & \textbf{2.347} & \textbf{3.176} & -0.894 & 0.18 & \textbf{8.838} & --- &  &  &  &  &  &  &  &  &  &  &  &  &  &  &  &  &  &  &  \\ 
  27. TAE2 & -0.764 & 0.531 & -0.458 & \textbf{1.963} & \textbf{-2.873} & -1.868 & -1.086 & -0.361 & 0.947 & \textbf{2.084} & 1.489 & 0.647 & 0.587 & -1.378 & 0.563 & -1.038 & 0.676 & -0.008 & \textbf{-2.491} & 0.48 & -1.197 & \textbf{-2.21} & \textbf{-4.728} & \textbf{-3.156} & \textbf{3.297} & -1.331 & --- &  &  &  &  &  &  &  &  &  &  &  &  &  &  &  &  &  &  \\ 
  28. TAE3 & -0.129 & -0.1 & 1.002 & 0.66 & -0.375 & 0.161 & \textbf{2.706} & -0.676 & -0.153 & -1.149 & -1.101 & -1.805 & 0.376 & -1.887 & -0.547 & -0.608 & -0.159 & -1.767 & -1.648 & 1.077 & 0.472 & -1.732 & -0.973 & -1.105 & \textbf{4.049} & \textbf{-4.826} & \textbf{3.133} & --- &  &  &  &  &  &  &  &  &  &  &  &  &  &  &  &  &  \\ 
  29. TAE4 & -0.914 & -0.179 & 0.1 & -0.379 & 0.628 & 1.302 & 1.059 & -0.394 & -0.517 & -0.839 & -1 & -1.381 & \textbf{2.461} & -0.242 & 0.32 & 1.611 & 1.266 & \textbf{-3.697} & \textbf{-2.064} & \textbf{-2.586} & \textbf{-3.312} & \textbf{-2.032} & \textbf{-6.53} & \textbf{-5.266} & 0.915 & \textbf{-6.198} & \textbf{3.592} & \textbf{4.195} & --- &  &  &  &  &  &  &  &  &  &  &  &  &  &  &  &  \\ 
  30. STBI1 & -1.453 & -0.123 & 0.229 & -0.739 & -1.478 & -1.107 & \textbf{1.979} & -0.245 & -0.926 & 0.396 & 0.339 & \textbf{-2.131} & \textbf{2.332} & -0.418 & -0.25 & 0.538 & \textbf{2.019} & \textbf{4.989} & 1.588 & \textbf{-9.508} & \textbf{-9.724} & 1.827 & \textbf{-5.306} & \textbf{-6.595} & \textbf{2.908} & 1.291 & 0.227 & \textbf{-2.315} & \textbf{-2.204} & --- &  &  &  &  &  &  &  &  &  &  &  &  &  &  &  \\ 
  31. STBI2 & -1.108 & \textbf{2.592} & 1.392 & -0.888 & 0.601 & 0.496 & \textbf{2.731} & 0.828 & -0.755 & -0.356 & -1.354 & \textbf{-3.094} & \textbf{2.913} & 0.367 & -1.86 & 1.27 & \textbf{3.511} & \textbf{4.739} & \textbf{8.83} & \textbf{-3.247} & \textbf{-3.261} & 1.402 & \textbf{-3.349} & \textbf{-5.984} & 1.486 & \textbf{4.976} & -0.281 & -0.645 & 0.052 & -0.01 & --- &  &  &  &  &  &  &  &  &  &  &  &  &  &  \\ 
  32. STBI4 & 0.089 & -0.092 & -0.031 & 0.152 & \textbf{-2.156} & \textbf{-2.557} & 0.888 & 0.813 & -0.408 & 0.541 & 0.404 & -1.234 & 1.136 & \textbf{-3.257} & -0.816 & -1.115 & 0.075 & \textbf{8.546} & 0.44 & \textbf{-8.107} & \textbf{-9.305} & \textbf{5.361} & -0.636 & -1.492 & \textbf{5.696} & 1.108 & 1.425 & \textbf{-2.083} & \textbf{-3.648} & \textbf{4.571} & \textbf{-8.612} & --- &  &  &  &  &  &  &  &  &  &  &  &  &  \\ 
  33. TAR1 & -0.684 & -1.407 & 0.135 & 0.528 & -0.376 & 0.247 & \textbf{2.762} & -0.915 & -0.869 & -1.648 & -1.52 & -0.365 & 1.784 & -0.387 & 0.769 & 1.733 & \textbf{2.213} & \textbf{2.542} & \textbf{2.076} & -1.148 & -1.078 & 0.144 & -1.533 & \textbf{-2.222} & 1.699 & \textbf{-3.663} & 0.255 & 1.896 & \textbf{6.717} & 0.573 & \textbf{2.405} & -0.045 & --- &  &  &  &  &  &  &  &  &  &  &  &  \\ 
  34. TAR2 & 0.207 & 0.474 & 0.096 & 0.902 & \textbf{-2.55} & -1.872 & 0.697 & \textbf{2.012} & \textbf{1.988} & 0.485 & 0.768 & 0.583 & 0.791 & \textbf{-2.195} & 0.443 & 0.11 & 1.036 & \textbf{1.976} & 0.419 & \textbf{-2.248} & \textbf{-3.974} & 0.902 & \textbf{-2.447} & \textbf{-3.685} & 0.546 & \textbf{-7.328} & 0.555 & \textbf{-4.638} & 1.48 & 0.445 & 0.813 & 0.436 & \textbf{3.988} & --- &  &  &  &  &  &  &  &  &  &  &  \\ 
  35. TAR3 & -0.81 & 0.887 & -0.458 & 0.287 & -1.465 & 0.291 & 1.759 & 0.24 & -0.023 & -0.823 & -0.031 & -0.861 & -1.041 & \textbf{-3.434} & 0.376 & -0.89 & 0.069 & \textbf{2.09} & 0.637 & -0.418 & \textbf{-2.699} & \textbf{2.766} & -0.067 & -1.192 & \textbf{3.477} & 0.97 & 0.43 & \textbf{-2.648} & -0.115 & \textbf{-3.038} & -0.238 & -1.484 & \textbf{-11.359} & \textbf{5.093} & --- &  &  &  &  &  &  &  &  &  &  \\ 
  36. STFC1 & \textbf{-2.488} & \textbf{-2.79} & -0.877 & 1.076 & \textbf{-3.892} & \textbf{-2.527} & 0.941 & 0.012 & 1.056 & 0.673 & \textbf{1.964} & -0.094 & 0.13 & -0.075 & \textbf{2.196} & \textbf{2.589} & 1.493 & -0.829 & -1.004 & -1.103 & \textbf{-2.487} & 0.184 & -1.317 & -1.719 & -0.73 & -1.355 & \textbf{2.603} & -0.676 & \textbf{2.233} & \textbf{-3.16} & \textbf{-3.089} & \textbf{-4.042} & \textbf{-2.088} & \textbf{2.855} & -0.749 & --- &  &  &  &  &  &  &  &  &  \\ 
  37. STFC2 & -1.848 & -1.112 & -0.97 & -1.257 & \textbf{-1.997} & \textbf{-3.004} & 1.702 & -0.712 & 0.557 & 0.076 & -1.065 & -0.771 & 0.422 & -1.818 & \textbf{2.375} & 1.177 & 0.769 & -1.556 & -0.376 & -1.431 & -1.235 & -1.739 & \textbf{-2.403} & \textbf{-5.324} & -0.517 & 1.532 & \textbf{6.298} & \textbf{2.533} & \textbf{5.027} & \textbf{-2.485} & \textbf{-3.118} & \textbf{-4.813} & -1.167 & \textbf{5.416} & 1.441 & \textbf{10.41} & --- &  &  &  &  &  &  &  &  \\ 
  38. STFC3 & 1.148 & 1.741 & 0.806 & \textbf{2.175} & -0.252 & 0.241 & 1.557 & 0.738 & 1.06 & \textbf{-2.14} & -0.642 & -0.51 & -0.476 & \textbf{-2.253} & 0.062 & 0.618 & -0.773 & \textbf{2.698} & \textbf{2.977} & -1.579 & -0.705 & \textbf{2.641} & \textbf{3.298} & \textbf{2.882} & 1.181 & \textbf{-4.789} & \textbf{-3.704} & \textbf{-2.715} & \textbf{-2.44} & \textbf{5.32} & \textbf{2.794} & \textbf{2.302} & -1.689 & -0.816 & \textbf{-4.466} & -1.7 & \textbf{-5.714} & --- &  &  &  &  &  &  &  \\ 
  39. STFC4 & 1.616 & -0.028 & 0.957 & \textbf{2.006} & \textbf{2.138} & 1.845 & \textbf{2.723} & -0.055 & 1.359 & \textbf{-3.24} & -0.612 & 0.369 & 1.342 & \textbf{-2.925} & -1.17 & -1.205 & -0.124 & 1.446 & \textbf{3.021} & 0.313 & 1.777 & \textbf{2.115} & -0.67 & -0.156 & \textbf{2.583} & \textbf{-2.08} & \textbf{-4.141} & \textbf{-2.757} & -0.426 & \textbf{4.028} & \textbf{2.506} & 1.637 & 1.007 & 0.535 & -1.508 & \textbf{-6.779} & \textbf{-9.994} & -1.164 & --- &  &  &  &  &  &  \\ 
  40. STT1 & -0.45 & \textbf{-2} & -0.514 & \textbf{-3.288} & \textbf{-5.522} & \textbf{-4.837} & \textbf{-2.019} & 1.497 & 1.326 & \textbf{-2.102} & 1.661 & 0.347 & 1.618 & -1.236 & -0.425 & \textbf{3.153} & 1.735 & \textbf{4.285} & \textbf{4.409} & -1.396 & \textbf{2.236} & 1.86 & -0.092 & 0.752 & \textbf{3.224} & \textbf{2.462} & \textbf{-3.661} & -1.574 & \textbf{-4.607} & 0.958 & \textbf{2.853} & \textbf{3.449} & -1.594 & \textbf{-6.833} & -0.532 & \textbf{-5.561} & \textbf{-5.186} & 1.887 & 1.376 & --- &  &  &  &  &  \\ 
  41. STT2 & -0.658 & \textbf{-2.176} & \textbf{-2.052} & \textbf{-2.533} & \textbf{-3.583} & \textbf{-4.555} & -0.477 & 1.808 & \textbf{1.987} & \textbf{-2.913} & 0.635 & -0.503 & 0.253 & -1.779 & -0.306 & 1.419 & 0.378 & 1.393 & 1.673 & \textbf{-4.544} & -1.676 & -1.831 & \textbf{-4.174} & \textbf{-1.996} & \textbf{2.078} & -0.143 & \textbf{-5.331} & \textbf{-2.298} & \textbf{-5.98} & -1.148 & 1.446 & 1.333 & \textbf{-2.83} & \textbf{-7.989} & -1.059 & \textbf{-4.078} & \textbf{-4.403} & \textbf{3.088} & 1.937 & \textbf{7.602} & --- &  &  &  &  \\ 
  42. STT3 & 0.805 & 1.555 & 1.868 & -1.245 & \textbf{3.114} & \textbf{3.098} & \textbf{3.676} & -0.408 & 0.203 & \textbf{-3.886} & -0.992 & -0.936 & 0.518 & -1.657 & -1.757 & 1.173 & 0.091 & \textbf{-2.63} & -1.49 & \textbf{-3.088} & \textbf{-2.834} & \textbf{-2.578} & \textbf{-3.5} & -1.924 & -0.219 & 1.716 & \textbf{-5.65} & -1.779 & -1.683 & \textbf{-4.037} & \textbf{-2.119} & \textbf{-3.769} & -0.725 & \textbf{-3.659} & \textbf{3.288} & \textbf{-4.901} & \textbf{-4.919} & 1.007 & \textbf{3.435} & \textbf{-5.929} & \textbf{2.259} & --- &  &  &  \\ 
  43. STT4 & \textbf{3.143} & 1.849 & 1.051 & 0.753 & 1.474 & 1.903 & \textbf{2.916} & 1.791 & 1.892 & \textbf{-3.191} & -0.293 & -0.636 & \textbf{3.396} & -0.834 & -1.688 & 0.844 & -0.058 & -1.217 & 0.38 & \textbf{-2.065} & 1.17 & -0.777 & \textbf{-2.163} & \textbf{-2.319} & -1.114 & \textbf{6.8} & \textbf{-2.64} & 0.91 & 0.38 & \textbf{-2.958} & 0.788 & \textbf{-4.021} & 1.836 & \textbf{-3.244} & \textbf{4.209} & \textbf{-3.881} & \textbf{-2.682} & \textbf{3.541} & \textbf{4.609} & \textbf{-9.636} & \textbf{-5.454} & \textbf{9.651} & --- &  &  \\ 
  44. STT5 & 1.185 & 1.242 & 0.219 & -1.067 & \textbf{3.41} & \textbf{3.725} & \textbf{4.699} & -1.057 & -1.307 & \textbf{-3.582} & \textbf{-4.9} & \textbf{-5.143} & \textbf{-1.974} & -0.797 & \textbf{-3.699} & -0.431 & -0.513 & \textbf{-4.237} & \textbf{1.989} & \textbf{-2.359} & 0.025 & \textbf{-4.994} & \textbf{-4.69} & \textbf{-4.54} & \textbf{-2.579} & \textbf{2.992} & \textbf{-2.095} & 0.709 & 0.686 & -1.414 & \textbf{2.759} & \textbf{-3.941} & -0.129 & \textbf{-3.158} & -1.065 & \textbf{-3.641} & \textbf{-2.185} & 0.163 & 1.879 & \textbf{-3.391} & -1.718 & \textbf{7.407} & \textbf{7.571} & --- &  \\ 
  45. STT6 & -0.297 & 0.889 & 1.851 & -0.034 & \textbf{-2.636} & \textbf{-3.036} & 0.471 & \textbf{2.617} & \textbf{3.365} & 0.115 & \textbf{2.317} & \textbf{2.296} & -0.509 & -1.261 & \textbf{2.338} & 1.603 & 1.5 & \textbf{4.47} & 1.8 & -0.566 & -0.949 & \textbf{6.289} & \textbf{5.045} & \textbf{3.275} & \textbf{5.109} & \textbf{4.679} & \textbf{3.642} & \textbf{4.368} & \textbf{2.23} & 0.39 & 0.408 & \textbf{3.006} & \textbf{4.807} & \textbf{3.011} & \textbf{8.409} & 1.122 & 1.04 & \textbf{7.43} & \textbf{5.611} & \textbf{-11.042} & \textbf{-10.205} & \textbf{-6.206} & \textbf{-6.203} & \textbf{-2.592} & --- \\ 
   \bottomrule
\multicolumn{46}{c}{\emph{Note:} Statistically significant residuals ($\mathsf{abs} > 1.96$) are marked}\\
\end{tabular}
\endgroup
}
\end{table*}
}
\newcommand{\residualsMMstd}{
\begin{table*}[ht]
\centering
\caption{Standardized residuals of the fitted model} 
\label{tab:residualsMMstd}
\begingroup\footnotesize
\begin{tabular}{rlllllllllllllllllllllllllllllllllllllllllllll}
  \toprule
 & MKF1 & MKF2 & MKF3 & MKF4 & MKTS1 & MKTS2 & MKTS3 & ctrl1 & ctrl2 & ctrl3 & awa1 & awa2 & awa3 & coll1 & coll2 & coll3 & coll4 & STPE1 & STPE2 & STPE3 & STPE4 & TAU1 & TAU2 & TAU3 & TAU4 & TAE1 & TAE2 & TAE3 & TAE4 & STBI1 & STBI2 & STBI4 & TAR1 & TAR2 & TAR3 & STFC1 & STFC2 & STFC3 & STFC4 & STT1 & STT2 & STT3 & STT4 & STT5 & STT6 \\ 
  \midrule
MKF1 &  &  &  &  &  &  &  &  &  &  &  &  &  &  &  &  &  &  &  &  &  &  &  &  &  &  &  &  &  &  &  &  &  &  &  &  &  &  &  &  &  &  &  &  &  \\ 
  MKF2 & 1.34 &  &  &  &  &  &  &  &  &  &  &  &  &  &  &  &  &  &  &  &  &  &  &  &  &  &  &  &  &  &  &  &  &  &  &  &  &  &  &  &  &  &  &  &  \\ 
  MKF3 & 1.03 & 1.16 &  &  &  &  &  &  &  &  &  &  &  &  &  &  &  &  &  &  &  &  &  &  &  &  &  &  &  &  &  &  &  &  &  &  &  &  &  &  &  &  &  &  &  \\ 
  MKF4 & -0.88 & \textbf{-2.18} & -1.5 &  &  &  &  &  &  &  &  &  &  &  &  &  &  &  &  &  &  &  &  &  &  &  &  &  &  &  &  &  &  &  &  &  &  &  &  &  &  &  &  &  &  \\ 
  MKTS1 & -1.34 & -1.85 & \textbf{-3.1} & \textbf{2.42} &  &  &  &  &  &  &  &  &  &  &  &  &  &  &  &  &  &  &  &  &  &  &  &  &  &  &  &  &  &  &  &  &  &  &  &  &  &  &  &  &  \\ 
  MKTS2 & -0.43 & -1.86 & \textbf{-2.72} & \textbf{2.68} & \textbf{4.01} &  &  &  &  &  &  &  &  &  &  &  &  &  &  &  &  &  &  &  &  &  &  &  &  &  &  &  &  &  &  &  &  &  &  &  &  &  &  &  &  \\ 
  MKTS3 & 0.89 & -0.36 & -0.15 & \textbf{4.79} & \textbf{-2.78} & \textbf{-3.04} &  &  &  &  &  &  &  &  &  &  &  &  &  &  &  &  &  &  &  &  &  &  &  &  &  &  &  &  &  &  &  &  &  &  &  &  &  &  &  \\ 
  ctrl1 & \textbf{1.99} & 1.64 & 1.42 & 1.55 & \textbf{2.68} & \textbf{2.99} & 1.89 &  &  &  &  &  &  &  &  &  &  &  &  &  &  &  &  &  &  &  &  &  &  &  &  &  &  &  &  &  &  &  &  &  &  &  &  &  &  \\ 
  ctrl2 & \textbf{-2.51} & -0.66 & 0.18 & 1.55 & -0.27 & 1.06 & 0.82 & \textbf{6.55} &  &  &  &  &  &  &  &  &  &  &  &  &  &  &  &  &  &  &  &  &  &  &  &  &  &  &  &  &  &  &  &  &  &  &  &  &  \\ 
  ctrl3 & -1.41 & -1.04 & -1.29 & \textbf{-2.38} & \textbf{-4.5} & \textbf{-3.51} & -0.62 & \textbf{-5.76} & \textbf{-4.94} &  &  &  &  &  &  &  &  &  &  &  &  &  &  &  &  &  &  &  &  &  &  &  &  &  &  &  &  &  &  &  &  &  &  &  &  \\ 
  awa1 & \textbf{-2.17} & 0.15 & -0.49 & 0.56 & \textbf{-2} & -0.65 & 0.64 & -0.84 & -0.55 & \textbf{3.95} &  &  &  &  &  &  &  &  &  &  &  &  &  &  &  &  &  &  &  &  &  &  &  &  &  &  &  &  &  &  &  &  &  &  &  \\ 
  awa2 & -1.29 & -0.61 & -0.76 & 1.71 & -0.68 & 0.58 & 1.37 & -0.7 & 0.03 & \textbf{4.02} & \textbf{5.65} &  &  &  &  &  &  &  &  &  &  &  &  &  &  &  &  &  &  &  &  &  &  &  &  &  &  &  &  &  &  &  &  &  &  \\ 
  awa3 & 0.26 & 0.63 & -0.21 & \textbf{2.11} & -0.45 & 0.37 & 0.79 & \textbf{-3.8} & \textbf{-2.63} & 1.03 & \textbf{-4.92} & \textbf{-4.31} &  &  &  &  &  &  &  &  &  &  &  &  &  &  &  &  &  &  &  &  &  &  &  &  &  &  &  &  &  &  &  &  &  \\ 
  coll1 & -1.3 & -1.5 & -1.17 & -1.64 & -1.21 & -0.74 & -0.54 & \textbf{-6.84} & \textbf{-5.55} & \textbf{4.93} & \textbf{-5.59} & \textbf{-3.98} & \textbf{2.72} &  &  &  &  &  &  &  &  &  &  &  &  &  &  &  &  &  &  &  &  &  &  &  &  &  &  &  &  &  &  &  &  \\ 
  coll2 & 0.9 & 0.67 & 0.2 & 0.67 & -0.44 & 0.02 & 1.26 & -1.8 & -1.8 & \textbf{7.43} & -1.88 & 0.11 & \textbf{3.91} & \textbf{3.7} &  &  &  &  &  &  &  &  &  &  &  &  &  &  &  &  &  &  &  &  &  &  &  &  &  &  &  &  &  &  &  \\ 
  coll3 & 1.46 & 0.9 & 1.05 & 0.18 & -0.14 & -0.37 & 0.99 & \textbf{-3.97} & \textbf{-4.66} & \textbf{6.84} & \textbf{-5.23} & \textbf{-3.23} & \textbf{4.41} & \textbf{2.12} & -1.18 &  &  &  &  &  &  &  &  &  &  &  &  &  &  &  &  &  &  &  &  &  &  &  &  &  &  &  &  &  &  \\ 
  coll4 & -0.2 & 0.33 & -0.72 & 0.28 & -0.58 & 1.34 & 0.59 & \textbf{-2.92} & \textbf{-2.25} & \textbf{8.49} & -0.85 & -0.66 & \textbf{6.33} & \textbf{-1.98} & \textbf{-5.76} & 0.91 &  &  &  &  &  &  &  &  &  &  &  &  &  &  &  &  &  &  &  &  &  &  &  &  &  &  &  &  &  \\ 
  STPE1 & 0.6 & 0.46 & 1.68 & 1.47 & -1.03 & -0.6 & \textbf{2.68} & \textbf{2.46} & 1.12 & \textbf{3.66} & \textbf{4.22} & \textbf{2.83} & \textbf{2.33} & 0.84 & \textbf{2.57} & \textbf{2.71} & \textbf{4.72} &  &  &  &  &  &  &  &  &  &  &  &  &  &  &  &  &  &  &  &  &  &  &  &  &  &  &  &  \\ 
  STPE2 & -0.15 & -0.6 & 1.4 & -1.23 & -0.28 & -0.06 & 1.22 & 1.01 & -0.99 & 1.22 & -0.43 & \textbf{-2.2} & \textbf{2.91} & -0.24 & 0.45 & 0.78 & \textbf{3.06} & \textbf{-2.85} &  &  &  &  &  &  &  &  &  &  &  &  &  &  &  &  &  &  &  &  &  &  &  &  &  &  &  \\ 
  STPE3 & 0.22 & 0.21 & -1.7 & -0.16 & -1.16 & -0.59 & 1.71 & -1.31 & \textbf{-3.06} & 0.97 & -1.6 & \textbf{-3.93} & -0.44 & \textbf{-2.85} & -1.62 & \textbf{-3.2} & -1.17 & \textbf{-14.01} & \textbf{-2.71} &  &  &  &  &  &  &  &  &  &  &  &  &  &  &  &  &  &  &  &  &  &  &  &  &  &  \\ 
  STPE4 & -0.26 & -0.55 & -0.72 & -0.99 & -1.45 & -1.06 & 0.95 & \textbf{-2.33} & \textbf{-3.56} & -0.08 & \textbf{-2.43} & \textbf{-3.65} & 0.46 & \textbf{-2.47} & \textbf{-2.33} & \textbf{-2.77} & -0.34 & \textbf{-12.45} & \textbf{-3.68} & \textbf{12.11} &  &  &  &  &  &  &  &  &  &  &  &  &  &  &  &  &  &  &  &  &  &  &  &  &  \\ 
  TAU1 & -0.83 & 0.26 & 0.34 & -0.41 & -0.64 & -0.94 & 1.38 & \textbf{2.19} & 1.56 & \textbf{1.99} & \textbf{2.28} & 1.23 & -0.13 & -1.33 & \textbf{3.14} & 1.69 & \textbf{3.46} & \textbf{8.55} & 1.85 & \textbf{-2.11} & \textbf{-3.25} &  &  &  &  &  &  &  &  &  &  &  &  &  &  &  &  &  &  &  &  &  &  &  &  \\ 
  TAU2 & -1.38 & 0.82 & -0.14 & 0.03 & -1.5 & \textbf{-2.14} & 0.61 & 0.34 & 0.09 & -0.65 & 0.92 & -0.41 & \textbf{-2.74} & \textbf{-3.85} & 0.6 & 0.36 & 1.19 & \textbf{5.19} & \textbf{-2.02} & \textbf{-4.76} & \textbf{-5.07} & \textbf{-2.2} &  &  &  &  &  &  &  &  &  &  &  &  &  &  &  &  &  &  &  &  &  &  &  \\ 
  TAU3 & -1.1 & -0.42 & -0.38 & 1.17 & -0.36 & -1.72 & 1.39 & -0.67 & -1.2 & -1.08 & 1.11 & 1.15 & \textbf{-2.27} & \textbf{-4.73} & -0.35 & -1.69 & 0.23 & \textbf{2.98} & \textbf{-4.48} & \textbf{-7.79} & \textbf{-7.89} & -1.88 & \textbf{9.06} &  &  &  &  &  &  &  &  &  &  &  &  &  &  &  &  &  &  &  &  &  &  \\ 
  TAU4 & 0.51 & 0.46 & 0.97 & 0.07 & 1.39 & 1.11 & 0.67 & -0.87 & -1.36 & -0.53 & 0.68 & 0.89 & -1.6 & \textbf{-2.37} & 0.65 & 0.85 & \textbf{2.21} & \textbf{6.59} & \textbf{3.23} & 0.44 & -0.88 & -1.72 & \textbf{-8.96} & \textbf{-5.57} &  &  &  &  &  &  &  &  &  &  &  &  &  &  &  &  &  &  &  &  &  \\ 
  TAE1 & -0.54 & 0.49 & -0.68 & -0.42 & -0.41 & -1.21 & \textbf{2.11} & 0.23 & 0.73 & 0.08 & -1.91 & -1.7 & \textbf{2.62} & -0.63 & 1.46 & 0.11 & 0.41 & \textbf{3.35} & \textbf{4.26} & \textbf{3.91} & \textbf{2.35} & \textbf{3.18} & -0.89 & 0.18 & \textbf{8.84} &  &  &  &  &  &  &  &  &  &  &  &  &  &  &  &  &  &  &  &  \\ 
  TAE2 & -0.76 & 0.53 & -0.46 & \textbf{1.96} & \textbf{-2.87} & -1.87 & -1.09 & -0.36 & 0.95 & \textbf{2.08} & 1.49 & 0.65 & 0.59 & -1.38 & 0.56 & -1.04 & 0.68 & -0.01 & \textbf{-2.49} & 0.48 & -1.2 & \textbf{-2.21} & \textbf{-4.73} & \textbf{-3.16} & \textbf{3.3} & -1.33 &  &  &  &  &  &  &  &  &  &  &  &  &  &  &  &  &  &  &  \\ 
  TAE3 & -0.13 & -0.1 & 1 & 0.66 & -0.37 & 0.16 & \textbf{2.71} & -0.68 & -0.15 & -1.15 & -1.1 & -1.8 & 0.38 & -1.89 & -0.55 & -0.61 & -0.16 & -1.77 & -1.65 & 1.08 & 0.47 & -1.73 & -0.97 & -1.1 & \textbf{4.05} & \textbf{-4.83} & \textbf{3.13} &  &  &  &  &  &  &  &  &  &  &  &  &  &  &  &  &  &  \\ 
  TAE4 & -0.91 & -0.18 & 0.1 & -0.38 & 0.63 & 1.3 & 1.06 & -0.39 & -0.52 & -0.84 & -1 & -1.38 & \textbf{2.46} & -0.24 & 0.32 & 1.61 & 1.27 & \textbf{-3.7} & \textbf{-2.06} & \textbf{-2.59} & \textbf{-3.31} & \textbf{-2.03} & \textbf{-6.53} & \textbf{-5.27} & 0.92 & \textbf{-6.2} & \textbf{3.59} & \textbf{4.2} &  &  &  &  &  &  &  &  &  &  &  &  &  &  &  &  &  \\ 
  STBI1 & -1.45 & -0.12 & 0.23 & -0.74 & -1.48 & -1.11 & \textbf{1.98} & -0.24 & -0.93 & 0.4 & 0.34 & \textbf{-2.13} & \textbf{2.33} & -0.42 & -0.25 & 0.54 & \textbf{2.02} & \textbf{4.99} & 1.59 & \textbf{-9.51} & \textbf{-9.72} & 1.83 & \textbf{-5.31} & \textbf{-6.59} & \textbf{2.91} & 1.29 & 0.23 & \textbf{-2.32} & \textbf{-2.2} &  &  &  &  &  &  &  &  &  &  &  &  &  &  &  &  \\ 
  STBI2 & -1.11 & \textbf{2.59} & 1.39 & -0.89 & 0.6 & 0.5 & \textbf{2.73} & 0.83 & -0.76 & -0.36 & -1.35 & \textbf{-3.09} & \textbf{2.91} & 0.37 & -1.86 & 1.27 & \textbf{3.51} & \textbf{4.74} & \textbf{8.83} & \textbf{-3.25} & \textbf{-3.26} & 1.4 & \textbf{-3.35} & \textbf{-5.98} & 1.49 & \textbf{4.98} & -0.28 & -0.64 & 0.05 & -0.01 &  &  &  &  &  &  &  &  &  &  &  &  &  &  &  \\ 
  STBI4 & 0.09 & -0.09 & -0.03 & 0.15 & \textbf{-2.16} & \textbf{-2.56} & 0.89 & 0.81 & -0.41 & 0.54 & 0.4 & -1.23 & 1.14 & \textbf{-3.26} & -0.82 & -1.12 & 0.08 & \textbf{8.55} & 0.44 & \textbf{-8.11} & \textbf{-9.31} & \textbf{5.36} & -0.64 & -1.49 & \textbf{5.7} & 1.11 & 1.43 & \textbf{-2.08} & \textbf{-3.65} & \textbf{4.57} & \textbf{-8.61} &  &  &  &  &  &  &  &  &  &  &  &  &  &  \\ 
  TAR1 & -0.68 & -1.41 & 0.14 & 0.53 & -0.38 & 0.25 & \textbf{2.76} & -0.91 & -0.87 & -1.65 & -1.52 & -0.37 & 1.78 & -0.39 & 0.77 & 1.73 & \textbf{2.21} & \textbf{2.54} & \textbf{2.08} & -1.15 & -1.08 & 0.14 & -1.53 & \textbf{-2.22} & 1.7 & \textbf{-3.66} & 0.26 & 1.9 & \textbf{6.72} & 0.57 & \textbf{2.41} & -0.04 &  &  &  &  &  &  &  &  &  &  &  &  &  \\ 
  TAR2 & 0.21 & 0.47 & 0.1 & 0.9 & \textbf{-2.55} & -1.87 & 0.7 & \textbf{2.01} & \textbf{1.99} & 0.49 & 0.77 & 0.58 & 0.79 & \textbf{-2.19} & 0.44 & 0.11 & 1.04 & \textbf{1.98} & 0.42 & \textbf{-2.25} & \textbf{-3.97} & 0.9 & \textbf{-2.45} & \textbf{-3.68} & 0.55 & \textbf{-7.33} & 0.55 & \textbf{-4.64} & 1.48 & 0.45 & 0.81 & 0.44 & \textbf{3.99} &  &  &  &  &  &  &  &  &  &  &  &  \\ 
  TAR3 & -0.81 & 0.89 & -0.46 & 0.29 & -1.46 & 0.29 & 1.76 & 0.24 & -0.02 & -0.82 & -0.03 & -0.86 & -1.04 & \textbf{-3.43} & 0.38 & -0.89 & 0.07 & \textbf{2.09} & 0.64 & -0.42 & \textbf{-2.7} & \textbf{2.77} & -0.07 & -1.19 & \textbf{3.48} & 0.97 & 0.43 & \textbf{-2.65} & -0.11 & \textbf{-3.04} & -0.24 & -1.48 & \textbf{-11.36} & \textbf{5.09} &  &  &  &  &  &  &  &  &  &  &  \\ 
  STFC1 & \textbf{-2.49} & \textbf{-2.79} & -0.88 & 1.08 & \textbf{-3.89} & \textbf{-2.53} & 0.94 & 0.01 & 1.06 & 0.67 & \textbf{1.96} & -0.09 & 0.13 & -0.07 & \textbf{2.2} & \textbf{2.59} & 1.49 & -0.83 & -1 & -1.1 & \textbf{-2.49} & 0.18 & -1.32 & -1.72 & -0.73 & -1.36 & \textbf{2.6} & -0.68 & \textbf{2.23} & \textbf{-3.16} & \textbf{-3.09} & \textbf{-4.04} & \textbf{-2.09} & \textbf{2.85} & -0.75 &  &  &  &  &  &  &  &  &  &  \\ 
  STFC2 & -1.85 & -1.11 & -0.97 & -1.26 & \textbf{-2} & \textbf{-3} & 1.7 & -0.71 & 0.56 & 0.08 & -1.07 & -0.77 & 0.42 & -1.82 & \textbf{2.38} & 1.18 & 0.77 & -1.56 & -0.38 & -1.43 & -1.24 & -1.74 & \textbf{-2.4} & \textbf{-5.32} & -0.52 & 1.53 & \textbf{6.3} & \textbf{2.53} & \textbf{5.03} & \textbf{-2.48} & \textbf{-3.12} & \textbf{-4.81} & -1.17 & \textbf{5.42} & 1.44 & \textbf{10.41} &  &  &  &  &  &  &  &  &  \\ 
  STFC3 & 1.15 & 1.74 & 0.81 & \textbf{2.18} & -0.25 & 0.24 & 1.56 & 0.74 & 1.06 & \textbf{-2.14} & -0.64 & -0.51 & -0.48 & \textbf{-2.25} & 0.06 & 0.62 & -0.77 & \textbf{2.7} & \textbf{2.98} & -1.58 & -0.71 & \textbf{2.64} & \textbf{3.3} & \textbf{2.88} & 1.18 & \textbf{-4.79} & \textbf{-3.7} & \textbf{-2.71} & \textbf{-2.44} & \textbf{5.32} & \textbf{2.79} & \textbf{2.3} & -1.69 & -0.82 & \textbf{-4.47} & -1.7 & \textbf{-5.71} &  &  &  &  &  &  &  &  \\ 
  STFC4 & 1.62 & -0.03 & 0.96 & \textbf{2.01} & \textbf{2.14} & 1.85 & \textbf{2.72} & -0.05 & 1.36 & \textbf{-3.24} & -0.61 & 0.37 & 1.34 & \textbf{-2.93} & -1.17 & -1.2 & -0.12 & 1.45 & \textbf{3.02} & 0.31 & 1.78 & \textbf{2.11} & -0.67 & -0.16 & \textbf{2.58} & \textbf{-2.08} & \textbf{-4.14} & \textbf{-2.76} & -0.43 & \textbf{4.03} & \textbf{2.51} & 1.64 & 1.01 & 0.54 & -1.51 & \textbf{-6.78} & \textbf{-9.99} & -1.16 &  &  &  &  &  &  &  \\ 
  STT1 & -0.45 & \textbf{-2} & -0.51 & \textbf{-3.29} & \textbf{-5.52} & \textbf{-4.84} & \textbf{-2.02} & 1.5 & 1.33 & \textbf{-2.1} & 1.66 & 0.35 & 1.62 & -1.24 & -0.42 & \textbf{3.15} & 1.74 & \textbf{4.29} & \textbf{4.41} & -1.4 & \textbf{2.24} & 1.86 & -0.09 & 0.75 & \textbf{3.22} & \textbf{2.46} & \textbf{-3.66} & -1.57 & \textbf{-4.61} & 0.96 & \textbf{2.85} & \textbf{3.45} & -1.59 & \textbf{-6.83} & -0.53 & \textbf{-5.56} & \textbf{-5.19} & 1.89 & 1.38 &  &  &  &  &  &  \\ 
  STT2 & -0.66 & \textbf{-2.18} & \textbf{-2.05} & \textbf{-2.53} & \textbf{-3.58} & \textbf{-4.55} & -0.48 & 1.81 & \textbf{1.99} & \textbf{-2.91} & 0.63 & -0.5 & 0.25 & -1.78 & -0.31 & 1.42 & 0.38 & 1.39 & 1.67 & \textbf{-4.54} & -1.68 & -1.83 & \textbf{-4.17} & \textbf{-2} & \textbf{2.08} & -0.14 & \textbf{-5.33} & \textbf{-2.3} & \textbf{-5.98} & -1.15 & 1.45 & 1.33 & \textbf{-2.83} & \textbf{-7.99} & -1.06 & \textbf{-4.08} & \textbf{-4.4} & \textbf{3.09} & 1.94 & \textbf{7.6} &  &  &  &  &  \\ 
  STT3 & 0.8 & 1.56 & 1.87 & -1.24 & \textbf{3.11} & \textbf{3.1} & \textbf{3.68} & -0.41 & 0.2 & \textbf{-3.89} & -0.99 & -0.94 & 0.52 & -1.66 & -1.76 & 1.17 & 0.09 & \textbf{-2.63} & -1.49 & \textbf{-3.09} & \textbf{-2.83} & \textbf{-2.58} & \textbf{-3.5} & -1.92 & -0.22 & 1.72 & \textbf{-5.65} & -1.78 & -1.68 & \textbf{-4.04} & \textbf{-2.12} & \textbf{-3.77} & -0.72 & \textbf{-3.66} & \textbf{3.29} & \textbf{-4.9} & \textbf{-4.92} & 1.01 & \textbf{3.44} & \textbf{-5.93} & \textbf{2.26} &  &  &  &  \\ 
  STT4 & \textbf{3.14} & 1.85 & 1.05 & 0.75 & 1.47 & 1.9 & \textbf{2.92} & 1.79 & 1.89 & \textbf{-3.19} & -0.29 & -0.64 & \textbf{3.4} & -0.83 & -1.69 & 0.84 & -0.06 & -1.22 & 0.38 & \textbf{-2.07} & 1.17 & -0.78 & \textbf{-2.16} & \textbf{-2.32} & -1.11 & \textbf{6.8} & \textbf{-2.64} & 0.91 & 0.38 & \textbf{-2.96} & 0.79 & \textbf{-4.02} & 1.84 & \textbf{-3.24} & \textbf{4.21} & \textbf{-3.88} & \textbf{-2.68} & \textbf{3.54} & \textbf{4.61} & \textbf{-9.64} & \textbf{-5.45} & \textbf{9.65} &  &  &  \\ 
  STT5 & 1.18 & 1.24 & 0.22 & -1.07 & \textbf{3.41} & \textbf{3.72} & \textbf{4.7} & -1.06 & -1.31 & \textbf{-3.58} & \textbf{-4.9} & \textbf{-5.14} & \textbf{-1.97} & -0.8 & \textbf{-3.7} & -0.43 & -0.51 & \textbf{-4.24} & \textbf{1.99} & \textbf{-2.36} & 0.02 & \textbf{-4.99} & \textbf{-4.69} & \textbf{-4.54} & \textbf{-2.58} & \textbf{2.99} & \textbf{-2.1} & 0.71 & 0.69 & -1.41 & \textbf{2.76} & \textbf{-3.94} & -0.13 & \textbf{-3.16} & -1.07 & \textbf{-3.64} & \textbf{-2.19} & 0.16 & 1.88 & \textbf{-3.39} & -1.72 & \textbf{7.41} & \textbf{7.57} &  &  \\ 
  STT6 & -0.3 & 0.89 & 1.85 & -0.03 & \textbf{-2.64} & \textbf{-3.04} & 0.47 & \textbf{2.62} & \textbf{3.37} & 0.12 & \textbf{2.32} & \textbf{2.3} & -0.51 & -1.26 & \textbf{2.34} & 1.6 & 1.5 & \textbf{4.47} & 1.8 & -0.57 & -0.95 & \textbf{6.29} & \textbf{5.04} & \textbf{3.28} & \textbf{5.11} & \textbf{4.68} & \textbf{3.64} & \textbf{4.37} & \textbf{2.23} & 0.39 & 0.41 & \textbf{3.01} & \textbf{4.81} & \textbf{3.01} & \textbf{8.41} & 1.12 & 1.04 & \textbf{7.43} & \textbf{5.61} & \textbf{-11.04} & \textbf{-10.2} & \textbf{-6.21} & \textbf{-6.2} & \textbf{-2.59} &  \\ 
   \bottomrule
\end{tabular}
\endgroup
\end{table*}
}
\newcommand{\fitTrust}{
\begin{table*}[ht]
\centering
\caption{Fit measures for the WLSMV-estimation of the correlational SEM.} 
\label{tab:fitTrust}
\begingroup\footnotesize
\begin{tabular}{rrrrrrrrrr}
  \toprule
$\chi^2$ & $\mathit{df}$ & $p_{\chi^2}$ & \textsf{CFI} & \textsf{TFI} & \textsf{RMSEA} & \textsf{LL} & \textsf{UL} & $p_{\mathsf{rmsea}}$ & \textsf{SRMR} \\ 
  \midrule
4034.26 & 897.00 & $<.001$ & 0.96 & 0.96 & 0.07 & 0.06 & 0.07 & $<.001$ & 0.05 \\ 
   \bottomrule
\end{tabular}
\endgroup
\end{table*}
}
\newcommand{\fitTrustCausal}{
\begin{table*}[ht]
\centering
\caption{Fit measures for WLSMV-estimation of the causal SEM.} 
\label{tab:fitTrustCausal}
\begingroup\footnotesize
\begin{tabular}{rrrrrrrrrr}
  \toprule
$\chi^2$ & $\mathit{df}$ & $p_{\chi^2}$ & \textsf{CFI} & \textsf{TFI} & \textsf{RMSEA} & \textsf{LL} & \textsf{UL} & $p_{\mathsf{rmsea}}$ & \textsf{SRMR} \\ 
  \midrule
3965.99 & 1319.00 & $<.001$ & 0.97 & 0.98 & 0.05 & 0.05 & 0.05 & $\phantom{<}.590$ & 0.05 \\ 
   \bottomrule
\end{tabular}
\endgroup
\end{table*}
}

\newcommand{\loadingsFit}{
\begin{table}[ht]
\centering
\caption{Factor loadings of fitted correlational path model.} 
\label{tab:loadingsFit}
\begingroup\footnotesize
\begin{tabular}{llrrrrr}
  \toprule
Latent Factor & Indicator & $\lambda$ & $\vari{SE}$ & $Z$ & $p$ & $\beta$ \\ 
  \midrule
fit & MKF1 & 1.00 & 0.00 &  &  & 0.81 \\ 
  fit & MKF2 & 1.02 & 0.04 & 24.39 & $<.001$ & 0.82 \\ 
  fit & MKF3 & 0.97 & 0.04 & 21.98 & $<.001$ & 0.78 \\ 
  fit & MKF4 & 0.92 & 0.04 & 20.58 & $<.001$ & 0.75 \\ 
  ts & MKTS1 & 1.00 & 0.00 &  &  & 0.90 \\ 
  ts & MKTS2 & 1.06 & 0.03 & 40.60 & $<.001$ & 0.96 \\ 
  ts & MKTS3 & 0.84 & 0.02 & 39.29 & $<.001$ & 0.76 \\ 
  ctrl & ctrl1 & 1.00 & 0.00 &  &  & 0.76 \\ 
  ctrl & ctrl2 & 1.00 & 0.05 & 18.73 & $<.001$ & 0.76 \\ 
  ctrl & ctrl3 & 0.81 & 0.06 & 13.83 & $<.001$ & 0.61 \\ 
  aware & awa1 & 1.00 & 0.00 &  &  & 0.78 \\ 
  aware & awa2 & 1.01 & 0.06 & 17.62 & $<.001$ & 0.80 \\ 
  aware & awa3 & 0.83 & 0.05 & 16.09 & $<.001$ & 0.65 \\ 
  collect & coll1 & 1.00 & 0.00 &  &  & 0.82 \\ 
  collect & coll2 & 0.92 & 0.02 & 39.38 & $<.001$ & 0.76 \\ 
  collect & coll3 & 1.15 & 0.02 & 56.91 & $<.001$ & 0.95 \\ 
  collect & coll4 & 1.06 & 0.02 & 62.13 & $<.001$ & 0.88 \\ 
  primuse & STPE1 & 1.00 & 0.00 &  &  & 0.91 \\ 
  primuse & STPE2 & 0.95 & 0.02 & 56.63 & $<.001$ & 0.87 \\ 
  primuse & STPE3 & 0.97 & 0.02 & 58.06 & $<.001$ & 0.88 \\ 
  primuse & STPE4 & 0.98 & 0.02 & 58.33 & $<.001$ & 0.89 \\ 
  privuse & TAU1 & 1.00 & 0.00 &  &  & 0.93 \\ 
  privuse & TAU2 & 1.04 & 0.01 & 138.05 & $<.001$ & 0.97 \\ 
  privuse & TAU3 & 1.03 & 0.01 & 140.16 & $<.001$ & 0.95 \\ 
  privuse & TAU4 & 0.97 & 0.01 & 101.93 & $<.001$ & 0.90 \\ 
  ease & TAE1 & 1.00 & 0.00 &  &  & 0.90 \\ 
  ease & TAE2 & 0.85 & 0.02 & 42.95 & $<.001$ & 0.77 \\ 
  ease & TAE3 & 0.97 & 0.02 & 50.74 & $<.001$ & 0.87 \\ 
  ease & TAE4 & 0.94 & 0.02 & 49.42 & $<.001$ & 0.85 \\ 
  bi & STBI1 & 1.00 & 0.00 &  &  & 0.95 \\ 
  bi & STBI2 & 0.95 & 0.01 & 100.68 & $<.001$ & 0.91 \\ 
  bi & STBI4 & 0.96 & 0.01 & 103.12 & $<.001$ & 0.92 \\ 
  results & TAR1 & 1.00 & 0.00 &  &  & 0.86 \\ 
  results & TAR2 & 0.88 & 0.02 & 40.84 & $<.001$ & 0.76 \\ 
  results & TAR3 & 1.00 & 0.02 & 44.49 & $<.001$ & 0.86 \\ 
  fc & STFC1 & 1.00 & 0.00 &  &  & 0.63 \\ 
  fc & STFC2 & 1.19 & 0.04 & 30.28 & $<.001$ & 0.75 \\ 
  fc & STFC3 & 1.14 & 0.05 & 24.80 & $<.001$ & 0.72 \\ 
  fc & STFC4 & 1.02 & 0.05 & 21.61 & $<.001$ & 0.64 \\ 
  trust & STT1 & 1.00 & 0.00 &  &  & 0.97 \\ 
  trust & STT2 & 1.00 & 0.01 & 117.49 & $<.001$ & 0.96 \\ 
  trust & STT3 & 0.79 & 0.02 & 50.81 & $<.001$ & 0.76 \\ 
  trust & STT4 & 0.86 & 0.01 & 67.79 & $<.001$ & 0.83 \\ 
  trust & STT5 & 0.65 & 0.02 & 31.88 & $<.001$ & 0.63 \\ 
  trust & STT6 & 0.84 & 0.02 & 54.08 & $<.001$ & 0.81 \\ 
   \bottomrule
\end{tabular}
\endgroup
\end{table}
}
\newcommand{\loadingsFitCFA}{
\begin{table}[ht]
\centering
\caption{Factor loadings of the WLSMV-estimated CFA of the measurement model.} 
\label{tab:loadingsFitCFA}
\begingroup\footnotesize
\begin{tabular}{llrrrrr}
  \toprule
Latent Factor & Indicator & $\lambda$ & $\vari{SE}$ & $Z$ & $p$ & $\beta$ \\ 
  \midrule
fit & MKF1 & 1.00 & 0.00 &  &  & 0.81 \\ 
  fit & MKF2 & 1.02 & 0.04 & 24.43 & $<.001$ & 0.83 \\ 
  fit & MKF3 & 0.97 & 0.04 & 21.94 & $<.001$ & 0.78 \\ 
  fit & MKF4 & 0.92 & 0.04 & 20.52 & $<.001$ & 0.75 \\ 
  ts & MKTS1 & 1.00 & 0.00 &  &  & 0.90 \\ 
  ts & MKTS2 & 1.06 & 0.03 & 40.70 & $<.001$ & 0.96 \\ 
  ts & MKTS3 & 0.84 & 0.02 & 39.08 & $<.001$ & 0.76 \\ 
  ctrl & ctrl1 & 1.00 & 0.00 &  &  & 0.79 \\ 
  ctrl & ctrl2 & 0.99 & 0.06 & 17.85 & $<.001$ & 0.78 \\ 
  ctrl & ctrl3 & 0.81 & 0.06 & 13.63 & $<.001$ & 0.64 \\ 
  aware & awa1 & 1.00 & 0.00 &  &  & 0.82 \\ 
  aware & awa2 & 1.03 & 0.06 & 16.18 & $<.001$ & 0.84 \\ 
  aware & awa3 & 0.86 & 0.05 & 15.70 & $<.001$ & 0.70 \\ 
  collect & coll1 & 1.00 & 0.00 &  &  & 0.82 \\ 
  collect & coll2 & 0.92 & 0.02 & 39.25 & $<.001$ & 0.76 \\ 
  collect & coll3 & 1.15 & 0.02 & 56.36 & $<.001$ & 0.95 \\ 
  collect & coll4 & 1.06 & 0.02 & 61.20 & $<.001$ & 0.88 \\ 
  primuse & STPE1 & 1.00 & 0.00 &  &  & 0.91 \\ 
  primuse & STPE2 & 0.95 & 0.02 & 56.72 & $<.001$ & 0.87 \\ 
  primuse & STPE3 & 0.97 & 0.02 & 58.04 & $<.001$ & 0.88 \\ 
  primuse & STPE4 & 0.98 & 0.02 & 58.36 & $<.001$ & 0.89 \\ 
  privuse & TAU1 & 1.00 & 0.00 &  &  & 0.93 \\ 
  privuse & TAU2 & 1.04 & 0.01 & 138.23 & $<.001$ & 0.97 \\ 
  privuse & TAU3 & 1.03 & 0.01 & 140.53 & $<.001$ & 0.95 \\ 
  privuse & TAU4 & 0.97 & 0.01 & 102.03 & $<.001$ & 0.90 \\ 
  ease & TAE1 & 1.00 & 0.00 &  &  & 0.90 \\ 
  ease & TAE2 & 0.85 & 0.02 & 42.94 & $<.001$ & 0.77 \\ 
  ease & TAE3 & 0.97 & 0.02 & 50.76 & $<.001$ & 0.87 \\ 
  ease & TAE4 & 0.94 & 0.02 & 49.38 & $<.001$ & 0.85 \\ 
  bi & STBI1 & 1.00 & 0.00 &  &  & 0.95 \\ 
  bi & STBI2 & 0.95 & 0.01 & 100.82 & $<.001$ & 0.91 \\ 
  bi & STBI4 & 0.96 & 0.01 & 103.36 & $<.001$ & 0.92 \\ 
  results & TAR1 & 1.00 & 0.00 &  &  & 0.86 \\ 
  results & TAR2 & 0.88 & 0.02 & 40.69 & $<.001$ & 0.76 \\ 
  results & TAR3 & 1.00 & 0.02 & 44.44 & $<.001$ & 0.87 \\ 
  fc & STFC1 & 1.00 & 0.00 &  &  & 0.69 \\ 
  fc & STFC2 & 1.19 & 0.04 & 30.50 & $<.001$ & 0.83 \\ 
  fc & STFC3 & 1.14 & 0.04 & 25.46 & $<.001$ & 0.79 \\ 
  fc & STFC4 & 1.02 & 0.05 & 22.04 & $<.001$ & 0.71 \\ 
  trust & STT1 & 1.00 & 0.00 &  &  & 0.97 \\ 
  trust & STT2 & 1.00 & 0.01 & 117.14 & $<.001$ & 0.96 \\ 
  trust & STT3 & 0.79 & 0.02 & 51.18 & $<.001$ & 0.77 \\ 
  trust & STT4 & 0.86 & 0.01 & 68.35 & $<.001$ & 0.83 \\ 
  trust & STT5 & 0.65 & 0.02 & 32.22 & $<.001$ & 0.63 \\ 
  trust & STT6 & 0.84 & 0.02 & 54.06 & $<.001$ & 0.81 \\ 
   \bottomrule
\end{tabular}
\endgroup
\end{table}
}
\newcommand{\loadingsFitCFAreliability}{
\begin{table*}[tbp]
\centering
\caption{Factor loadings of the WLSMV-estimated CFA of the measurement model.} 
\label{tab:loadingsFitCFAreliability}
\begingroup\footnotesize
\begin{tabular}{lllrrrrrrrrrrrr}
  \toprule
 \multirow{2}{*}{Factor} & \multirow{2}{*}{Indicator} & \multicolumn{4}{c}{Factor Loading} & \multicolumn{4}{c}{Standardized Solution} & \multicolumn{5}{c}{Reliability} \\ 
 \cmidrule(lr){3-6} \cmidrule(lr){7-10} \cmidrule(lr){11-15} 
  & & $\lambda$ & $\vari{SE}_{\lambda}$ & $Z_{\lambda}$ & $p_{\lambda}$ & $\beta$ & $\vari{SE}_{\beta}$ & $Z_{\beta}$ & $p_{\beta}$ & $R^2$ & $\mathit{AVE}$ & $\alpha$ & $\omega$ & $\mathit{S/\!N}_\omega$ \\
  \midrule
 fit & MKF1 & $1.00^+$ &  &  &  & 0.81 & 0.02 & 35.05 & $<.001$ & 0.66 & 0.63 & 0.83 & 0.82 & 4.63 \\ 
   & MKF2 & $1.02\phantom{^+}$ & 0.04 & 24.43 & $<.001$ & 0.83 & 0.02 & 37.90 & $<.001$ & 0.68 &  &  &  &  \\ 
   & MKF3 & $0.97\phantom{^+}$ & 0.04 & 21.94 & $<.001$ & 0.78 & 0.02 & 32.12 & $<.001$ & 0.61 &  &  &  &  \\ 
   & MKF4 & $0.92\phantom{^+}$ & 0.04 & 20.52 & $<.001$ & 0.75 & 0.03 & 26.67 & $<.001$ & 0.56 &  &  &  &  \\ 
  ts & MKTS1 & $1.00^+$ &  &  &  & 0.90 & 0.01 & 73.09 & $<.001$ & 0.82 & 0.78 & 0.87 & 0.90 & 8.67 \\ 
   & MKTS2 & $1.06\phantom{^+}$ & 0.03 & 40.70 & $<.001$ & 0.96 & 0.01 & 78.95 & $<.001$ & 0.93 &  &  &  &  \\ 
   & MKTS3 & $0.84\phantom{^+}$ & 0.02 & 39.08 & $<.001$ & 0.76 & 0.02 & 42.75 & $<.001$ & 0.58 &  &  &  &  \\ 
  ctrl & ctrl1 & $1.00^+$ &  &  &  & 0.79 & 0.03 & 30.02 & $<.001$ & 0.62 & 0.54 & 0.65 & 0.73 & 2.76 \\ 
   & ctrl2 & $0.99\phantom{^+}$ & 0.06 & 17.85 & $<.001$ & 0.78 & 0.03 & 30.46 & $<.001$ & 0.61 &  &  &  &  \\ 
   & ctrl3 & $0.81\phantom{^+}$ & 0.06 & 13.63 & $<.001$ & 0.64 & 0.04 & 17.38 & $<.001$ & 0.40 &  &  &  &  \\ 
  aware & awa1 & $1.00^+$ &  &  &  & 0.82 & 0.03 & 27.64 & $<.001$ & 0.67 & 0.62 & 0.66 & 0.73 & 2.70 \\ 
   & awa2 & $1.03\phantom{^+}$ & 0.06 & 16.18 & $<.001$ & 0.84 & 0.03 & 28.32 & $<.001$ & 0.70 &  &  &  &  \\ 
   & awa3 & $0.86\phantom{^+}$ & 0.05 & 15.70 & $<.001$ & 0.70 & 0.03 & 20.79 & $<.001$ & 0.49 &  &  &  &  \\ 
  collect & coll1 & $1.00^+$ &  &  &  & 0.82 & 0.01 & 61.96 & $<.001$ & 0.68 & 0.73 & 0.89 & 0.89 & 8.45 \\ 
   & coll2 & $0.92\phantom{^+}$ & 0.02 & 39.25 & $<.001$ & 0.76 & 0.02 & 48.45 & $<.001$ & 0.57 &  &  &  &  \\ 
   & coll3 & $1.15\phantom{^+}$ & 0.02 & 56.36 & $<.001$ & 0.95 & 0.01 & 128.34 & $<.001$ & 0.90 &  &  &  &  \\ 
   & coll4 & $1.06\phantom{^+}$ & 0.02 & 61.20 & $<.001$ & 0.88 & 0.01 & 88.81 & $<.001$ & 0.77 &  &  &  &  \\ 
  primuse & STPE1 & $1.00^+$ &  &  &  & 0.91 & 0.01 & 73.64 & $<.001$ & 0.83 & 0.79 & 0.90 & 0.92 & 11.93 \\ 
   & STPE2 & $0.95\phantom{^+}$ & 0.02 & 56.72 & $<.001$ & 0.87 & 0.01 & 82.70 & $<.001$ & 0.75 &  &  &  &  \\ 
   & STPE3 & $0.97\phantom{^+}$ & 0.02 & 58.04 & $<.001$ & 0.88 & 0.01 & 95.98 & $<.001$ & 0.78 &  &  &  &  \\ 
   & STPE4 & $0.98\phantom{^+}$ & 0.02 & 58.36 & $<.001$ & 0.89 & 0.01 & 99.00 & $<.001$ & 0.79 &  &  &  &  \\ 
  privuse & TAU1 & $1.00^+$ &  &  &  & 0.93 & 0.01 & 151.88 & $<.001$ & 0.86 & 0.88 & 0.94 & 0.95 & 17.30 \\ 
   & TAU2 & $1.04\phantom{^+}$ & 0.01 & 138.23 & $<.001$ & 0.97 & 0.00 & 223.09 & $<.001$ & 0.93 &  &  &  &  \\ 
   & TAU3 & $1.03\phantom{^+}$ & 0.01 & 140.53 & $<.001$ & 0.95 & 0.00 & 205.54 & $<.001$ & 0.91 &  &  &  &  \\ 
   & TAU4 & $0.97\phantom{^+}$ & 0.01 & 102.03 & $<.001$ & 0.90 & 0.01 & 106.90 & $<.001$ & 0.81 &  &  &  &  \\ 
  ease & TAE1 & $1.00^+$ &  &  &  & 0.90 & 0.01 & 75.75 & $<.001$ & 0.81 & 0.72 & 0.88 & 0.89 & 8.12 \\ 
   & TAE2 & $0.85\phantom{^+}$ & 0.02 & 42.94 & $<.001$ & 0.77 & 0.02 & 50.63 & $<.001$ & 0.59 &  &  &  &  \\ 
   & TAE3 & $0.97\phantom{^+}$ & 0.02 & 50.76 & $<.001$ & 0.87 & 0.01 & 71.99 & $<.001$ & 0.76 &  &  &  &  \\ 
   & TAE4 & $0.94\phantom{^+}$ & 0.02 & 49.38 & $<.001$ & 0.85 & 0.01 & 66.73 & $<.001$ & 0.72 &  &  &  &  \\ 
  bi & STBI1 & $1.00^+$ &  &  &  & 0.95 & 0.01 & 189.61 & $<.001$ & 0.91 & 0.86 & 0.93 & 0.93 & 13.56 \\ 
   & STBI2 & $0.95\phantom{^+}$ & 0.01 & 100.82 & $<.001$ & 0.91 & 0.01 & 122.42 & $<.001$ & 0.83 &  &  &  &  \\ 
   & STBI4 & $0.96\phantom{^+}$ & 0.01 & 103.36 & $<.001$ & 0.92 & 0.01 & 126.51 & $<.001$ & 0.84 &  &  &  &  \\ 
  results & TAR1 & $1.00^+$ &  &  &  & 0.86 & 0.01 & 65.62 & $<.001$ & 0.74 & 0.69 & 0.84 & 0.85 & 5.51 \\ 
   & TAR2 & $0.88\phantom{^+}$ & 0.02 & 40.69 & $<.001$ & 0.76 & 0.02 & 49.45 & $<.001$ & 0.58 &  &  &  &  \\ 
   & TAR3 & $1.00\phantom{^+}$ & 0.02 & 44.44 & $<.001$ & 0.87 & 0.01 & 61.83 & $<.001$ & 0.75 &  &  &  &  \\ 
  fc & STFC1 & $1.00^+$ &  &  &  & 0.69 & 0.02 & 34.95 & $<.001$ & 0.48 & 0.57 & 0.78 & 0.82 & 4.54 \\ 
   & STFC2 & $1.19\phantom{^+}$ & 0.04 & 30.50 & $<.001$ & 0.83 & 0.02 & 52.77 & $<.001$ & 0.69 &  &  &  &  \\ 
   & STFC3 & $1.14\phantom{^+}$ & 0.04 & 25.46 & $<.001$ & 0.79 & 0.02 & 38.96 & $<.001$ & 0.63 &  &  &  &  \\ 
   & STFC4 & $1.02\phantom{^+}$ & 0.05 & 22.04 & $<.001$ & 0.71 & 0.02 & 29.26 & $<.001$ & 0.50 &  &  &  &  \\ 
  trust & STT1 & $1.00^+$ &  &  &  & 0.97 & 0.00 & 202.29 & $<.001$ & 0.93 & 0.70 & 0.90 & 0.91 & 10.13 \\ 
   & STT2 & $1.00\phantom{^+}$ & 0.01 & 117.14 & $<.001$ & 0.96 & 0.00 & 203.71 & $<.001$ & 0.93 &  &  &  &  \\ 
   & STT3 & $0.79\phantom{^+}$ & 0.02 & 51.18 & $<.001$ & 0.77 & 0.02 & 50.90 & $<.001$ & 0.59 &  &  &  &  \\ 
   & STT4 & $0.86\phantom{^+}$ & 0.01 & 68.35 & $<.001$ & 0.83 & 0.01 & 68.59 & $<.001$ & 0.69 &  &  &  &  \\ 
   & STT5 & $0.65\phantom{^+}$ & 0.02 & 32.22 & $<.001$ & 0.63 & 0.02 & 32.02 & $<.001$ & 0.40 &  &  &  &  \\ 
   & STT6 & $0.84\phantom{^+}$ & 0.02 & 54.06 & $<.001$ & 0.81 & 0.01 & 55.24 & $<.001$ & 0.65 &  &  &  &  \\ 
   \bottomrule
\multicolumn{14}{c}{\emph{Note:} $^+$ fixed parameter; the standardized solution is STDALL}\\
\end{tabular}
\endgroup
\end{table*}
}

\newcommand{\fullRegressionsFit}{
\begin{table}[ht]
\centering
\caption{Regression coefficients of the causal SEM.} 
\label{tab:fullRegressionsFit}
\begingroup\footnotesize
\begin{tabular}{lcrrrrr}
  \toprule
Relation & $H$ & $B$ & $\vari{SE}$ & $Z$ & $p$ & $\beta$ \\ 
  \midrule
$\mathsf{trust}\sim{}\mathsf{fc}$*** & $H_{7}$ & 1.256 & 0.049 & 25.506 & $<.001$ & 0.791 \\ 
  $\mathsf{trust}\sim{}\mathsf{d\_gov}$ & $H_{C, 1}$ & -0.014 & 0.154 & -0.091 & $\phantom{<}.928$ & -0.005 \\ 
  $\mathsf{trust}\sim{}\mathsf{d\_company}$ & $H_{C, 1}$ & 0.030 & 0.155 & 0.190 & $\phantom{<}.849$ & 0.011 \\ 
  $\mathsf{trust}\sim{}\mathsf{d\_uni}$ & $H_{C, 1}$ & 0.221 & 0.159 & 1.389 & $\phantom{<}.165$ & 0.081 \\ 
  $\mathsf{trust}\sim{}\mathsf{benefits}$ & $H_{C, 2}$ & -0.144 & 0.100 & -1.440 & $\phantom{<}.150$ & -0.061 \\ 
  $\mathsf{trust}\sim{}\mathsf{usage}$ & $H_{C, 3}$ & -0.089 & 0.100 & -0.888 & $\phantom{<}.375$ & -0.039 \\ 
  $\mathsf{trust}\sim{}\mathsf{simplicity}$ & $H_{C, 4}$ & 0.029 & 0.100 & 0.289 & $\phantom{<}.773$ & 0.013 \\ 
  $\mathsf{trust}\sim{}\mathsf{people}$ & $H_{C, 5}$ & -0.034 & 0.100 & -0.338 & $\phantom{<}.736$ & -0.015 \\ 
  $\mathsf{trust}\sim{}\mathsf{d\_fullsupport}$ & $H_{C, 6}$ & 0.167 & 0.119 & 1.410 & $\phantom{<}.159$ & 0.067 \\ 
  $\mathsf{trust}\sim{}\mathsf{d\_contact}$* & $H_{C, 6}$ & 0.281 & 0.121 & 2.319 & $\phantom{<}.020$ & 0.103 \\ 
  $\mathsf{primuse}\sim{}\mathsf{ease}$ & $H_{14}$ & 0.114 & 0.060 & 1.886 & $\phantom{<}.059$ & 0.113 \\ 
  $\mathsf{primuse}\sim{}\mathsf{results}$** & $H_{9}$ & 0.177 & 0.063 & 2.817 & $\phantom{<}.005$ & 0.164 \\ 
  $\mathsf{primuse}\sim{}\mathsf{trust}$*** & $H_{12}$ & 0.398 & 0.037 & 10.893 & $<.001$ & 0.420 \\ 
  $\mathsf{primuse}\sim{}\mathsf{ctrl}$*** & $H_{3}$ & -0.787 & 0.150 & -5.240 & $<.001$ & -0.648 \\ 
  $\mathsf{primuse}\sim{}\mathsf{aware}$*** & $H_{3}$ & 0.922 & 0.181 & 5.082 & $<.001$ & 0.766 \\ 
  $\mathsf{primuse}\sim{}\mathsf{collect}$ & $H_{3}$ & -0.070 & 0.074 & -0.946 & $\phantom{<}.344$ & -0.062 \\ 
  $\mathsf{primuse}\sim{}\mathsf{fit}$ & $H_{5}$ & 0.050 & 0.062 & 0.801 & $\phantom{<}.423$ & 0.044 \\ 
  $\mathsf{primuse}\sim{}\mathsf{ts}$** & $H_{5}$ & -0.174 & 0.057 & -3.027 & $\phantom{<}.002$ & -0.170 \\ 
  $\mathsf{primuse}\sim{}\mathsf{d\_gov}$ & $H_{C, 1}$ & -0.175 & 0.112 & -1.562 & $\phantom{<}.118$ & -0.071 \\ 
  $\mathsf{primuse}\sim{}\mathsf{d\_company}$* & $H_{C, 1}$ & -0.250 & 0.112 & -2.238 & $\phantom{<}.025$ & -0.095 \\ 
  $\mathsf{primuse}\sim{}\mathsf{d\_uni}$ & $H_{C, 1}$ & -0.133 & 0.122 & -1.092 & $\phantom{<}.275$ & -0.052 \\ 
  $\mathsf{primuse}\sim{}\mathsf{benefits}$ & $H_{C, 2}$ & 0.008 & 0.077 & 0.102 & $\phantom{<}.919$ & 0.003 \\ 
  $\mathsf{primuse}\sim{}\mathsf{usage}$ & $H_{C, 3}$ & -0.112 & 0.075 & -1.500 & $\phantom{<}.134$ & -0.052 \\ 
  $\mathsf{primuse}\sim{}\mathsf{simplicity}$* & $H_{C, 4}$ & 0.203 & 0.079 & 2.562 & $\phantom{<}.010$ & 0.096 \\ 
  $\mathsf{primuse}\sim{}\mathsf{people}$ & $H_{C, 5}$ & 0.112 & 0.075 & 1.504 & $\phantom{<}.132$ & 0.054 \\ 
  $\mathsf{primuse}\sim{}\mathsf{d\_fullsupport}$ & $H_{C, 6}$ & -0.116 & 0.095 & -1.221 & $\phantom{<}.222$ & -0.049 \\ 
  $\mathsf{primuse}\sim{}\mathsf{d\_contact}$ & $H_{C, 6}$ & -0.096 & 0.089 & -1.077 & $\phantom{<}.282$ & -0.037 \\ 
  $\mathsf{privuse}\sim{}\mathsf{ease}$*** & $H_{15}$ & 0.165 & 0.050 & 3.336 & $<.001$ & 0.164 \\ 
  $\mathsf{privuse}\sim{}\mathsf{results}$** & $H_{10}$ & 0.187 & 0.065 & 2.857 & $\phantom{<}.004$ & 0.172 \\ 
  $\mathsf{privuse}\sim{}\mathsf{trust}$*** & $H_{13}$ & 0.452 & 0.039 & 11.517 & $<.001$ & 0.475 \\ 
  $\mathsf{privuse}\sim{}\mathsf{ctrl}$*** & $H_{4}$ & -0.989 & 0.213 & -4.652 & $<.001$ & -0.811 \\ 
  $\mathsf{privuse}\sim{}\mathsf{aware}$*** & $H_{4}$ & 1.356 & 0.266 & 5.106 & $<.001$ & 1.122 \\ 
  $\mathsf{privuse}\sim{}\mathsf{collect}$** & $H_{4}$ & -0.342 & 0.108 & -3.164 & $\phantom{<}.002$ & -0.304 \\ 
  $\mathsf{privuse}\sim{}\mathsf{fit}$ & $H_{6}$ & 0.036 & 0.081 & 0.450 & $\phantom{<}.653$ & 0.032 \\ 
  $\mathsf{privuse}\sim{}\mathsf{ts}$* & $H_{6}$ & -0.159 & 0.071 & -2.247 & $\phantom{<}.025$ & -0.154 \\ 
  $\mathsf{privuse}\sim{}\mathsf{d\_gov}$ & $H_{C, 1}$ & -0.014 & 0.098 & -0.140 & $\phantom{<}.888$ & -0.006 \\ 
  $\mathsf{privuse}\sim{}\mathsf{d\_company}$ & $H_{C, 1}$ & -0.058 & 0.102 & -0.569 & $\phantom{<}.570$ & -0.022 \\ 
  $\mathsf{privuse}\sim{}\mathsf{d\_uni}$ & $H_{C, }$ & -0.036 & 0.115 & -0.312 & $\phantom{<}.755$ & -0.014 \\ 
  $\mathsf{privuse}\sim{}\mathsf{benefits}$ & $H_{C, 2}$ & 0.021 & 0.069 & 0.303 & $\phantom{<}.762$ & 0.009 \\ 
  $\mathsf{privuse}\sim{}\mathsf{usage}$ & $H_{C, 3}$ & 0.056 & 0.067 & 0.828 & $\phantom{<}.408$ & 0.026 \\ 
  $\mathsf{privuse}\sim{}\mathsf{simplicity}$ & $H_{C, 4}$ & 0.069 & 0.071 & 0.977 & $\phantom{<}.328$ & 0.032 \\ 
  $\mathsf{privuse}\sim{}\mathsf{people}$ & $H_{C, 5}$ & -0.054 & 0.068 & -0.791 & $\phantom{<}.429$ & -0.026 \\ 
  $\mathsf{privuse}\sim{}\mathsf{d\_fullsupport}$ & $H_{C, 6}$ & -0.095 & 0.082 & -1.148 & $\phantom{<}.251$ & -0.040 \\ 
  $\mathsf{privuse}\sim{}\mathsf{d\_contact}$ & $H_{C, 6}$ & -0.050 & 0.085 & -0.591 & $\phantom{<}.555$ & -0.019 \\ 
  $\mathsf{ease}\sim{}\mathsf{results}$*** & $H_{11}$ & 0.487 & 0.042 & 11.706 & $<.001$ & 0.452 \\ 
  $\mathsf{ease}\sim{}\mathsf{fc}$*** & $H_{8}$ & 0.649 & 0.062 & 10.551 & $<.001$ & 0.433 \\ 
  $\mathsf{ease}\sim{}\mathsf{d\_gov}$ & $H_{C, 1}$ & 0.201 & 0.142 & 1.414 & $\phantom{<}.157$ & 0.082 \\ 
  $\mathsf{ease}\sim{}\mathsf{d\_company}$ & $H_{C, 1}$ & 0.167 & 0.148 & 1.122 & $\phantom{<}.262$ & 0.063 \\ 
  $\mathsf{ease}\sim{}\mathsf{d\_uni}$ & $H_{C, 1}$ & 0.117 & 0.156 & 0.753 & $\phantom{<}.451$ & 0.046 \\ 
  $\mathsf{ease}\sim{}\mathsf{benefits}$ & $H_{C, 2}$ & 0.095 & 0.102 & 0.933 & $\phantom{<}.351$ & 0.043 \\ 
  $\mathsf{ease}\sim{}\mathsf{usage}$ & $H_{C, 3}$ & -0.067 & 0.102 & -0.659 & $\phantom{<}.510$ & -0.031 \\ 
  $\mathsf{ease}\sim{}\mathsf{simplicity}$*** & $H_{C, 4}$ & 0.423 & 0.093 & 4.573 & $<.001$ & 0.201 \\ 
  $\mathsf{ease}\sim{}\mathsf{people}$ & $H_{C, 5}$ & 0.012 & 0.092 & 0.130 & $\phantom{<}.897$ & 0.006 \\ 
  $\mathsf{ease}\sim{}\mathsf{d\_fullsupport}$ & $H_{C, 6}$ & 0.133 & 0.112 & 1.189 & $\phantom{<}.234$ & 0.057 \\ 
  $\mathsf{ease}\sim{}\mathsf{d\_contact}$** & $H_{C, 6}$ & 0.311 & 0.109 & 2.849 & $\phantom{<}.004$ & 0.120 \\ 
  $\mathsf{bi}\sim{}\mathsf{primuse}$*** & $H_{16}$ & 0.687 & 0.051 & 13.511 & $<.001$ & 0.662 \\ 
  $\mathsf{bi}\sim{}\mathsf{privuse}$*** & $H_{17}$ & 0.460 & 0.071 & 6.494 & $<.001$ & 0.446 \\ 
  $\mathsf{bi}\sim{}\mathsf{ease}$ & $H_{18}$ & -0.005 & 0.054 & -0.092 & $\phantom{<}.927$ & -0.005 \\ 
  $\mathsf{download}\sim{}\mathsf{bi}$ & $H_{19}$ & 0.050 & 0.061 & 0.807 & $\phantom{<}.420$ & 0.048 \\ 
  $\mathsf{bi}\sim{}\mathsf{ctrl}$*** & $H_{1}$ & 1.152 & 0.221 & 5.204 & $<.001$ & 0.914 \\ 
  $\mathsf{bi}\sim{}\mathsf{aware}$*** & $H_{1}$ & -1.431 & 0.270 & -5.304 & $<.001$ & -1.146 \\ 
  $\mathsf{bi}\sim{}\mathsf{collect}$*** & $H_{1}$ & 0.478 & 0.109 & 4.385 & $<.001$ & 0.410 \\ 
  $\mathsf{bi}\sim{}\mathsf{fit}$ & $H_{2}$ & -0.035 & 0.083 & -0.420 & $\phantom{<}.675$ & -0.029 \\ 
  $\mathsf{bi}\sim{}\mathsf{ts}$ & $H_{2}$ & 0.140 & 0.073 & 1.912 & $\phantom{<}.056$ & 0.132 \\ 
   \bottomrule
\end{tabular}
\endgroup
\end{table}
}
\newcommand{\selectedRegressionsFit}{
\begin{table}[ht]
\centering
\caption{Selected coefficients of the correlational SEM, $p < .15$.} 
\label{tab:selectedRegressionsFit}
\begingroup\footnotesize
\begin{tabular}{lcrrrr}
  \toprule
Relation & $H$ & $B$ & $\vari{SE}$ & $p$ & $\beta$ \\ 
  \midrule
$\mathsf{trust}\sim{}\mathsf{fc}$*** & $H_{7}$ & 1.201 & 0.047 & $<.001$ & 0.782 \\ 
  $\mathsf{primuse}\sim{}\mathsf{ease}$* & $H_{14}$ & 0.120 & 0.060 & $\phantom{<}.047$ & 0.119 \\ 
  $\mathsf{primuse}\sim{}\mathsf{results}$** & $H_{9}$ & 0.167 & 0.062 & $\phantom{<}.007$ & 0.157 \\ 
  $\mathsf{primuse}\sim{}\mathsf{trust}$*** & $H_{12}$ & 0.391 & 0.035 & $<.001$ & 0.414 \\ 
  $\mathsf{primuse}\sim{}\mathsf{ctrl}$*** & $H_{3}$ & -0.700 & 0.128 & $<.001$ & -0.585 \\ 
  $\mathsf{primuse}\sim{}\mathsf{aware}$*** & $H_{3}$ & 0.819 & 0.149 & $<.001$ & 0.704 \\ 
  $\mathsf{primuse}\sim{}\mathsf{ts}$*** & $H_{5}$ & -0.178 & 0.054 & $<.001$ & -0.177 \\ 
  $\mathsf{privuse}\sim{}\mathsf{ease}$** & $H_{15}$ & 0.159 & 0.050 & $\phantom{<}.001$ & 0.155 \\ 
  $\mathsf{privuse}\sim{}\mathsf{results}$** & $H_{10}$ & 0.195 & 0.062 & $\phantom{<}.002$ & 0.181 \\ 
  $\mathsf{privuse}\sim{}\mathsf{trust}$*** & $H_{13}$ & 0.441 & 0.036 & $<.001$ & 0.461 \\ 
  $\mathsf{privuse}\sim{}\mathsf{ctrl}$*** & $H_{4}$ & -0.783 & 0.159 & $<.001$ & -0.645 \\ 
  $\mathsf{privuse}\sim{}\mathsf{aware}$*** & $H_{4}$ & 1.074 & 0.192 & $<.001$ & 0.910 \\ 
  $\mathsf{privuse}\sim{}\mathsf{collect}$** & $H_{4}$ & -0.258 & 0.084 & $\phantom{<}.002$ & -0.230 \\ 
  $\mathsf{privuse}\sim{}\mathsf{ts}$* & $H_{6}$ & -0.143 & 0.060 & $\phantom{<}.017$ & -0.140 \\ 
  $\mathsf{ease}\sim{}\mathsf{results}$*** & $H_{11}$ & 0.499 & 0.042 & $<.001$ & 0.475 \\ 
  $\mathsf{ease}\sim{}\mathsf{fc}$*** & $H_{8}$ & 0.627 & 0.061 & $<.001$ & 0.438 \\ 
  $\mathsf{bi}\sim{}\mathsf{primuse}$*** & $H_{16}$ & 0.713 & 0.051 & $<.001$ & 0.682 \\ 
  $\mathsf{bi}\sim{}\mathsf{privuse}$*** & $H_{17}$ & 0.438 & 0.064 & $<.001$ & 0.425 \\ 
  $\mathsf{bi}\sim{}\mathsf{ctrl}$*** & $H_{1}$ & 0.998 & 0.181 & $<.001$ & 0.797 \\ 
  $\mathsf{bi}\sim{}\mathsf{aware}$*** & $H_{1}$ & -1.223 & 0.213 & $<.001$ & -1.006 \\ 
  $\mathsf{bi}\sim{}\mathsf{collect}$*** & $H_{1}$ & 0.411 & 0.089 & $<.001$ & 0.355 \\ 
  $\mathsf{bi}\sim{}\mathsf{ts}$* & $H_{2}$ & 0.142 & 0.066 & $\phantom{<}.031$ & 0.135 \\ 
   \bottomrule
\end{tabular}
\endgroup
\end{table}
}
\newcommand{\selectedRegressionsFitCausal}{
\begin{table}[ht]
\centering
\caption{Selected coefficients of the causal SEM, $p < .15$.} 
\label{tab:selectedRegressionsFitCausal}
\begingroup\footnotesize
\begin{tabular}{lcrrrr}
  \toprule
Relation & $H$ & $B$ & $\vari{SE}$ & $p$ & $\beta$ \\ 
  \midrule
$\mathsf{trust}\sim{}\mathsf{benefits}$ & $H_{C, 2}$ & -0.144 & 0.100 & $\phantom{<}.150$ & -0.061 \\ 
  $\mathsf{trust}\sim{}\mathsf{d\_contact}$* & $H_{C, 6}$ & 0.281 & 0.121 & $\phantom{<}.020$ & 0.103 \\ 
  $\mathsf{primuse}\sim{}\mathsf{d\_gov}$ & $H_{C, 1}$ & -0.175 & 0.112 & $\phantom{<}.118$ & -0.071 \\ 
  $\mathsf{primuse}\sim{}\mathsf{d\_company}$* & $H_{C, 1}$ & -0.250 & 0.112 & $\phantom{<}.025$ & -0.095 \\ 
  $\mathsf{primuse}\sim{}\mathsf{usage}$ & $H_{C, 3}$ & -0.112 & 0.075 & $\phantom{<}.134$ & -0.052 \\ 
  $\mathsf{primuse}\sim{}\mathsf{simplicity}$* & $H_{C, 4}$ & 0.203 & 0.079 & $\phantom{<}.010$ & 0.096 \\ 
  $\mathsf{primuse}\sim{}\mathsf{people}$ & $H_{C, 5}$ & 0.112 & 0.075 & $\phantom{<}.132$ & 0.054 \\ 
  $\mathsf{ease}\sim{}\mathsf{simplicity}$*** & $H_{C, 4}$ & 0.423 & 0.093 & $<.001$ & 0.201 \\ 
  $\mathsf{ease}\sim{}\mathsf{d\_contact}$** & $H_{C, 6}$ & 0.311 & 0.109 & $\phantom{<}.004$ & 0.120 \\ 
   \bottomrule
\end{tabular}
\endgroup
\end{table}
}
\newcommand{\selectedRegressionsFitCausalByH}{
\begin{table}[ht]
\centering
\caption{Selected coefficients of the causal SEM, $p < .15$.} 
\label{tab:selectedRegressionsFitCausalByH}
\begingroup\footnotesize
\begin{tabular}{lcrrrr}
  \toprule
Relation & $H$ & $B$ & $\vari{SE}$ & $p$ & $\beta$ \\ 
  \midrule
$\mathsf{primuse}\sim{}\mathsf{d\_gov}$ & $H_{C, 1}$ & -0.175 & 0.112 & $\phantom{<}.118$ & -0.071 \\ 
  $\mathsf{primuse}\sim{}\mathsf{d\_company}$* & $H_{C, 1}$ & -0.250 & 0.112 & $\phantom{<}.025$ & -0.095 \\ 
  $\mathsf{trust}\sim{}\mathsf{benefits}$ & $H_{C, 2}$ & -0.144 & 0.100 & $\phantom{<}.150$ & -0.061 \\ 
  $\mathsf{primuse}\sim{}\mathsf{usage}$ & $H_{C, 3}$ & -0.112 & 0.075 & $\phantom{<}.134$ & -0.052 \\ 
  $\mathsf{primuse}\sim{}\mathsf{simplicity}$* & $H_{C, 4}$ & 0.203 & 0.079 & $\phantom{<}.010$ & 0.096 \\ 
  $\mathsf{ease}\sim{}\mathsf{simplicity}$*** & $H_{C, 4}$ & 0.423 & 0.093 & $<.001$ & 0.201 \\ 
  $\mathsf{primuse}\sim{}\mathsf{people}$ & $H_{C, 5}$ & 0.112 & 0.075 & $\phantom{<}.132$ & 0.054 \\ 
  $\mathsf{trust}\sim{}\mathsf{d\_contact}$* & $H_{C, 6}$ & 0.281 & 0.121 & $\phantom{<}.020$ & 0.103 \\ 
  $\mathsf{ease}\sim{}\mathsf{d\_contact}$** & $H_{C, 6}$ & 0.311 & 0.109 & $\phantom{<}.004$ & 0.120 \\ 
   \bottomrule
\end{tabular}
\endgroup
\end{table}
}
\newcommand{\selectedRegressionsFitCausalBySize}{
\begin{table}[ht]
\centering
\caption{Selected coefficients of the causal SEM, $p < .15$.} 
\label{tab:selectedRegressionsFitCausalBySize}
\begingroup\footnotesize
\begin{tabular}{lcrrrr}
  \toprule
Relation & $H$ & $B$ & $\vari{SE}$ & $p$ & $\beta$ \\ 
  \midrule
$\mathsf{primuse}\sim{}\mathsf{d\_company}$* & $H_{C, 1}$ & -0.250 & 0.112 & $\phantom{<}.025$ & -0.095 \\ 
  $\mathsf{primuse}\sim{}\mathsf{d\_gov}$ & $H_{C, 1}$ & -0.175 & 0.112 & $\phantom{<}.118$ & -0.071 \\ 
  $\mathsf{trust}\sim{}\mathsf{benefits}$ & $H_{C, 2}$ & -0.144 & 0.100 & $\phantom{<}.150$ & -0.061 \\ 
  $\mathsf{primuse}\sim{}\mathsf{usage}$ & $H_{C, 3}$ & -0.112 & 0.075 & $\phantom{<}.134$ & -0.052 \\ 
  $\mathsf{primuse}\sim{}\mathsf{people}$ & $H_{C, 5}$ & 0.112 & 0.075 & $\phantom{<}.132$ & 0.054 \\ 
  $\mathsf{primuse}\sim{}\mathsf{simplicity}$* & $H_{C, 4}$ & 0.203 & 0.079 & $\phantom{<}.010$ & 0.096 \\ 
  $\mathsf{trust}\sim{}\mathsf{d\_contact}$* & $H_{C, 6}$ & 0.281 & 0.121 & $\phantom{<}.020$ & 0.103 \\ 
  $\mathsf{ease}\sim{}\mathsf{d\_contact}$** & $H_{C, 6}$ & 0.311 & 0.109 & $\phantom{<}.004$ & 0.120 \\ 
  $\mathsf{ease}\sim{}\mathsf{simplicity}$*** & $H_{C, 4}$ & 0.423 & 0.093 & $<.001$ & 0.201 \\ 
   \bottomrule
\end{tabular}
\endgroup
\end{table}
}
\newcommand{\selectedRegressionsFitTAM}{
\begin{table}[ht]
\centering
\caption{Regression coefficients of the correlational SEM.} 
\label{tab:selectedRegressionsFitTAM}
\begingroup\footnotesize
\begin{tabular}{lcrrrr}
  \toprule
Relation & $H$ & $B$ & $\vari{SE}$ & $p$ & $\beta$ \\ 
  \midrule
$\mathsf{trust}\sim{}\mathsf{fc}$*** & $H_{7}$ & 1.201 & 0.047 & $<.001$ & 0.782 \\ 
  $\mathsf{primuse}\sim{}\mathsf{ease}$* & $H_{14}$ & 0.120 & 0.060 & $\phantom{<}.047$ & 0.119 \\ 
  $\mathsf{primuse}\sim{}\mathsf{results}$** & $H_{9}$ & 0.167 & 0.062 & $\phantom{<}.007$ & 0.157 \\ 
  $\mathsf{primuse}\sim{}\mathsf{trust}$*** & $H_{12}$ & 0.391 & 0.035 & $<.001$ & 0.414 \\ 
  $\mathsf{primuse}\sim{}\mathsf{ctrl}$*** & $H_{3}$ & -0.700 & 0.128 & $<.001$ & -0.585 \\ 
  $\mathsf{primuse}\sim{}\mathsf{aware}$*** & $H_{3}$ & 0.819 & 0.149 & $<.001$ & 0.704 \\ 
  $\mathsf{primuse}\sim{}\mathsf{collect}$ & $H_{3}$ & -0.048 & 0.065 & $\phantom{<}.455$ & -0.044 \\ 
  $\mathsf{primuse}\sim{}\mathsf{fit}$ & $H_{5}$ & 0.081 & 0.057 & $\phantom{<}.156$ & 0.072 \\ 
  $\mathsf{primuse}\sim{}\mathsf{ts}$*** & $H_{5}$ & -0.178 & 0.054 & $<.001$ & -0.177 \\ 
  $\mathsf{privuse}\sim{}\mathsf{ease}$** & $H_{15}$ & 0.159 & 0.050 & $\phantom{<}.001$ & 0.155 \\ 
  $\mathsf{privuse}\sim{}\mathsf{results}$** & $H_{10}$ & 0.195 & 0.062 & $\phantom{<}.002$ & 0.181 \\ 
  $\mathsf{privuse}\sim{}\mathsf{trust}$*** & $H_{13}$ & 0.441 & 0.036 & $<.001$ & 0.461 \\ 
  $\mathsf{privuse}\sim{}\mathsf{ctrl}$*** & $H_{4}$ & -0.783 & 0.159 & $<.001$ & -0.645 \\ 
  $\mathsf{privuse}\sim{}\mathsf{aware}$*** & $H_{4}$ & 1.074 & 0.192 & $<.001$ & 0.910 \\ 
  $\mathsf{privuse}\sim{}\mathsf{collect}$** & $H_{4}$ & -0.258 & 0.084 & $\phantom{<}.002$ & -0.230 \\ 
  $\mathsf{privuse}\sim{}\mathsf{fit}$ & $H_{6}$ & 0.084 & 0.067 & $\phantom{<}.209$ & 0.074 \\ 
  $\mathsf{privuse}\sim{}\mathsf{ts}$* & $H_{6}$ & -0.143 & 0.060 & $\phantom{<}.017$ & -0.140 \\ 
  $\mathsf{ease}\sim{}\mathsf{results}$*** & $H_{11}$ & 0.499 & 0.042 & $<.001$ & 0.475 \\ 
  $\mathsf{ease}\sim{}\mathsf{fc}$*** & $H_{8}$ & 0.627 & 0.061 & $<.001$ & 0.438 \\ 
  $\mathsf{bi}\sim{}\mathsf{primuse}$*** & $H_{16}$ & 0.713 & 0.051 & $<.001$ & 0.682 \\ 
  $\mathsf{bi}\sim{}\mathsf{privuse}$*** & $H_{17}$ & 0.438 & 0.064 & $<.001$ & 0.425 \\ 
  $\mathsf{bi}\sim{}\mathsf{ease}$ & $H_{18}$ & 0.009 & 0.055 & $\phantom{<}.869$ & 0.008 \\ 
  $\mathsf{bi}\sim{}\mathsf{ctrl}$*** & $H_{1}$ & 0.998 & 0.181 & $<.001$ & 0.797 \\ 
  $\mathsf{bi}\sim{}\mathsf{aware}$*** & $H_{1}$ & -1.223 & 0.213 & $<.001$ & -1.006 \\ 
  $\mathsf{bi}\sim{}\mathsf{collect}$*** & $H_{1}$ & 0.411 & 0.089 & $<.001$ & 0.355 \\ 
  $\mathsf{bi}\sim{}\mathsf{fit}$ & $H_{2}$ & -0.077 & 0.072 & $\phantom{<}.289$ & -0.065 \\ 
  $\mathsf{bi}\sim{}\mathsf{ts}$* & $H_{2}$ & 0.142 & 0.066 & $\phantom{<}.031$ & 0.135 \\ 
   \bottomrule
\end{tabular}
\endgroup
\end{table}
}
\newcommand{\selectedRegressionsFitTAMbyH}{
\begin{table}[ht]
\centering
\caption{Regression coefficients of the correlational SEM.} 
\label{tab:selectedRegressionsFitTAMbyH}
\begingroup\footnotesize
\begin{tabular}{lcrrrr}
  \toprule
Relation & $H$ & $B$ & $\vari{SE}$ & $p$ & $\beta$ \\ 
  \midrule
$\mathsf{bi}\sim{}\mathsf{ctrl}$*** & $H_{1}$ & 0.998 & 0.181 & $<.001$ & 0.797 \\ 
  $\mathsf{bi}\sim{}\mathsf{aware}$*** & $H_{1}$ & -1.223 & 0.213 & $<.001$ & -1.006 \\ 
  $\mathsf{bi}\sim{}\mathsf{collect}$*** & $H_{1}$ & 0.411 & 0.089 & $<.001$ & 0.355 \\ 
  $\mathsf{bi}\sim{}\mathsf{fit}$ & $H_{2}$ & -0.077 & 0.072 & $\phantom{<}.289$ & -0.065 \\ 
  $\mathsf{bi}\sim{}\mathsf{ts}$* & $H_{2}$ & 0.142 & 0.066 & $\phantom{<}.031$ & 0.135 \\ 
  $\mathsf{primuse}\sim{}\mathsf{ctrl}$*** & $H_{3}$ & -0.700 & 0.128 & $<.001$ & -0.585 \\ 
  $\mathsf{primuse}\sim{}\mathsf{aware}$*** & $H_{3}$ & 0.819 & 0.149 & $<.001$ & 0.704 \\ 
  $\mathsf{primuse}\sim{}\mathsf{collect}$ & $H_{3}$ & -0.048 & 0.065 & $\phantom{<}.455$ & -0.044 \\ 
  $\mathsf{privuse}\sim{}\mathsf{ctrl}$*** & $H_{4}$ & -0.783 & 0.159 & $<.001$ & -0.645 \\ 
  $\mathsf{privuse}\sim{}\mathsf{aware}$*** & $H_{4}$ & 1.074 & 0.192 & $<.001$ & 0.910 \\ 
  $\mathsf{privuse}\sim{}\mathsf{collect}$** & $H_{4}$ & -0.258 & 0.084 & $\phantom{<}.002$ & -0.230 \\ 
  $\mathsf{primuse}\sim{}\mathsf{fit}$ & $H_{5}$ & 0.081 & 0.057 & $\phantom{<}.156$ & 0.072 \\ 
  $\mathsf{primuse}\sim{}\mathsf{ts}$*** & $H_{5}$ & -0.178 & 0.054 & $<.001$ & -0.177 \\ 
  $\mathsf{privuse}\sim{}\mathsf{fit}$ & $H_{6}$ & 0.084 & 0.067 & $\phantom{<}.209$ & 0.074 \\ 
  $\mathsf{privuse}\sim{}\mathsf{ts}$* & $H_{6}$ & -0.143 & 0.060 & $\phantom{<}.017$ & -0.140 \\ 
  $\mathsf{trust}\sim{}\mathsf{fc}$*** & $H_{7}$ & 1.201 & 0.047 & $<.001$ & 0.782 \\ 
  $\mathsf{ease}\sim{}\mathsf{fc}$*** & $H_{8}$ & 0.627 & 0.061 & $<.001$ & 0.438 \\ 
  $\mathsf{primuse}\sim{}\mathsf{results}$** & $H_{9}$ & 0.167 & 0.062 & $\phantom{<}.007$ & 0.157 \\ 
  $\mathsf{privuse}\sim{}\mathsf{results}$** & $H_{10}$ & 0.195 & 0.062 & $\phantom{<}.002$ & 0.181 \\ 
  $\mathsf{ease}\sim{}\mathsf{results}$*** & $H_{11}$ & 0.499 & 0.042 & $<.001$ & 0.475 \\ 
  $\mathsf{primuse}\sim{}\mathsf{trust}$*** & $H_{12}$ & 0.391 & 0.035 & $<.001$ & 0.414 \\ 
  $\mathsf{privuse}\sim{}\mathsf{trust}$*** & $H_{13}$ & 0.441 & 0.036 & $<.001$ & 0.461 \\ 
  $\mathsf{primuse}\sim{}\mathsf{ease}$* & $H_{14}$ & 0.120 & 0.060 & $\phantom{<}.047$ & 0.119 \\ 
  $\mathsf{privuse}\sim{}\mathsf{ease}$** & $H_{15}$ & 0.159 & 0.050 & $\phantom{<}.001$ & 0.155 \\ 
  $\mathsf{bi}\sim{}\mathsf{primuse}$*** & $H_{16}$ & 0.713 & 0.051 & $<.001$ & 0.682 \\ 
  $\mathsf{bi}\sim{}\mathsf{privuse}$*** & $H_{17}$ & 0.438 & 0.064 & $<.001$ & 0.425 \\ 
  $\mathsf{bi}\sim{}\mathsf{ease}$ & $H_{18}$ & 0.009 & 0.055 & $\phantom{<}.869$ & 0.008 \\ 
   \bottomrule
\end{tabular}
\endgroup
\end{table}
}
\newcommand{\selectedRegressionsFitTAMsig}{
\begin{table}[ht]
\centering
\caption{Selected coefficients of the correlational SEM, $p < .15$.} 
\label{tab:selectedRegressionsFitTAMsig}
\begingroup\footnotesize
\begin{tabular}{lcrrrr}
  \toprule
Relation & $H$ & $B$ & $\vari{SE}$ & $p$ & $\beta$ \\ 
  \midrule
$\mathsf{trust}\sim{}\mathsf{fc}$*** & $H_{7}$ & 1.201 & 0.047 & $<.001$ & 0.782 \\ 
  $\mathsf{primuse}\sim{}\mathsf{ease}$* & $H_{14}$ & 0.120 & 0.060 & $\phantom{<}.047$ & 0.119 \\ 
  $\mathsf{primuse}\sim{}\mathsf{results}$** & $H_{9}$ & 0.167 & 0.062 & $\phantom{<}.007$ & 0.157 \\ 
  $\mathsf{primuse}\sim{}\mathsf{trust}$*** & $H_{12}$ & 0.391 & 0.035 & $<.001$ & 0.414 \\ 
  $\mathsf{primuse}\sim{}\mathsf{ctrl}$*** & $H_{3}$ & -0.700 & 0.128 & $<.001$ & -0.585 \\ 
  $\mathsf{primuse}\sim{}\mathsf{aware}$*** & $H_{3}$ & 0.819 & 0.149 & $<.001$ & 0.704 \\ 
  $\mathsf{primuse}\sim{}\mathsf{ts}$*** & $H_{5}$ & -0.178 & 0.054 & $<.001$ & -0.177 \\ 
  $\mathsf{privuse}\sim{}\mathsf{ease}$** & $H_{15}$ & 0.159 & 0.050 & $\phantom{<}.001$ & 0.155 \\ 
  $\mathsf{privuse}\sim{}\mathsf{results}$** & $H_{10}$ & 0.195 & 0.062 & $\phantom{<}.002$ & 0.181 \\ 
  $\mathsf{privuse}\sim{}\mathsf{trust}$*** & $H_{13}$ & 0.441 & 0.036 & $<.001$ & 0.461 \\ 
  $\mathsf{privuse}\sim{}\mathsf{ctrl}$*** & $H_{4}$ & -0.783 & 0.159 & $<.001$ & -0.645 \\ 
  $\mathsf{privuse}\sim{}\mathsf{aware}$*** & $H_{4}$ & 1.074 & 0.192 & $<.001$ & 0.910 \\ 
  $\mathsf{privuse}\sim{}\mathsf{collect}$** & $H_{4}$ & -0.258 & 0.084 & $\phantom{<}.002$ & -0.230 \\ 
  $\mathsf{privuse}\sim{}\mathsf{ts}$* & $H_{6}$ & -0.143 & 0.060 & $\phantom{<}.017$ & -0.140 \\ 
  $\mathsf{ease}\sim{}\mathsf{results}$*** & $H_{11}$ & 0.499 & 0.042 & $<.001$ & 0.475 \\ 
  $\mathsf{ease}\sim{}\mathsf{fc}$*** & $H_{8}$ & 0.627 & 0.061 & $<.001$ & 0.438 \\ 
  $\mathsf{bi}\sim{}\mathsf{primuse}$*** & $H_{16}$ & 0.713 & 0.051 & $<.001$ & 0.682 \\ 
  $\mathsf{bi}\sim{}\mathsf{privuse}$*** & $H_{17}$ & 0.438 & 0.064 & $<.001$ & 0.425 \\ 
  $\mathsf{bi}\sim{}\mathsf{ctrl}$*** & $H_{1}$ & 0.998 & 0.181 & $<.001$ & 0.797 \\ 
  $\mathsf{bi}\sim{}\mathsf{aware}$*** & $H_{1}$ & -1.223 & 0.213 & $<.001$ & -1.006 \\ 
  $\mathsf{bi}\sim{}\mathsf{collect}$*** & $H_{1}$ & 0.411 & 0.089 & $<.001$ & 0.355 \\ 
  $\mathsf{bi}\sim{}\mathsf{ts}$* & $H_{2}$ & 0.142 & 0.066 & $\phantom{<}.031$ & 0.135 \\ 
   \bottomrule
\end{tabular}
\endgroup
\end{table}
}
\newcommand{\selectedRegressionsFitTAMbySize}{
\begin{table}[ht]
\centering
\caption{Regression coefficients of the correlational SEM, by size} 
\label{tab:selectedRegressionsFitTAMbySize}
\begingroup\footnotesize
\begin{tabular}{lcrrrr}
  \toprule
Relation & $H$ & $B$ & $\vari{SE}$ & $p$ & $\beta$ \\ 
  \midrule
$\mathsf{bi}\sim{}\mathsf{aware}$*** & $H_{1}$ & -1.223 & 0.213 & $<.001$ & -1.006 \\ 
  $\mathsf{privuse}\sim{}\mathsf{ctrl}$*** & $H_{4}$ & -0.783 & 0.159 & $<.001$ & -0.645 \\ 
  $\mathsf{primuse}\sim{}\mathsf{ctrl}$*** & $H_{3}$ & -0.700 & 0.128 & $<.001$ & -0.585 \\ 
  $\mathsf{privuse}\sim{}\mathsf{collect}$** & $H_{4}$ & -0.258 & 0.084 & $\phantom{<}.002$ & -0.230 \\ 
  $\mathsf{primuse}\sim{}\mathsf{ts}$*** & $H_{5}$ & -0.178 & 0.054 & $<.001$ & -0.177 \\ 
  $\mathsf{privuse}\sim{}\mathsf{ts}$* & $H_{6}$ & -0.143 & 0.060 & $\phantom{<}.017$ & -0.140 \\ 
  $\mathsf{bi}\sim{}\mathsf{fit}$ & $H_{2}$ & -0.077 & 0.072 & $\phantom{<}.289$ & -0.065 \\ 
  $\mathsf{primuse}\sim{}\mathsf{collect}$ & $H_{3}$ & -0.048 & 0.065 & $\phantom{<}.455$ & -0.044 \\ 
  $\mathsf{bi}\sim{}\mathsf{ease}$ & $H_{18}$ & 0.009 & 0.055 & $\phantom{<}.869$ & 0.008 \\ 
  $\mathsf{primuse}\sim{}\mathsf{fit}$ & $H_{5}$ & 0.081 & 0.057 & $\phantom{<}.156$ & 0.072 \\ 
  $\mathsf{privuse}\sim{}\mathsf{fit}$ & $H_{6}$ & 0.084 & 0.067 & $\phantom{<}.209$ & 0.074 \\ 
  $\mathsf{primuse}\sim{}\mathsf{ease}$* & $H_{14}$ & 0.120 & 0.060 & $\phantom{<}.047$ & 0.119 \\ 
  $\mathsf{bi}\sim{}\mathsf{ts}$* & $H_{2}$ & 0.142 & 0.066 & $\phantom{<}.031$ & 0.135 \\ 
  $\mathsf{privuse}\sim{}\mathsf{ease}$** & $H_{15}$ & 0.159 & 0.050 & $\phantom{<}.001$ & 0.155 \\ 
  $\mathsf{primuse}\sim{}\mathsf{results}$** & $H_{9}$ & 0.167 & 0.062 & $\phantom{<}.007$ & 0.157 \\ 
  $\mathsf{privuse}\sim{}\mathsf{results}$** & $H_{10}$ & 0.195 & 0.062 & $\phantom{<}.002$ & 0.181 \\ 
  $\mathsf{bi}\sim{}\mathsf{collect}$*** & $H_{1}$ & 0.411 & 0.089 & $<.001$ & 0.355 \\ 
  $\mathsf{primuse}\sim{}\mathsf{trust}$*** & $H_{12}$ & 0.391 & 0.035 & $<.001$ & 0.414 \\ 
  $\mathsf{bi}\sim{}\mathsf{privuse}$*** & $H_{17}$ & 0.438 & 0.064 & $<.001$ & 0.425 \\ 
  $\mathsf{ease}\sim{}\mathsf{fc}$*** & $H_{8}$ & 0.627 & 0.061 & $<.001$ & 0.438 \\ 
  $\mathsf{privuse}\sim{}\mathsf{trust}$*** & $H_{13}$ & 0.441 & 0.036 & $<.001$ & 0.461 \\ 
  $\mathsf{ease}\sim{}\mathsf{results}$*** & $H_{11}$ & 0.499 & 0.042 & $<.001$ & 0.475 \\ 
  $\mathsf{bi}\sim{}\mathsf{primuse}$*** & $H_{16}$ & 0.713 & 0.051 & $<.001$ & 0.682 \\ 
  $\mathsf{primuse}\sim{}\mathsf{aware}$*** & $H_{3}$ & 0.819 & 0.149 & $<.001$ & 0.704 \\ 
  $\mathsf{trust}\sim{}\mathsf{fc}$*** & $H_{7}$ & 1.201 & 0.047 & $<.001$ & 0.782 \\ 
  $\mathsf{bi}\sim{}\mathsf{ctrl}$*** & $H_{1}$ & 0.998 & 0.181 & $<.001$ & 0.797 \\ 
  $\mathsf{privuse}\sim{}\mathsf{aware}$*** & $H_{4}$ & 1.074 & 0.192 & $<.001$ & 0.910 \\ 
   \bottomrule
\end{tabular}
\endgroup
\end{table}
}

\newcommand{\privacyComparison}{
\begin{table}[ht]
\centering
\caption{Comparison of privacy concern in indirect and direct modelling.} 
\label{tab:privacyComparison}
\begingroup\footnotesize
\begin{tabular}{crrrrrrrr}
  \toprule
Relation & $B$ & $\vari{SE}$ & $p$ & $\beta$ & $B_d$ & $\vari{SE}_d$ & $p_d$ & $\beta_d$ \\ 
  \midrule
$\mathsf{primuse}\sim{}\mathsf{ctrl}$ & -0.314 & 0.069 & $<.001$ & -0.269 & -0.700 & 0.128 & $<.001$ & -0.585 \\ 
  $\mathsf{primuse}\sim{}\mathsf{aware}$ & 0.340 & 0.078 & $<.001$ & 0.303 & 0.819 & 0.149 & $<.001$ & 0.704 \\ 
  $\mathsf{primuse}\sim{}\mathsf{collect}$ & 0.136 & 0.044 & $\phantom{<}.002$ & 0.123 & -0.048 & 0.065 & $\phantom{<}.455$ & -0.044 \\ 
  $\mathsf{privuse}\sim{}\mathsf{ctrl}$ & -0.247 & 0.069 & $<.001$ & -0.208 & -0.783 & 0.159 & $<.001$ & -0.645 \\ 
  $\mathsf{privuse}\sim{}\mathsf{aware}$ & 0.390 & 0.078 & $<.001$ & 0.340 & 1.074 & 0.192 & $<.001$ & 0.910 \\ 
  $\mathsf{privuse}\sim{}\mathsf{collect}$ & -0.017 & 0.041 & $\phantom{<}.677$ & -0.015 & -0.258 & 0.084 & $\phantom{<}.002$ & -0.230 \\ 
  $\mathsf{bi}\sim{}\mathsf{ctrl}$ &  &  & NA &  & 0.998 & 0.181 & $<.001$ & 0.797 \\ 
  $\mathsf{bi}\sim{}\mathsf{aware}$ &  &  & NA &  & -1.223 & 0.213 & $<.001$ & -1.006 \\ 
  $\mathsf{bi}\sim{}\mathsf{collect}$ &  &  & NA &  & 0.411 & 0.089 & $<.001$ & 0.355 \\ 
   \bottomrule
\end{tabular}
\endgroup
\end{table}
}

\newcommand{\lavaanPlot}{
\begin{sidewaysfigure}[p]
{\centering
\includegraphics[keepaspectratio,width=\textwidth]{lavaanplot_fitTrust}\caption{Measurement and Structural Model}
\label{fig:plot.lavaan}
}
\end{sidewaysfigure}
}
\newcommand{\lavaanPlotSig}{
\begin{sidewaysfigure}[p]
{\centering
\includegraphics[keepaspectratio,width=\textwidth]{lavaanplot_fitTrust_sig}\caption{Signifiant paths of the Structural Equation Model}
\label{fig:plot.lavaan}
}
\end{sidewaysfigure}
}
\newcommand{\lavaanPlotRev}{
\begin{sidewaysfigure}[p]
{\centering
\includegraphics[keepaspectratio,width=\textwidth]{lavaanplot_fitTrust_rev}\caption{Refined Measurement and Structural Model}
\label{fig:plot.lavaan.rev}
}
\end{sidewaysfigure}
}
\newcommand{\lavaanPlotRevSig}{
\begin{sidewaysfigure}[p]
{\centering
\includegraphics[keepaspectratio,width=\textwidth]{lavaanplot_fitTrust_rev_sig}\caption{Signifiant paths of the Refined Structural Equation Model}
\label{fig:plot.lavaan}
}
\end{sidewaysfigure}
}




\newcommand{\instumentReliabilityTable}{
\begin{table*}[ht]
\centering
\caption{Pre-model reliability metrics of the instruments used.} 
\label{tab:instruments.reliability}
\begingroup\footnotesize
\begin{tabular}{lrrrrrrlrr}
  \toprule
Construct & $\alpha$ & $\mathsf{LL}(\alpha)$ & $\mathsf{UL}(\alpha)$ & $\lambda_{6}$ & $\mathsf{avg}(r)$ & $r_{\mathsf{i/w}}$ & Bartlett $p$ & \textsf{KMO} & \textsf{MSA} \\ 
  \midrule
IUIPC Control & 0.65 & 0.61 & 0.69 & 0.60 & 0.38 &  & $<.001$ & 0.58 &  \\ 
  \qquad ctrl1 &  &  &  &  &  & 0.82 &  &  & 0.55 \\ 
  \qquad ctrl2 &  &  &  &  &  & 0.82 &  &  & 0.55 \\ 
  \qquad ctrl3 &  &  &  &  &  & 0.66 &  &  & 0.79 \\ 
  IUIPC Awareness & 0.69 & 0.65 & 0.73 & 0.63 & 0.43 &  & $<.001$ & 0.61 &  \\ 
  \qquad awa1 &  &  &  &  &  & 0.77 &  &  & 0.59 \\ 
  \qquad awa2 &  &  &  &  &  & 0.78 &  &  & 0.58 \\ 
  \qquad awa3 &  &  &  &  &  & 0.79 &  &  & 0.78 \\ 
  IUIPC Collection & 0.89 & 0.87 & 0.90 & 0.86 & 0.66 &  & $<.001$ & 0.82 &  \\ 
  \qquad coll1 &  &  &  &  &  & 0.87 &  &  & 0.86 \\ 
  \qquad coll2 &  &  &  &  &  & 0.79 &  &  & 0.90 \\ 
  \qquad coll3 &  &  &  &  &  & 0.92 &  &  & 0.75 \\ 
  \qquad coll4 &  &  &  &  &  & 0.87 &  &  & 0.81 \\ 
  Faith in Technology & 0.83 & 0.81 & 0.85 & 0.79 & 0.55 &  & $<.001$ & 0.81 &  \\ 
  \qquad MKF1 &  &  &  &  &  & 0.83 &  &  & 0.79 \\ 
  \qquad MKF2 &  &  &  &  &  & 0.85 &  &  & 0.79 \\ 
  \qquad MKF3 &  &  &  &  &  & 0.81 &  &  & 0.82 \\ 
  \qquad MKF4 &  &  &  &  &  & 0.77 &  &  & 0.85 \\ 
  Trusting Stance & 0.87 & 0.85 & 0.88 & 0.84 & 0.69 &  & $<.001$ & 0.69 &  \\ 
  \qquad MKTS1 &  &  &  &  &  & 0.92 &  &  & 0.66 \\ 
  \qquad MKTS2 &  &  &  &  &  & 0.93 &  &  & 0.64 \\ 
  \qquad MKTS3 &  &  &  &  &  & 0.82 &  &  & 0.86 \\ 
  Subjective Norm & 0.95 & 0.94 & 0.95 & 0.93 & 0.86 &  & $<.001$ & 0.76 &  \\ 
  \qquad STSI1 &  &  &  &  &  & 0.94 &  &  & 0.82 \\ 
  \qquad STSI2 &  &  &  &  &  & 0.96 &  &  & 0.71 \\ 
  \qquad STSI3 &  &  &  &  &  & 0.95 &  &  & 0.77 \\ 
  Facilitating Conditions & 0.79 & 0.76 & 0.81 & 0.76 & 0.48 &  & $<.001$ & 0.73 &  \\ 
  \qquad STFC1 &  &  &  &  &  & 0.81 &  &  & 0.69 \\ 
  \qquad STFC2 &  &  &  &  &  & 0.85 &  &  & 0.69 \\ 
  \qquad STFC3 &  &  &  &  &  & 0.76 &  &  & 0.78 \\ 
  \qquad STFC4 &  &  &  &  &  & 0.69 &  &  & 0.78 \\ 
  Results Demonstrability & 0.84 & 0.82 & 0.86 & 0.79 & 0.64 &  & $<.001$ & 0.72 &  \\ 
  \qquad TAR1 &  &  &  &  &  & 0.86 &  &  & 0.75 \\ 
  \qquad TAR2 &  &  &  &  &  & 0.89 &  &  & 0.68 \\ 
  \qquad TAR3 &  &  &  &  &  & 0.87 &  &  & 0.74 \\ 
  Trust & 0.91 & 0.90 & 0.92 & 0.90 & 0.62 &  & $<.001$ & 0.87 &  \\ 
  \qquad STT1 &  &  &  &  &  & 0.87 &  &  & 0.81 \\ 
  \qquad STT2 &  &  &  &  &  & 0.89 &  &  & 0.80 \\ 
  \qquad STT3 &  &  &  &  &  & 0.83 &  &  & 0.91 \\ 
  \qquad STT4 &  &  &  &  &  & 0.86 &  &  & 0.90 \\ 
  \qquad STT5 &  &  &  &  &  & 0.76 &  &  & 0.93 \\ 
  \qquad STT6 &  &  &  &  &  & 0.73 &  &  & 0.96 \\ 
  Primary Usefulness & 0.90 & 0.89 & 0.91 & 0.89 & 0.69 &  & $<.001$ & 0.78 &  \\ 
  \qquad STPE1 &  &  &  &  &  & 0.82 &  &  & 0.82 \\ 
  \qquad STPE2 &  &  &  &  &  & 0.89 &  &  & 0.82 \\ 
  \qquad STPE3 &  &  &  &  &  & 0.90 &  &  & 0.75 \\ 
  \qquad STPE4 &  &  &  &  &  & 0.90 &  &  & 0.75 \\ 
  Privacy Usefulness & 0.94 & 0.94 & 0.95 & 0.93 & 0.81 &  & $<.001$ & 0.86 &  \\ 
  \qquad TAU1 &  &  &  &  &  & 0.92 &  &  & 0.89 \\ 
  \qquad TAU2 &  &  &  &  &  & 0.94 &  &  & 0.82 \\ 
  \qquad TAU3 &  &  &  &  &  & 0.94 &  &  & 0.82 \\ 
  \qquad TAU4 &  &  &  &  &  & 0.90 &  &  & 0.91 \\ 
  Ease of Use & 0.89 & 0.87 & 0.90 & 0.86 & 0.66 &  & $<.001$ & 0.84 &  \\ 
  \qquad TAE1 &  &  &  &  &  & 0.87 &  &  & 0.86 \\ 
  \qquad TAE2 &  &  &  &  &  & 0.85 &  &  & 0.86 \\ 
  \qquad TAE3 &  &  &  &  &  & 0.87 &  &  & 0.81 \\ 
  \qquad TAE4 &  &  &  &  &  & 0.86 &  &  & 0.82 \\ 
  Behavioral Intention & 0.95 & 0.95 & 0.96 & 0.95 & 0.84 &  & $<.001$ & 0.85 &  \\ 
  \qquad STBI1 &  &  &  &  &  & 0.96 &  &  & 0.84 \\ 
  \qquad STBI2 &  &  &  &  &  & 0.90 &  &  & 0.90 \\ 
  \qquad STBI3 &  &  &  &  &  & 0.96 &  &  & 0.81 \\ 
  \qquad STBI4 &  &  &  &  &  & 0.93 &  &  & 0.85 \\ 
   \bottomrule
\end{tabular}
\endgroup
\end{table*}
}

\newcommand{\fmFITwls}{
\begin{table*}[ht]
\centering
\caption{Scaled fit measures of the WLSMV-estimated Faith-in-Technology measurement model} 
\label{tab:fmFITwls}
\begingroup\footnotesize
\begin{tabular}{rrrrrrrrrr}
  \toprule
$\chi^2$ & $\mathit{df}$ & $p_{\chi^2}$ & \textsf{CFI} & \textsf{TFI} & \textsf{RMSEA} & \textsf{LL} & \textsf{UL} & $p_{\mathsf{rmsea}}$ & \textsf{SRMR} \\ 
  \midrule
81.10 & 13.00 & $<.001$ & 0.99 & 0.99 & 0.08 & 0.06 & 0.10 & $\phantom{<}.001$ & 0.03 \\ 
   \bottomrule
\end{tabular}
\endgroup
\end{table*}
}
\newcommand{\residualsFITwls}{
\begin{table*}[tbp]
\centering\caption{Residuals of the WLSMV-estimated Faith-in-Technology measurement model}
\label{tab:residualsFITwls}
\captionsetup{position=top}
\subfloat[Correlation residuals]{
\label{tab:residualsFITwlscor}
\centering
\begingroup\footnotesize
\begin{tabular}{rlllllll}
  \toprule
 & 1 & 2 & 3 & 4 & 5 & 6 & 7 \\ 
  \midrule
1. MKF1 & --- &  &  &  &  &  &  \\ 
  2. MKF2 & 0.018 & --- &  &  &  &  &  \\ 
  3. MKF3 & 0.014 & 0.033 & --- &  &  &  &  \\ 
  4. MKF4 & -0.028 & -0.042 & -0.03 & --- &  &  &  \\ 
  5. MKTS1 & -0.032 & -0.032 & -0.055 & 0.046 & --- &  &  \\ 
  6. MKTS2 & -0.018 & -0.034 & -0.05 & 0.052 & 0.004 & --- &  \\ 
  7. MKTS3 & 0.023 & 0.007 & 0.011 & \textbf{0.115} & -0.011 & -0.014 & --- \\ 
   \bottomrule
\multicolumn{8}{c}{\emph{Note:} Correlation residuals in absolute $> 0.1$ are marked}\\
\end{tabular}
\endgroup
}

\subfloat[Covariance residuals]{%
\label{tab:residualsFITwlsraw}
\centering
\begingroup\footnotesize
\begin{tabular}{rlllllll}
  \toprule
 & 1 & 2 & 3 & 4 & 5 & 6 & 7 \\ 
  \midrule
1. MKF1 & --- &  &  &  &  &  &  \\ 
  2. MKF2 & 0.018 & --- &  &  &  &  &  \\ 
  3. MKF3 & 0.014 & 0.033 & --- &  &  &  &  \\ 
  4. MKF4 & -0.028 & -0.042 & -0.03 & --- &  &  &  \\ 
  5. MKTS1 & -0.032 & -0.032 & -0.055 & 0.046 & --- &  &  \\ 
  6. MKTS2 & -0.018 & -0.034 & -0.05 & 0.052 & 0.004 & --- &  \\ 
  7. MKTS3 & 0.023 & 0.007 & 0.011 & 0.115 & -0.011 & -0.014 & --- \\ 
   \bottomrule
\multicolumn{8}{c}{\emph{Note:} Standardized residuals are not available for estimator WLSMVS}\\
\end{tabular}
\endgroup
}
\end{table*}
}
\newcommand{\loadingsFITwls}{
\begin{table*}[tbp]
\centering
\caption{Loadings of the WLSMV-estimated Faith-in-Technology measurement model} 
\label{tab:loadingsFITwls}
\begingroup\footnotesize
\begin{tabular}{lllrrrrrrrrrrrr}
  \toprule
 \multirow{2}{*}{Factor} & \multirow{2}{*}{Indicator} & \multicolumn{4}{c}{Factor Loading} & \multicolumn{4}{c}{Standardized Solution} & \multicolumn{5}{c}{Reliability} \\ 
 \cmidrule(lr){3-6} \cmidrule(lr){7-10} \cmidrule(lr){11-15} 
  & & $\lambda$ & $\vari{SE}_{\lambda}$ & $Z_{\lambda}$ & $p_{\lambda}$ & $\beta$ & $\vari{SE}_{\beta}$ & $Z_{\beta}$ & $p_{\beta}$ & $R^2$ & $\mathit{AVE}$ & $\alpha$ & $\omega$ & $\mathit{S/\!N}_\omega$ \\
  \midrule
 fit & MKF1 & $1.00^+$ &  &  &  & 0.82 & 0.02 & 49.38 & $<.001$ & 0.68 & 0.63 & 0.83 & 0.82 & 4.61 \\ 
   & MKF2 & $1.00\phantom{^+}$ & 0.03 & 36.79 & $<.001$ & 0.82 & 0.02 & 48.32 & $<.001$ & 0.67 &  &  &  &  \\ 
   & MKF3 & $0.94\phantom{^+}$ & 0.03 & 32.81 & $<.001$ & 0.78 & 0.02 & 41.76 & $<.001$ & 0.60 &  &  &  &  \\ 
   & MKF4 & $0.90\phantom{^+}$ & 0.03 & 30.86 & $<.001$ & 0.74 & 0.02 & 35.80 & $<.001$ & 0.55 &  &  &  &  \\ 
  ts & MKTS1 & $1.00^+$ &  &  &  & 0.91 & 0.01 & 96.41 & $<.001$ & 0.82 & 0.77 & 0.87 & 0.89 & 8.44 \\ 
   & MKTS2 & $1.06\phantom{^+}$ & 0.02 & 58.00 & $<.001$ & 0.97 & 0.01 & 111.42 & $<.001$ & 0.93 &  &  &  &  \\ 
   & MKTS3 & $0.81\phantom{^+}$ & 0.02 & 49.25 & $<.001$ & 0.74 & 0.02 & 46.95 & $<.001$ & 0.54 &  &  &  &  \\ 
   \bottomrule
\multicolumn{14}{c}{\emph{Note:} $^+$ fixed parameter; the standardized solution is STDALL}\\
\end{tabular}
\endgroup
\end{table*}
}
\newcommand{\fmIUIPCwls}{
\begin{table*}[ht]
\centering
\caption{Scaled fit measures of the WLSMV-estimated IUIPC 1\textsuperscript{st}-order measurement model} 
\label{tab:fmIUIPCwls}
\begingroup\footnotesize
\begin{tabular}{rrrrrrrrrr}
  \toprule
$\chi^2$ & $\mathit{df}$ & $p_{\chi^2}$ & \textsf{CFI} & \textsf{TFI} & \textsf{RMSEA} & \textsf{LL} & \textsf{UL} & $p_{\mathsf{rmsea}}$ & \textsf{SRMR} \\ 
  \midrule
551.57 & 32.00 & $<.001$ & 0.96 & 0.94 & 0.14 & 0.13 & 0.15 & $<.001$ & 0.09 \\ 
   \bottomrule
\end{tabular}
\endgroup
\end{table*}
}
\newcommand{\residualsIUIPCwls}{
\begin{table*}[tbp]
\centering\caption{Residuals of the WLSMV-estimated IUIPC 1\textsuperscript{st}-order measurement model}
\label{tab:residualsIUIPCwls}
\captionsetup{position=top}
\subfloat[Correlation residuals]{
\label{tab:residualsIUIPCwlscor}
\centering
\begingroup\footnotesize
\begin{tabular}{rllllllllll}
  \toprule
 & 1 & 2 & 3 & 4 & 5 & 6 & 7 & 8 & 9 & 10 \\ 
  \midrule
1. ctrl1 & --- &  &  &  &  &  &  &  &  &  \\ 
  2. ctrl2 & \textbf{0.103} & --- &  &  &  &  &  &  &  &  \\ 
  3. ctrl3 & \textbf{-0.143} & \textbf{-0.151} & --- &  &  &  &  &  &  &  \\ 
  4. awa1 & -0.007 & -0.015 & \textbf{0.1} & --- &  &  &  &  &  &  \\ 
  5. awa2 & -0.016 & -0.015 & 0.073 & 0.061 & --- &  &  &  &  &  \\ 
  6. awa3 & -0.072 & -0.064 & 0.021 & \textbf{-0.102} & \textbf{-0.106} & --- &  &  &  &  \\ 
  7. coll1 & \textbf{-0.156} & \textbf{-0.141} & \textbf{0.113} & \textbf{-0.173} & \textbf{-0.13} & 0.085 & --- &  &  &  \\ 
  8. coll2 & -0.037 & -0.048 & \textbf{0.176} & -0.06 & -0.006 & \textbf{0.121} & 0.029 & --- &  &  \\ 
  9. coll3 & -0.073 & \textbf{-0.105} & \textbf{0.141} & \textbf{-0.127} & -0.096 & \textbf{0.116} & 0.008 & -0.009 & --- &  \\ 
  10. coll4 & -0.056 & -0.054 & \textbf{0.171} & -0.02 & -0.027 & \textbf{0.159} & -0.007 & -0.031 & 0.003 & --- \\ 
   \bottomrule
\multicolumn{11}{c}{\emph{Note:} Correlation residuals in absolute $> 0.1$ are marked}\\
\end{tabular}
\endgroup
}

\subfloat[Covariance residuals]{%
\label{tab:residualsIUIPCwlsraw}
\centering
\begingroup\footnotesize
\begin{tabular}{rllllllllll}
  \toprule
 & 1 & 2 & 3 & 4 & 5 & 6 & 7 & 8 & 9 & 10 \\ 
  \midrule
1. ctrl1 & --- &  &  &  &  &  &  &  &  &  \\ 
  2. ctrl2 & 0.103 & --- &  &  &  &  &  &  &  &  \\ 
  3. ctrl3 & -0.143 & -0.151 & --- &  &  &  &  &  &  &  \\ 
  4. awa1 & -0.007 & -0.015 & 0.1 & --- &  &  &  &  &  &  \\ 
  5. awa2 & -0.016 & -0.015 & 0.073 & 0.061 & --- &  &  &  &  &  \\ 
  6. awa3 & -0.072 & -0.064 & 0.021 & -0.102 & -0.106 & --- &  &  &  &  \\ 
  7. coll1 & -0.156 & -0.141 & 0.113 & -0.173 & -0.13 & 0.085 & --- &  &  &  \\ 
  8. coll2 & -0.037 & -0.048 & 0.176 & -0.06 & -0.006 & 0.121 & 0.029 & --- &  &  \\ 
  9. coll3 & -0.073 & -0.105 & 0.141 & -0.127 & -0.096 & 0.116 & 0.008 & -0.009 & --- &  \\ 
  10. coll4 & -0.056 & -0.054 & 0.171 & -0.02 & -0.027 & 0.159 & -0.007 & -0.031 & 0.003 & --- \\ 
   \bottomrule
\multicolumn{11}{c}{\emph{Note:} Standardized residuals are not available for estimator WLSMVS}\\
\end{tabular}
\endgroup
}
\end{table*}
}
\newcommand{\loadingsIUIPCwls}{
\begin{table*}[tbp]
\centering
\caption{Loadings of the WLSMV-estimated IUIPC 1\textsuperscript{st}-order measurement model} 
\label{tab:loadingsIUIPCwls}
\begingroup\footnotesize
\begin{tabular}{lllrrrrrrrrrrrr}
  \toprule
 \multirow{2}{*}{Factor} & \multirow{2}{*}{Indicator} & \multicolumn{4}{c}{Factor Loading} & \multicolumn{4}{c}{Standardized Solution} & \multicolumn{5}{c}{Reliability} \\ 
 \cmidrule(lr){3-6} \cmidrule(lr){7-10} \cmidrule(lr){11-15} 
  & & $\lambda$ & $\vari{SE}_{\lambda}$ & $Z_{\lambda}$ & $p_{\lambda}$ & $\beta$ & $\vari{SE}_{\beta}$ & $Z_{\beta}$ & $p_{\beta}$ & $R^2$ & $\mathit{AVE}$ & $\alpha$ & $\omega$ & $\mathit{S/\!N}_\omega$ \\
  \midrule
 ctrl & ctrl1 & $1.00^+$ &  &  &  & 0.76 & 0.02 & 33.12 & $<.001$ & 0.58 & 0.55 & 0.65 & 0.74 & 2.79 \\ 
   & ctrl2 & $1.03\phantom{^+}$ & 0.05 & 19.73 & $<.001$ & 0.78 & 0.02 & 33.29 & $<.001$ & 0.61 &  &  &  &  \\ 
   & ctrl3 & $0.89\phantom{^+}$ & 0.05 & 19.24 & $<.001$ & 0.67 & 0.03 & 24.66 & $<.001$ & 0.45 &  &  &  &  \\ 
  aware & awa1 & $1.00^+$ &  &  &  & 0.81 & 0.02 & 34.65 & $<.001$ & 0.66 & 0.62 & 0.66 & 0.72 & 2.57 \\ 
   & awa2 & $1.06\phantom{^+}$ & 0.05 & 21.06 & $<.001$ & 0.86 & 0.02 & 35.52 & $<.001$ & 0.75 &  &  &  &  \\ 
   & awa3 & $0.83\phantom{^+}$ & 0.04 & 19.59 & $<.001$ & 0.67 & 0.03 & 23.16 & $<.001$ & 0.45 &  &  &  &  \\ 
  collect & coll1 & $1.00^+$ &  &  &  & 0.82 & 0.01 & 62.34 & $<.001$ & 0.68 & 0.73 & 0.89 & 0.89 & 8.44 \\ 
   & coll2 & $0.92\phantom{^+}$ & 0.02 & 40.39 & $<.001$ & 0.76 & 0.02 & 49.13 & $<.001$ & 0.57 &  &  &  &  \\ 
   & coll3 & $1.16\phantom{^+}$ & 0.02 & 56.68 & $<.001$ & 0.95 & 0.01 & 132.15 & $<.001$ & 0.91 &  &  &  &  \\ 
   & coll4 & $1.06\phantom{^+}$ & 0.02 & 61.79 & $<.001$ & 0.87 & 0.01 & 87.56 & $<.001$ & 0.76 &  &  &  &  \\ 
   \bottomrule
\multicolumn{14}{c}{\emph{Note:} $^+$ fixed parameter; the standardized solution is STDALL}\\
\end{tabular}
\endgroup
\end{table*}
}
\newcommand{\fmSTwls}{
\begin{table*}[ht]
\centering
\caption{Scaled fit measures of the WLSMV-estimated Perceived System Trustworthiness measurement model} 
\label{tab:fmSTwls}
\begingroup\footnotesize
\begin{tabular}{rrrrrrrrrr}
  \toprule
$\chi^2$ & $\mathit{df}$ & $p_{\chi^2}$ & \textsf{CFI} & \textsf{TFI} & \textsf{RMSEA} & \textsf{LL} & \textsf{UL} & $p_{\mathsf{rmsea}}$ & \textsf{SRMR} \\ 
  \midrule
1152.69 & 62.00 & $<.001$ & 0.97 & 0.96 & 0.15 & 0.14 & 0.15 & $<.001$ & 0.06 \\ 
   \bottomrule
\end{tabular}
\endgroup
\end{table*}
}
\newcommand{\residualsSTwls}{
\begin{table*}[tbp]
\centering\caption{Residuals of the WLSMV-estimated Perceived System Trustworthiness measurement model}
\label{tab:residualsSTwls}
\captionsetup{position=top}
\subfloat[Correlation residuals]{
\label{tab:residualsSTwlscor}
\centering
\begingroup\footnotesize
\begin{tabular}{rlllllllllllll}
  \toprule
 & 1 & 2 & 3 & 4 & 5 & 6 & 7 & 8 & 9 & 10 & 11 & 12 & 13 \\ 
  \midrule
1. TAR1 & --- &  &  &  &  &  &  &  &  &  &  &  &  \\ 
  2. TAR2 & 0.037 & --- &  &  &  &  &  &  &  &  &  &  &  \\ 
  3. TAR3 & -0.098 & 0.028 & --- &  &  &  &  &  &  &  &  &  &  \\ 
  4. STFC1 & -0.033 & 0.02 & -0.029 & --- &  &  &  &  &  &  &  &  &  \\ 
  5. STFC2 & -0.002 & 0.062 & 0.018 & \textbf{0.127} & --- &  &  &  &  &  &  &  &  \\ 
  6. STFC3 & -0.006 & -0.022 & -0.052 & -0.038 & -0.072 & --- &  &  &  &  &  &  &  \\ 
  7. STFC4 & 0.041 & 0.006 & -0.014 & \textbf{-0.129} & \textbf{-0.146} & 0.004 & --- &  &  &  &  &  &  \\ 
  8. STT1 & 0.003 & \textbf{-0.108} & 0.001 & -0.092 & -0.068 & 0.043 & 0.039 & --- &  &  &  &  &  \\ 
  9. STT2 & -0.027 & \textbf{-0.129} & -0.023 & -0.085 & -0.073 & 0.039 & 0.033 & 0.012 & --- &  &  &  &  \\ 
  10. STT3 & -0.009 & -0.084 & 0.027 & \textbf{-0.111} & -0.09 & 0.014 & 0.061 & -0.059 & -0.01 & --- &  &  &  \\ 
  11. STT4 & 0.041 & -0.068 & 0.058 & -0.076 & -0.042 & 0.061 & 0.093 & -0.065 & -0.065 & 0.094 & --- &  &  \\ 
  12. STT5 & -0.001 & -0.09 & -0.031 & \textbf{-0.114} & -0.056 & -0.001 & 0.035 & -0.05 & -0.053 & 0.083 & 0.098 & --- &  \\ 
  13. STT6 & \textbf{0.125} & 0.063 & \textbf{0.151} & 0.033 & 0.045 & \textbf{0.141} & \textbf{0.13} & -0.063 & -0.075 & -0.044 & -0.03 & -0.021 & --- \\ 
   \bottomrule
\multicolumn{14}{c}{\emph{Note:} Correlation residuals in absolute $> 0.1$ are marked}\\
\end{tabular}
\endgroup
}

\subfloat[Covariance residuals]{%
\label{tab:residualsSTwlsraw}
\centering
\begingroup\footnotesize
\begin{tabular}{rlllllllllllll}
  \toprule
 & 1 & 2 & 3 & 4 & 5 & 6 & 7 & 8 & 9 & 10 & 11 & 12 & 13 \\ 
  \midrule
1. TAR1 & --- &  &  &  &  &  &  &  &  &  &  &  &  \\ 
  2. TAR2 & 0.037 & --- &  &  &  &  &  &  &  &  &  &  &  \\ 
  3. TAR3 & -0.098 & 0.028 & --- &  &  &  &  &  &  &  &  &  &  \\ 
  4. STFC1 & -0.033 & 0.02 & -0.029 & --- &  &  &  &  &  &  &  &  &  \\ 
  5. STFC2 & -0.002 & 0.062 & 0.018 & 0.127 & --- &  &  &  &  &  &  &  &  \\ 
  6. STFC3 & -0.006 & -0.022 & -0.052 & -0.038 & -0.072 & --- &  &  &  &  &  &  &  \\ 
  7. STFC4 & 0.041 & 0.006 & -0.014 & -0.129 & -0.146 & 0.004 & --- &  &  &  &  &  &  \\ 
  8. STT1 & 0.003 & -0.108 & 0.001 & -0.092 & -0.068 & 0.043 & 0.039 & --- &  &  &  &  &  \\ 
  9. STT2 & -0.027 & -0.129 & -0.023 & -0.085 & -0.073 & 0.039 & 0.033 & 0.012 & --- &  &  &  &  \\ 
  10. STT3 & -0.009 & -0.084 & 0.027 & -0.111 & -0.09 & 0.014 & 0.061 & -0.059 & -0.01 & --- &  &  &  \\ 
  11. STT4 & 0.041 & -0.068 & 0.058 & -0.076 & -0.042 & 0.061 & 0.093 & -0.065 & -0.065 & 0.094 & --- &  &  \\ 
  12. STT5 & -0.001 & -0.09 & -0.031 & -0.114 & -0.056 & -0.001 & 0.035 & -0.05 & -0.053 & 0.083 & 0.098 & --- &  \\ 
  13. STT6 & 0.125 & 0.063 & 0.151 & 0.033 & 0.045 & 0.141 & 0.13 & -0.063 & -0.075 & -0.044 & -0.03 & -0.021 & --- \\ 
   \bottomrule
\multicolumn{14}{c}{\emph{Note:} Standardized residuals are not available for estimator WLSMVS}\\
\end{tabular}
\endgroup
}
\end{table*}
}
\newcommand{\loadingsSTwls}{
\begin{table*}[tbp]
\centering
\caption{Loadings of the WLSMV-estimated Perceived System Trustworthiness measurement model} 
\label{tab:loadingsSTwls}
\begingroup\footnotesize
\begin{tabular}{lllrrrrrrrrrrrr}
  \toprule
 \multirow{2}{*}{Factor} & \multirow{2}{*}{Indicator} & \multicolumn{4}{c}{Factor Loading} & \multicolumn{4}{c}{Standardized Solution} & \multicolumn{5}{c}{Reliability} \\ 
 \cmidrule(lr){3-6} \cmidrule(lr){7-10} \cmidrule(lr){11-15} 
  & & $\lambda$ & $\vari{SE}_{\lambda}$ & $Z_{\lambda}$ & $p_{\lambda}$ & $\beta$ & $\vari{SE}_{\beta}$ & $Z_{\beta}$ & $p_{\beta}$ & $R^2$ & $\mathit{AVE}$ & $\alpha$ & $\omega$ & $\mathit{S/\!N}_\omega$ \\
  \midrule
 results & TAR1 & $1.00^+$ &  &  &  & 0.83 & 0.02 & 54.73 & $<.001$ & 0.69 & 0.69 & 0.84 & 0.84 & 5.44 \\ 
   & TAR2 & $0.95\phantom{^+}$ & 0.03 & 37.84 & $<.001$ & 0.79 & 0.01 & 53.12 & $<.001$ & 0.62 &  &  &  &  \\ 
   & TAR3 & $1.04\phantom{^+}$ & 0.03 & 40.11 & $<.001$ & 0.87 & 0.01 & 58.73 & $<.001$ & 0.75 &  &  &  &  \\ 
  fc & STFC1 & $1.00^+$ &  &  &  & 0.72 & 0.02 & 39.64 & $<.001$ & 0.52 & 0.57 & 0.78 & 0.82 & 4.55 \\ 
   & STFC2 & $1.15\phantom{^+}$ & 0.04 & 32.23 & $<.001$ & 0.83 & 0.02 & 54.92 & $<.001$ & 0.69 &  &  &  &  \\ 
   & STFC3 & $1.08\phantom{^+}$ & 0.04 & 29.94 & $<.001$ & 0.78 & 0.02 & 41.95 & $<.001$ & 0.61 &  &  &  &  \\ 
   & STFC4 & $0.95\phantom{^+}$ & 0.04 & 24.04 & $<.001$ & 0.69 & 0.02 & 29.64 & $<.001$ & 0.47 &  &  &  &  \\ 
  trust & STT1 & $1.00^+$ &  &  &  & 0.95 & 0.01 & 177.83 & $<.001$ & 0.90 & 0.69 & 0.90 & 0.91 & 10.20 \\ 
   & STT2 & $1.03\phantom{^+}$ & 0.01 & 105.41 & $<.001$ & 0.98 & 0.00 & 204.78 & $<.001$ & 0.96 &  &  &  &  \\ 
   & STT3 & $0.83\phantom{^+}$ & 0.01 & 64.60 & $<.001$ & 0.79 & 0.01 & 62.30 & $<.001$ & 0.62 &  &  &  &  \\ 
   & STT4 & $0.87\phantom{^+}$ & 0.01 & 76.17 & $<.001$ & 0.83 & 0.01 & 75.62 & $<.001$ & 0.69 &  &  &  &  \\ 
   & STT5 & $0.68\phantom{^+}$ & 0.02 & 36.74 & $<.001$ & 0.65 & 0.02 & 36.21 & $<.001$ & 0.42 &  &  &  &  \\ 
   & STT6 & $0.79\phantom{^+}$ & 0.02 & 51.18 & $<.001$ & 0.75 & 0.01 & 50.37 & $<.001$ & 0.56 &  &  &  &  \\ 
   \bottomrule
\multicolumn{14}{c}{\emph{Note:} $^+$ fixed parameter; the standardized solution is STDALL}\\
\end{tabular}
\endgroup
\end{table*}
}
\newcommand{\fmTAMwls}{
\begin{table*}[ht]
\centering
\caption{Scaled fit measures of the WLSMV-estimated Perceived Technology Acceptance measurement model} 
\label{tab:fmTAMwls}
\begingroup\footnotesize
\begin{tabular}{rrrrrrrrrr}
  \toprule
$\chi^2$ & $\mathit{df}$ & $p_{\chi^2}$ & \textsf{CFI} & \textsf{TFI} & \textsf{RMSEA} & \textsf{LL} & \textsf{UL} & $p_{\mathsf{rmsea}}$ & \textsf{SRMR} \\ 
  \midrule
1487.82 & 84.00 & $<.001$ & 0.97 & 0.96 & 0.14 & 0.14 & 0.15 & $<.001$ & 0.05 \\ 
   \bottomrule
\end{tabular}
\endgroup
\end{table*}
}
\newcommand{\residualsTAMwls}{
\begin{table*}[tbp]
\centering\caption{Residuals of the WLSMV-estimated Perceived Technology Acceptance measurement model}
\label{tab:residualsTAMwls}
\captionsetup{position=top}
\subfloat[Correlation residuals]{
\label{tab:residualsTAMwlscor}
\centering
\begingroup\footnotesize
\begin{tabular}{rlllllllllllllll}
  \toprule
 & 1 & 2 & 3 & 4 & 5 & 6 & 7 & 8 & 9 & 10 & 11 & 12 & 13 & 14 & 15 \\ 
  \midrule
1. STPE1 & --- &  &  &  &  &  &  &  &  &  &  &  &  &  &  \\ 
  2. STPE2 & -0.009 & --- &  &  &  &  &  &  &  &  &  &  &  &  &  \\ 
  3. STPE3 & \textbf{-0.192} & -0.024 & --- &  &  &  &  &  &  &  &  &  &  &  &  \\ 
  4. STPE4 & \textbf{-0.186} & -0.029 & 0.075 & --- &  &  &  &  &  &  &  &  &  &  &  \\ 
  5. TAU1 & \textbf{0.119} & 0.03 & -0.035 & -0.052 & --- &  &  &  &  &  &  &  &  &  &  \\ 
  6. TAU2 & 0.073 & -0.025 & -0.069 & -0.077 & -0.008 & --- &  &  &  &  &  &  &  &  &  \\ 
  7. TAU3 & 0.044 & -0.057 & \textbf{-0.115} & \textbf{-0.122} & -0.008 & 0.017 & --- &  &  &  &  &  &  &  &  \\ 
  8. TAU4 & \textbf{0.105} & 0.05 & 0.003 & -0.012 & 0.001 & -0.046 & -0.024 & --- &  &  &  &  &  &  &  \\ 
  9. TAE1 & 0.055 & 0.055 & 0.047 & 0.03 & 0.039 & -0.009 & 0 & 0.093 & --- &  &  &  &  &  &  \\ 
  10. TAE2 & -0.001 & -0.038 & -0.003 & -0.026 & -0.03 & -0.061 & -0.051 & 0.038 & -0.022 & --- &  &  &  &  &  \\ 
  11. TAE3 & -0.029 & -0.024 & 0.007 & 0 & -0.028 & -0.019 & -0.025 & 0.051 & -0.051 & 0.02 & --- &  &  &  &  \\ 
  12. TAE4 & -0.051 & -0.021 & -0.037 & -0.043 & -0.017 & -0.072 & -0.068 & 0.026 & -0.058 & 0.05 & 0.047 & --- &  &  &  \\ 
  13. STBI1 & 0.048 & 0.015 & \textbf{-0.108} & \textbf{-0.106} & 0.018 & -0.054 & -0.071 & 0.037 & 0.013 & -0.005 & -0.039 & -0.019 & --- &  &  \\ 
  14. STBI2 & 0.054 & 0.091 & -0.04 & -0.034 & 0.021 & -0.036 & -0.069 & 0.028 & 0.061 & -0.007 & -0.011 & 0.013 & 0.003 & --- &  \\ 
  15. STBI4 & 0.089 & 0.007 & \textbf{-0.105} & \textbf{-0.118} & 0.056 & -0.01 & -0.019 & 0.073 & 0.014 & 0.012 & -0.035 & -0.038 & 0.012 & -0.046 & --- \\ 
   \bottomrule
\multicolumn{16}{c}{\emph{Note:} Correlation residuals in absolute $> 0.1$ are marked}\\
\end{tabular}
\endgroup
}

\subfloat[Covariance residuals]{%
\label{tab:residualsTAMwlsraw}
\centering
\begingroup\footnotesize
\begin{tabular}{rlllllllllllllll}
  \toprule
 & 1 & 2 & 3 & 4 & 5 & 6 & 7 & 8 & 9 & 10 & 11 & 12 & 13 & 14 & 15 \\ 
  \midrule
1. STPE1 & --- &  &  &  &  &  &  &  &  &  &  &  &  &  &  \\ 
  2. STPE2 & -0.009 & --- &  &  &  &  &  &  &  &  &  &  &  &  &  \\ 
  3. STPE3 & -0.192 & -0.024 & --- &  &  &  &  &  &  &  &  &  &  &  &  \\ 
  4. STPE4 & -0.186 & -0.029 & 0.075 & --- &  &  &  &  &  &  &  &  &  &  &  \\ 
  5. TAU1 & 0.119 & 0.03 & -0.035 & -0.052 & --- &  &  &  &  &  &  &  &  &  &  \\ 
  6. TAU2 & 0.073 & -0.025 & -0.069 & -0.077 & -0.008 & --- &  &  &  &  &  &  &  &  &  \\ 
  7. TAU3 & 0.044 & -0.057 & -0.115 & -0.122 & -0.008 & 0.017 & --- &  &  &  &  &  &  &  &  \\ 
  8. TAU4 & 0.105 & 0.05 & 0.003 & -0.012 & 0.001 & -0.046 & -0.024 & --- &  &  &  &  &  &  &  \\ 
  9. TAE1 & 0.055 & 0.055 & 0.047 & 0.03 & 0.039 & -0.009 & 0 & 0.093 & --- &  &  &  &  &  &  \\ 
  10. TAE2 & -0.001 & -0.038 & -0.003 & -0.026 & -0.03 & -0.061 & -0.051 & 0.038 & -0.022 & --- &  &  &  &  &  \\ 
  11. TAE3 & -0.029 & -0.024 & 0.007 & 0 & -0.028 & -0.019 & -0.025 & 0.051 & -0.051 & 0.02 & --- &  &  &  &  \\ 
  12. TAE4 & -0.051 & -0.021 & -0.037 & -0.043 & -0.017 & -0.072 & -0.068 & 0.026 & -0.058 & 0.05 & 0.047 & --- &  &  &  \\ 
  13. STBI1 & 0.048 & 0.015 & -0.108 & -0.106 & 0.018 & -0.054 & -0.071 & 0.037 & 0.013 & -0.005 & -0.039 & -0.019 & --- &  &  \\ 
  14. STBI2 & 0.054 & 0.091 & -0.04 & -0.034 & 0.021 & -0.036 & -0.069 & 0.028 & 0.061 & -0.007 & -0.011 & 0.013 & 0.003 & --- &  \\ 
  15. STBI4 & 0.089 & 0.007 & -0.105 & -0.118 & 0.056 & -0.01 & -0.019 & 0.073 & 0.014 & 0.012 & -0.035 & -0.038 & 0.012 & -0.046 & --- \\ 
   \bottomrule
\multicolumn{16}{c}{\emph{Note:} Standardized residuals are not available for estimator WLSMVS}\\
\end{tabular}
\endgroup
}
\end{table*}
}
\newcommand{\loadingsTAMwls}{
\begin{table*}[tbp]
\centering
\caption{Loadings of the WLSMV-estimated Perceived Technology Acceptance measurement model} 
\label{tab:loadingsTAMwls}
\begingroup\footnotesize
\begin{tabular}{lllrrrrrrrrrrrr}
  \toprule
 \multirow{2}{*}{Factor} & \multirow{2}{*}{Indicator} & \multicolumn{4}{c}{Factor Loading} & \multicolumn{4}{c}{Standardized Solution} & \multicolumn{5}{c}{Reliability} \\ 
 \cmidrule(lr){3-6} \cmidrule(lr){7-10} \cmidrule(lr){11-15} 
  & & $\lambda$ & $\vari{SE}_{\lambda}$ & $Z_{\lambda}$ & $p_{\lambda}$ & $\beta$ & $\vari{SE}_{\beta}$ & $Z_{\beta}$ & $p_{\beta}$ & $R^2$ & $\mathit{AVE}$ & $\alpha$ & $\omega$ & $\mathit{S/\!N}_\omega$ \\
  \midrule
 primuse & STPE1 & $1.00^+$ &  &  &  & 0.90 & 0.01 & 82.23 & $<.001$ & 0.81 & 0.79 & 0.90 & 0.92 & 11.66 \\ 
   & STPE2 & $0.96\phantom{^+}$ & 0.01 & 65.79 & $<.001$ & 0.86 & 0.01 & 89.23 & $<.001$ & 0.74 &  &  &  &  \\ 
   & STPE3 & $0.99\phantom{^+}$ & 0.01 & 68.51 & $<.001$ & 0.89 & 0.01 & 105.74 & $<.001$ & 0.80 &  &  &  &  \\ 
   & STPE4 & $1.00\phantom{^+}$ & 0.01 & 68.05 & $<.001$ & 0.89 & 0.01 & 108.97 & $<.001$ & 0.80 &  &  &  &  \\ 
  privuse & TAU1 & $1.00^+$ &  &  &  & 0.92 & 0.01 & 157.32 & $<.001$ & 0.85 & 0.87 & 0.94 & 0.94 & 16.96 \\ 
   & TAU2 & $1.05\phantom{^+}$ & 0.01 & 149.09 & $<.001$ & 0.97 & 0.00 & 229.29 & $<.001$ & 0.94 &  &  &  &  \\ 
   & TAU3 & $1.04\phantom{^+}$ & 0.01 & 146.62 & $<.001$ & 0.96 & 0.00 & 204.11 & $<.001$ & 0.92 &  &  &  &  \\ 
   & TAU4 & $0.96\phantom{^+}$ & 0.01 & 113.22 & $<.001$ & 0.89 & 0.01 & 110.62 & $<.001$ & 0.79 &  &  &  &  \\ 
  ease & TAE1 & $1.00^+$ &  &  &  & 0.90 & 0.01 & 78.05 & $<.001$ & 0.81 & 0.72 & 0.88 & 0.89 & 8.14 \\ 
   & TAE2 & $0.86\phantom{^+}$ & 0.02 & 45.66 & $<.001$ & 0.78 & 0.01 & 53.21 & $<.001$ & 0.60 &  &  &  &  \\ 
   & TAE3 & $0.98\phantom{^+}$ & 0.02 & 52.15 & $<.001$ & 0.88 & 0.01 & 73.86 & $<.001$ & 0.78 &  &  &  &  \\ 
   & TAE4 & $0.93\phantom{^+}$ & 0.02 & 49.31 & $<.001$ & 0.83 & 0.01 & 62.31 & $<.001$ & 0.69 &  &  &  &  \\ 
  bi & STBI1 & $1.00^+$ &  &  &  & 0.96 & 0.00 & 206.81 & $<.001$ & 0.92 & 0.86 & 0.93 & 0.93 & 13.49 \\ 
   & STBI2 & $0.94\phantom{^+}$ & 0.01 & 112.10 & $<.001$ & 0.90 & 0.01 & 129.47 & $<.001$ & 0.82 &  &  &  &  \\ 
   & STBI4 & $0.96\phantom{^+}$ & 0.01 & 109.04 & $<.001$ & 0.92 & 0.01 & 132.29 & $<.001$ & 0.85 &  &  &  &  \\ 
   \bottomrule
\multicolumn{14}{c}{\emph{Note:} $^+$ fixed parameter; the standardized solution is STDALL}\\
\end{tabular}
\endgroup
\end{table*}
}

\newcommand{\corLVCFA}{
\begin{table*}[ht]
\centering
\caption{Correlations of implied latent variables.} 
\label{tab:corLVCFA}
\begingroup\footnotesize
\begin{tabular}{rllllllllllll}
  \toprule
 & fit & ts & ctrl & aware & collect & primuse & privuse & ease & bi & results & fc & trust \\ 
  \midrule
fit & 1.00 &  &  &  &  &  &  &  &  &  &  &  \\ 
  ts & .57 & 1.00 &  &  &  &  &  &  &  &  &  &  \\ 
  ctrl & .20 & .11 & 1.00 &  &  &  &  &  &  &  &  &  \\ 
  aware & .24 & .11 & .66 & 1.00 &  &  &  &  &  &  &  &  \\ 
  collect & -.06 & -.16 & .35 & .50 & 1.00 &  &  &  &  &  &  &  \\ 
  primuse & .25 & .11 & .13 & .23 & .14 & 1.00 &  &  &  &  &  &  \\ 
  privuse & .38 & .24 & .23 & .26 & .04 & .60 & 1.00 &  &  &  &  &  \\ 
  ease & .33 & .27 & .13 & .20 & -.04 & .50 & .64 & 1.00 &  &  &  &  \\ 
  bi & .23 & .14 & .18 & .24 & .16 & .79 & .64 & .57 & 1.00 &  &  &  \\ 
  results & .27 & .22 & .21 & .26 & .03 & .51 & .67 & .82 & .60 & 1.00 &  &  \\ 
  fc & .30 & .21 & .19 & .20 & -.001 & .49 & .54 & .74 & .61 & .68 & 1.00 &  \\ 
  trust & .42 & .40 & .10 & .09 & -.16 & .54 & .68 & .64 & .66 & .64 & .63 & 1.00 \\ 
   \bottomrule
\end{tabular}
\endgroup
\end{table*}
}

\begin{DocumentVersionBootstrapping}
\end{DocumentVersionBootstrapping}

\newcommand{\ACSlong}{Attribute-Based Credential System (ACS)\xspace}
\newcommand{\ACSname}{Attribute-Based Credential System\xspace}
\newcommand{\ACSpllong}{Attribute-Based Credential Systems (ACS)\xspace}
\newcommand{\ACSplname}{Attribute-Based Credential Systems\xspace}
\newcommand{\ACS}{ACS\xspace}


\author{Rachel Crowder} 
\affiliation{%
  \institution{Newcastle University}
  \city{Newcastle upon Tyne}
  \country{United Kingdom}}

\author{George Price}
\affiliation{%
  \institution{Newcastle University}
  \city{Newcastle upon Tyne}
  \country{United Kingdom}}

\author{Thomas Gro{\ss}}
\affiliation{%
  \institution{Newcastle University}
  \city{Newcastle upon Tyne}
  \country{United Kingdom}}
  
\begin{DocumentVersionConference}
\title[Simply tell me how]{Simply tell me how---On Trustworthiness and Technology Acceptance of Attribute-Based Credentials}
\end{DocumentVersionConference}
\begin{DocumentVersionTR}
\title{Perceived Trustworthiness of Attribute-Based Credential Systems (Technical Report)\thanks{Open Science Framework: \protect\url{https://osf.io/w39bv/}}}
\end{DocumentVersionTR}

\input{abstract.tex}

\maketitle


\input{intro.tex}

\input{background.tex}

\input{related.tex}

\input{aims.tex}

\input{method.tex}

%
%




\section{Results}
\label{sec:results}

\subsection{Sample}
The initial sample consisted of $940$ participants, $473$ from the study on intrinsic properties, $467$ from the study on presentation properties.
Cases were removed from the sample without replacement based on two criteria:
\begin{inparaenum}[(i)]
  \item The participant did not complete the full survey;
  \item The participant failed more than one attention check, as pre-registered.
\end{inparaenum}
Table~\ref{tab:sample} shows the refinement to the final sample with a final $N = 812$.
We provide its demographic characteristics in Table~\ref{tab:demo}.

\begin{table}
\centering
\caption{Sample Refinement}
\label{tab:sample}
\begin{tabular}{lccccc}
\toprule
\multirow{2}{*}{Phase} & \multicolumn{2}{c}{Intrinsic} & \multicolumn{2}{c}{Presentation} &\multirow{2}{*}{Total}\\
\cmidrule(lr){2-3} \cmidrule(lr){4-5}
  & Excluded & Size & Excluded & Size & \\ 
\midrule
Starting Sample &   & $473$ & & $467$ & $940$\\
Incomplete & $58$ & $415$
   & $34$ & $433$ 
   & $848$\\
Duplicate & $25$ & $390$
   & $0$ & $433$
   & $823$\\
\textsf{FailedAC}$ > 1$ & $36$ & $379$
   & $0$ & $433$ 
   & $812$\\
\cmidrule(lr){6-6}
Final Sample & & & & & $812$ \\
\bottomrule
\end{tabular}
\end{table}

\input{demoTableOne.tex}

\subsection{Measurement Model Evaluation}
\label{sec:MM}
We first established a confirmatory factor analysis (CFA) of the measurement model, that is, of the indicator variables and latent variables considered.
We established this CFA on the variables specified in operationalization Table~\ref{tab:ops.m}\processifversion{DocumentVersionAppendix}{, including the expected covariances highlighted in Table~\ref{tab:covar} in Appendix~\ref{sec:covar}}.

We found that the measurement model showed a good global fit, documented in Table~\ref{tab:fitCFAMeasurement}. We inspected the residuals for local fit and were satisfied.
While we anticipated correlated residuals, based on observations on the questionnaire, we chose not to include those in the model \textit{ad post facto}.

\fitCFAMeasurement

Table~\ref{tab:loadingsFitCFAreliability} incorporates the WLSMV-estimated factor loadings of the measurement model. All loadings are estimated with high confidence and a $p$-value of $p < .001$.
In terms of reliability, the factors \textsf{ctrl} and \textsf{aware} show the lowest internal consistency, with $\omega = 0.73$ and a signal-to-noise ratio ($S/N_\omega \leq 2.76$). 
Privacy usefulness (\textsf{privuse} yields the greatest internal consistency, with $\omega = 0.95$).
\loadingsFitCFAreliability
Overall, we assess that the measurement model is valid and sufficiently reliable to continue our investigation with the correlational structural equation model.

\subsection{Correlational Model}
\label{sec:CorrM}
As second step, we evaluated the correlational path model, which incorporates the measurement model established in Section~\ref{sec:MM} and adds regression equations modelling the hypothesized relations introduced in Section~\ref{sec:aims}. This more complex model yielded a statistically significant improvement over the measurement model, $\chi^2(23) = 430.138, p < .001$. This already gave us confidence in the validity of the correlational model.

We established the correlational model itself in two steps: by first computing a model without the direct paths between attitudes and subjective norms to behavioral intention, that is, excluding the test of $H_1$ and $H_2$, and then adding these paths in model building. The hypothesized model incluing the paths modelling hypotheses $H_1$ and $H_2$ fits statistically significantly better than the indirect model: We rejected the equal-fit hypothesis, $\chi^2(5) = 78.199, p < .001$.

We depict the resulting structual path model in Figure~\ref{fig:semPlotModel}.  \processifversion{DocumentVersionAppendix}{Appendix~\ref{sec:full_model} also offers the full model, incl. the measurement model, in Figure~\ref{fig:semPlotModelFull}. All visualized models operate on fully standardized coefficients. The full regression fit is included in Appendix Table~\ref{tab:fullRegressionsFit}.}
\semPlotModel
We evaluate the global fit of the correlational model in Table~\ref{tab:fitTrust}, showing a decent fit. The scaled RMSEA of $0.066$ indicates potentially troublesome residuals. On the other hand, the SRMR of $0.0546934$ considered robust for the WLSMV estimation is defensible.
\fitTrust

\subsubsection{\ACS Trustworthiness \& Acceptance}
We report selected regression coefficients in Table~\ref{tab:selectedRegressionsFitTAMbyH} and discuss their consequences in the following paragraphs.

\selectedRegressionsFitTAMbyH

\begin{figure*}[tbp]
\centering
\includegraphics[keepaspectratio,width=.7\textwidth]{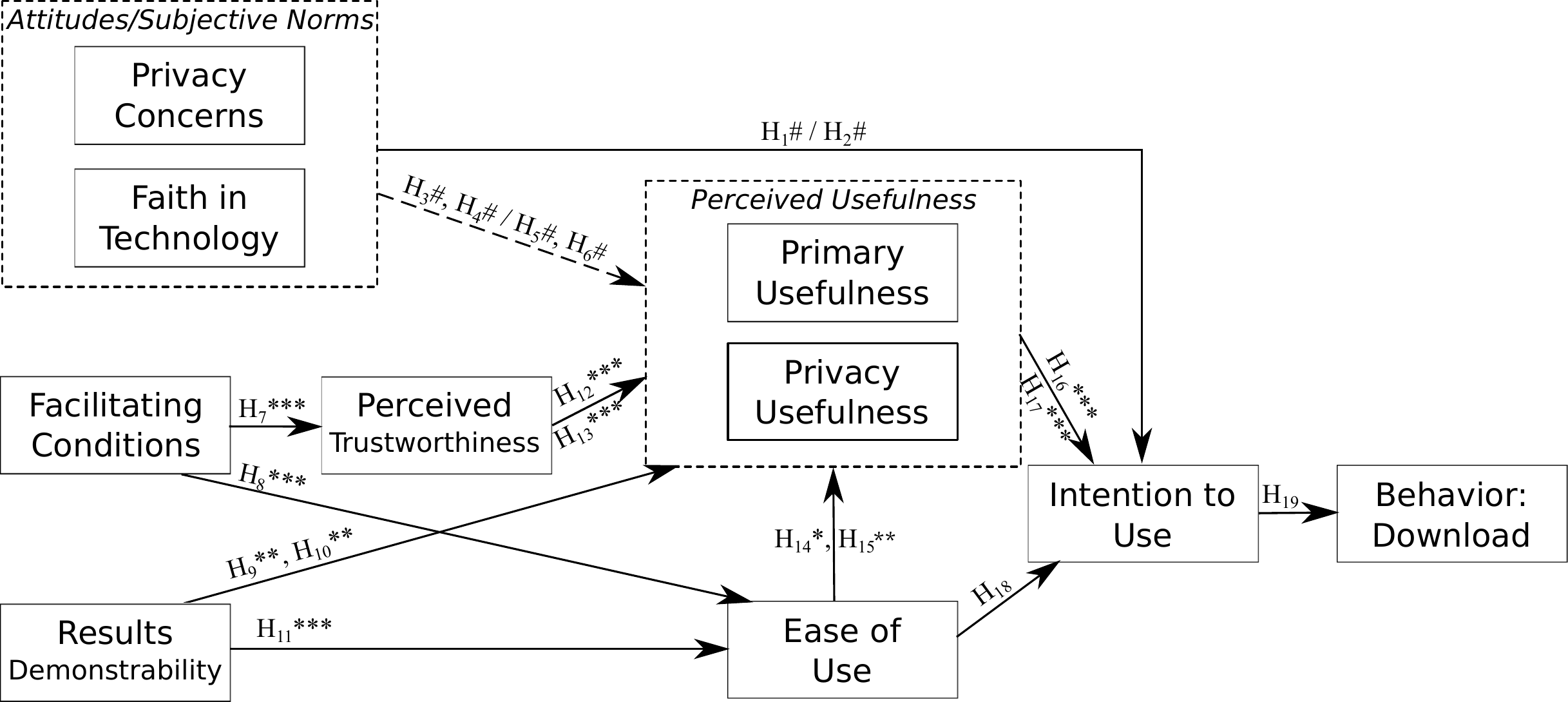}
\caption{Results on hypothesis testing.\\\normalfont{\emph{Note}: slash denotes different antecedents; comma denotes different consequences; \#: contradicting effects; *: $p < .05$, **: $p < .01$, ***: $p < .001$. }}
\label{fig:sem_aim_results}
\end{figure*}

Let us first consider the structural equation model on perceived trustworthiness and technology acceptance, that is, hypotheses $H_1$ thru $H_{18}$ from Section~\ref{sec:aims_tam}. Figure~\ref{fig:sem_aim_results} gives an overview of which hypotheses were retained, while Table~\ref{tab:selectedRegressionsFitTAMbyH} includes the corresponding near-significant or significant regression coefficients.

\subsubsection{Core Technology Acceptance Model}
As expected in core TAM, Primary Usefulness, Privacy Usefulness all impact behavioral intention positively. We thereby reject the null hypotheses corresponding to $H_{16}$ and $H_{17}$. Regarding $H_{18}$, the impact of ease of use on behavioral intention is statistically significant in the model without direct attitude/subjective-norm paths, but not so in the model with them.
Ease of Use has a statistically significant effect on both kinds of Usefulness. Hence, we reject the null hypotheses corresponding to $H_{14}$ and $H_{15}$.

\subsubsection{TAM Antecedents}
Perceived Trustworthiness has a statistically significant impact on both forms of Usefulness. Hence we reject the null hypotheses corresponding to $H_{12}$ and $H_{13}$.

Results Demonstrability has a statistically significantly positive impact on both forms of Usefulness as well as Ease of Use. We thereby reject the null hypotheses corresponding to $H_9$, $H_{10}$ and $H_{11}$.

Facilitating Conditions statistically significantly yielded a positive change in Perceived Trustworthiness and Ease of Use. We reject the null hypotheses corrresponding to $H_7$ and $H_8$.

\subsubsection{Attitudes \& Subjective Norms}
\input{privacy_comparison_total.tex}
We consider privacy concern first and present a comparison of the situation in indirect and direct modelling in Table~\ref{tab:privacyComparison}.
First, we observe that control and awareness have contradicting impact on their consequences. While \textsf{ctrl} impacts both forms of usefulness negatively, \textsf{aware} impacts them positively. The impact of collection is comparatively small.
We could not unequivably reject the null hypotheses corresponding to the hypotheses ($H_1$, $H_3$, $H_4$) on impact of privacy concerns (IUIPC control, awareness, and collection) on either Usefulness or Intention to Use. In the comparison between indirect and direct modelling, the direct effects of \textsf{ctrl}, \textsf{aware} and \textsf{collect} on behavioral intention negate the indirect effects through the perceived usefulness.

We find a similar situation with faith in technology (\textsf{fit}) and trusting stance (\textsf{ts}). Here, faith technology has a positive impact on usefulness and trusting stance a negative impact on usefulness. Likewise, the direct paths to behavioral intention negate these effects.

\subsubsection{Mediation Analysis}
We conduct a mediation analysis on the effects of privacy concern in Table~\ref{tab:privacyComparison}. The table compares the standardized coefficients $\beta$ of privacy concern factors with respect to their direct, indirect and total effect on behavioral intention. The left-side model, called ``indirect'' only incorporates the indirect effects of privacy concern through perceived usefulness. The right-side model, called ``direct'' incorporates a direct path from privacy concern factors to behavioral intention. Of the two models, the direct model is a statistically significantly better fit, $\chi^2(5) = 78.199, p < .001$. Consulting the table, we observe an inconsistent mediation, in which the direct effect contradicts the indirect effect mediated through perceived usefulness. The direct effects have a greater absolute magnitude than the sum of the indirect effects.

\subsection{Causal Model}
We compared the model with only indirect paths between attitudes and subjective norms to behavioral intention (excluding the test of $H_1$ and $H_2$) to the model incuding direct paths. We found that the model with direct paths improved the global fit statistically significantly, $\chi^2(5) = 83.791, p < .001$.
We show the global fit of the WLSMV-estimated causal path model in Table~\ref{tab:fitTrustCausal}. The causal model exhibits an excellent fit, better than the correlational model, evident in fulfilling the close-fit hypothesis.

\fitTrustCausal

We report that the regression coefficients of correlational and causal models are estimated to be essentially equal. The estimates reported in Table~\ref{tab:selectedRegressionsFitTAMbyH} still hold.
Hence, the overall model stayed stable with the addition of the causal conditions. We continue to shed light on the impact of the causal conditions.

\subsubsection{Impact of \ACS Properties}
As outlined in Section~\ref{sec:aims_causal}, we anticipated intrinsic and presentation properties experimentally manipulated to impact the antecedents of the core technology acceptance model. Table~\ref{tab:selectedRegressionsFitCausalByH} shows the corresponding near-significant and significant coefficients. We note that only the negative impact on perceived trustworthiness was statistically significant.

\selectedRegressionsFitCausalByH

\paragraph{Intrinsic Properties}
The intrinsic properties (provider, benefits and usage) were largely not statistically significant, apart from the provider being set to a company, which had a slight negative impact on primary usefulness. Hence, we failed to reject the null hypotheses corresponding to $H_{C, 1}$, $H_{C, 2}$ and $H_{C, 3}$.

\paragraph{Presentation Properties}
The presentation of the \ACS (simplicity, people, support) made a considerable difference. 
While the presence of people did not show a statistically significant impact, failing to reject the null hypothesis corresponding to $H_{C, 5}$, the other two interventions carried significant weight. In the support category it made statistically significant difference to have a contact option, affecting both trust and ease of use positively on the order of $\beta \approx 0.1$.
Hence, we rejected the null hypothesis corresponding to $H_{C, 6}$.

Simplicity had the greatest impact on the model, statistically significantly affecting primary usefulness ($\beta = 0.115$) and ease of use ($\beta = 0.2$). We thereby rejected the null hypothesis corresponding to $H_{C, 4}$.


\input{discussion.tex}

\input{limitations.tex}

\input{conclusion.tex}

\input{acknowledgments.tex}

\balance

\bibliographystyle{acm}
\bibliography{methods_resources,./sem,./qdesign,trustworthiness,psychometry,privacy,CryptoLibrary}

\end{document}

%% file: abstract.tex
\begin{abstract}
{
\ACSpllong have been long proposed as privacy-preserving means of attribute-based authentication, yet neither been considered particularly usable nor found wide-spread adoption, to date. 
%
To establish what variables drive the adoption of \ACS as a usable security and privacy technology, we investigated how intrinsic and presentation properties impact their perceived trustworthiness and behavioral intent to adopt them.
%
We conducted two confirmatory, fractional-factorial, between-subject, random-controlled trials with a total UK-representative sample of $N = 812$ participants. Each participant inspected one of 24 variants of Anonymous Credential System Web site, which encoded a combination of three intrinsic factors (\textsf{provider}, \textsf{usage}, \textsf{benefits}) and three presentation factors (\textsf{simplicity}, presence of \textsf{people}, level of available \textsf{support}). 
Participants stated their privacy and faith-in-technology subjective norms before the trial. After having completed the Web site inspection, they reported on the perceived trustworthiness, the technology adoption readiness, and their behavioral intention to follow through.
%
We established a robust covariance-based structural equation model of the perceived trustworthiness and technology acceptance, showing that communicating facilitating conditions as well as demonstrating results drive the overall acceptance and behavioral intent. Of the manipulated causal variables, communicating with simplicity and on the everyday usage had the greatest and most consistently positive impact on the overall technology acceptance.
%
After earlier correlational empirical research on \ACS technology acceptance, ours is the first research showing cause-effect relations in a structural latent factor model with substantial sample size.}
\end{abstract}
\keywords{attribute-based credentials, anonymous credential systems, perceived trustworthiness, technology acceptance}


%% file: intro.tex
\section{Introduction}

\ACSpllong and Anonymous Credential Systems as a specialization offer a great potential for the privacy protection of users in a wide range of scenarios. They have received attention from industry behemoths, such as IBM and Microsoft with their respective Identity Mixer (IDEMIX) and U-Prove implementations. They found some adoption in, for instance, in the IRMA Card~\cite{vullers2013efficient,de2014towards} of Radboud University or the collaboration between Microsoft Research and Signal, headed for deployment in the Signal client~\cite{chase2019signal}. 
However, in the global scheme of things, \ACS failed to reach broad adoption.

The complex technical and socio-economic challenges and herculean task of building a sustainable and thriving \ACS eco system notwithstanding, it remains an open question whether users would actually widely adopt \ACS granted their availability.
Furthermore, it is an open question what human factors would consistently drive such an adoption.

Prior empirical correlational research by Benenson et al.~\cite{benenson2015user} with $N=30$ university students from Patras supports that \ACS technology adoption could be modelled with the Technology Acceptance Model (TAM 2.0). Similarly, recent work by Harborth and Pape~\cite{harborth2018examining} showed that a TAM-inspired PLS path model with $N=141$ users could fit the adoption of the mix-network \textsf{JonDoNym} quite well. Even though the latter result is not on \ACS themselves, its model is instructive vis-{\`a}-vis of the \ACS analysis of Benenson et al.~\cite{benenson2015user}.

Still, to this date, we are missing a robust structural latent variable model for the perceived trustworthiness and technology acceptance of \ACSplname. 
Prior research having only been correlational, no research to date has establish actual case-effect relations which factors would drive \ACS adoption. As the given models were focused on modest extensions of core TAM, we were missing important factors such as facilitating conditions for perceived trustworthiness or results demonstrability for technology acceptance.

\paragraph*{Aims.} We ask the research question what makes a sound latent-variable model of \ACS technology acceptance.
We further ask which intrinsic properties of an \ACS or its provider and which presentation properties of its delivery channel will impact perceived trustworthiness and eventual technology acceptance.

\paragraph*{Our Contributions.}
\begin{inparaenum}[(i)]
\item We report the first large-scale study with $N=812$ participants representative of the UK population to establish a robust model of perceived trustworthiness and technology acceptance of \ACSplname.
\item This study is the first to establish a cause-and-effect random controlled trial with systematic manipulation of independent variables influencing perceived trustworthiness and technology adoption in a between-subjects fractional-factorial design.
\item We offer the first confirmatory latent-variable covariance-based structural equation model (CB-SEM) of these factors, the first model to include attitudes and subjective norms as covariates as well as facilitating conditions and results demonstrability as antecedents.
\end{inparaenum}


%% file: background.tex
\section{Background}
This work investigates human factors in relation to \ACSpllong, using the Technology Acceptance Model (TAM) and measures of perceived trustworthiness. We introduce these areas in turn.

\subsection{\ACSplname}
Originally conceptualized by David Chaum~\cite{chaum1985security}, \ACSpllong bear the promise of offering authentication without identification. We especially focus on Anonymous Credential Schemes as one form with strong privacy-preserving properties. One strand of development starting from Brands' construction~\cite{brands2000rethinking} became the \ACS Microsoft U-Prove. Another based on one of the first breakthrough constructions by Camenisch and Lysyanskaya~\cite{camenisch2002signature} was developed into the IBM Identity Mixer (IDEMIX) \ACS~\cite{camenisch2002design}. The latter was subsequently extended with methods for revocation~\cite{camenisch2002dynamic}, efficient attribute encoding~\cite{camenisch2008efficient}, and smart card enablement~\cite{bichsel2009anonymous}, all technical advancements towards the adoption in government-certified identity cards and passports. Later advancements were made in elliptic-curve instantiations~\cite{CL04,TG20a}

More recent advancements include the development of the IRMA Card~\cite{vullers2013efficient,de2014towards}, a smart card version of \ACS with selective disclosure as well as the ACS deployment for Signal to achieve privacy-preserving authentication of Signal groups~\cite{chase2019signal}. Furthermore, \ACS were expanded to model graph signatures~\cite{Gross15}, relational anonymous credential schemes~\cite{GroTan2023RelationalIACR,TaSfGr2020} and confidentiality-preserving security assurance~\cite{Gross14ccsw}, for which the nomology in this paper likely also holds.

\subsection{Technology Acceptance Model}
Originally developed by Davis~\cite{davis1985technology,davis1989user} as an adaptation of the Theory of Reasoned Action (TRA), the Technology Acceptance Model (TAM) offers a framework to explain computer usage behavior. It has a considerable history of scrutiny~\cite{mathieson1991predicting,szajna1996empirical,lee2003technology}. It has been extended is with a range of antecedents such as subjective norms and results demonstrability in TAM 2.0~\cite{venkatesh2000theoretical} as well as with constructs for trust and risk~\cite{pavlou2003consumer}. It was further extended to the Extended Unified Theory of Acceptance and Use of Technology (UTAUT2)~\cite{venkatesh2012consumer}, which was in turn used as a foundation for perceived trustworthiness scales.

TAM and its variants have been used and extended for the modeling of a range of privacy-enhancing technologies, such as \ACSplname~\cite{benenson2015user} and Mix-Networks~\cite{harborth2018examining}, which we shall examine in related works. 

\subsection{Measuring Perceived Trustworthiness}
The Cambridge Dictionary defined trustworthiness as ``the quality or fact of being able to be trusted.''
We investigated a range of options to measure perceived trustworthiness, especially with the context of online and computer systems. The investigated instruments are largely from the areas of marketing and e-commerce.

In our investigation, we found a distinction between a perceived trustworthiness of provider (of a service or an artifact) and the perceived trustworthiness of a system (or artifact).

In terms of Perceived Provider Trustworthiness, Lee and Turban~\cite{lee2001trust} focused on trust in online merchants, distinguishing between merchant properties, medium properties and contextual properties. Gefen~\cite{gefen2002reflections} considered different dimensions of trust and trustworthiness for electronic commerce.
B{\"u}ttner and G{\"o}ritz~\cite{buttner2008perceived} discussed measures of trust for online shops. Their scale includes ability, benevolence, integrity, and predictability, the former three having been used by Lee and Turban and Gefen, as well.

For Perceived System Trustworthiness on the other hand, Corritore et al.~\cite{corritore2005measuring} investigated the perceived trustworthiness of Web sites, and considered as constructs honesty and reputation and risk. Bart et al.~\cite{bart2005drivers} considered Web sites, evaluating factors such as privacy and security or absence of errors as antecedents of trust.

In a separate line of work, McKnight et al.~\cite{mcknight2011trust} investigated trust in specific technologies, including general faith in technology and trusting stance, which we adopted to measure covariates.

Alalwan et al.\cite{alalwan2017factors} extended the UTAUT2 with notions of trust, which used a trust scale by Gefen. Due to the affinity to the Technology Acceptance Model, this scale was one of the foundations of our measurements of perceived trustworthiness.

\subsection{Measuring Privacy Concern}
There exist a range of instruments measuring privacy concern, well documented for instance by Preibusch~\cite{preibusch2013guide}. For this  study, we chose Internet Users Information Privacy Concern (IUIPC)~\cite{malhotra2004internet} as the privacy concern instrument. IUIPC has received scrutiny in terms of its validity and reliability~\cite{Gross2020IUIPCPETS}. While the instrument was reported to show good pedigree in terms of content validity, there were weaknesses found wrt. the reliability of the subscales control and awareness, which led to the proposal of an eight-item brief scale called IUIPC-8~\cite{Gross2023IUIPC-8}.

\subsection{Modelling Non-Normal, Ordinal Data}
In a reflective measurement model, the indicators are endogenous, that is, caused by the latent factors.
Standard maximum likelihood (ML) estimation assumes the multivariate normality for the joint population distribution of the endogenous variables, given the exogenous variables~\cite{kline2015principles,kline2012assumptions}. This can only hold for continuous variables. Consequently, analyzing ordinal data, such as obtained from Likert scales from self-report instruments introduced above, with ML estimation may yield inaccurate results~\cite{distefano2002impact} and the community has discussed a range of appropriate alternatives~\cite{bovaird2012measurement}. To make the point, Liddell and Kruschke~\cite{liddell2018analyzing} illustrated misrepresentations that can occur when ordinal data is analyzed in metric models.

To establish accurate models on non-normal, ordinal data, we turn to diagonally weighted least square estimation with robust standard errors and a mean- and variance adjusted test statistic (WLSMV). In general, WLSMV models are more complex to comprehend, because they operate on a probit-estimation and thresholds for choosing a level on an indicator. They also require a greater sample size than ML models. Furthermore, Shi et al.~\cite{shi2020assessing} cautioned that the RMSEA fit index could be inaccurate especially for larger models with more than $p=20$ covariance observations. They advocated SRMR as a fit estimate that stays accurate especially with a larger sample size ($N \geq 500$). We will take these considerations into account in our global fit evaluation.

%% file: related.tex
\section{Related Works}

There are two main related works to consider for the perceived trustworthiness and technology acceptance of \ACSpllong. 

\paragraph*{Perceived Trustworthiness \& Technology Acceptance of  \ACS} 
First, Beneson et al.\cite{benenson2015user} investigated the perceived trustworthiness and technology acceptance of  \ACS after a preliminary examination of the subject~\cite{benenson2014user}. They issued ACS smart card to distributed systems students of Patras University and received $N=30$ observations from their questionnaires. 

In terms of research design, Beneson et al.\cite{benenson2015user} specified a new TAM model and diligently offered the reliability statistics for the new instruments. They distinguished primary and secondary task in keeping with the observation in usable security that those are processed differently. We follow their lead in this respect.

Benenson et al. included risk and trust in their instruments, similar to Pavlou's extension of TAM~\cite{pavlou2003consumer}, however only used a single item for these two constructs. We perceived single-item constructs as unreliable and fraught with statistical perils (e.g., needing to be considered ordinal and not interval scale) and, thereby, opted for more comprehensive scales.
Benenson et al. also included items on situation awareness that are not found in the different versions of TAM.
We also observed that the items on perceived anonymity and usefulness of the secondary task seem rather similar, which might challenge their validity.

Benenson et al. concluded that ``the sample size (30 participants) is prohibitively small for deeper statistical analysis such as multiple regressions or structural equation modeling;'' we agree to this commendably cautious assessment. Hence, their study only included a preliminary---yet instructive---correlation analysis.

\paragraph*{Trustworthiness and Technology Acceptance of JonDoNym}

Second, Harborth and Pape~\cite{harborth2018examining} established a partial least square structural equation model (PLS-SEM) with SmartPLS on a sample of \textsf{JonDonym}\footnote{\url{https://anonymous-proxy-servers.net}} users. \textsf{JonDonym} is not an  \ACS, but a mix-net service which was founded on the Java Anon Proxy (JAP).

Harborth and Pape established a structural equation on $N=141$ questionnaire responses from anonymous users.
Following the guidance by Hair et al.~\cite{hair2017pls,hair2016primer}, we note that PLS path modeling focuses on maximizing the variance explained and on theory development. The competing covariance-based SEM (CB-SEM) is more appropriate for confirmatory research. Even though PLS-SEM has been frequently criticized in the field, a number of researchers came eloquently to its defence~\cite{hair2017pls,sarstedt2016estimation}. As a considerable advantage, we point out that PLS-SEM is generally non-parametric and robust in face of distribution problems. It can typically also cope with smaller sample sizes than CB-SEM, \textit{ceteris paribus}.

We noticed in this research that the users seem to have been anonymous and that no demographics were provided. While this raises the obvious question of the characteristics of the sample, there is a subtler point that anonymous users may not be distinguished as unique and that, therefore, the independence of observations may have been jeopardized.

%% file: aims.tex
\section{Aims}
\label{sec:aims}

\begin{figure*}[tb]
\centering
\includegraphics[keepaspectratio,width=.8\textwidth]{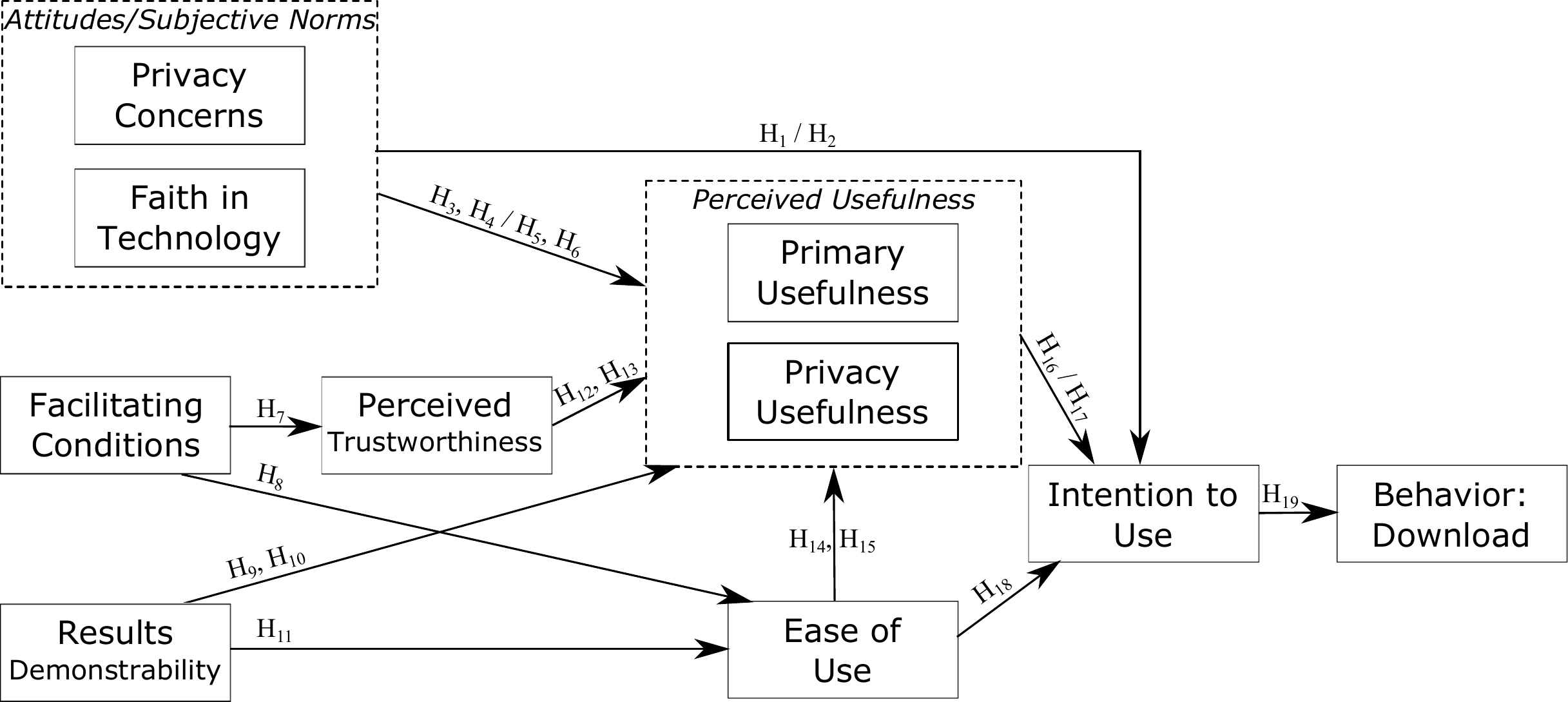}
\caption{Aimed at structural equation model based.\\\normalfont{\emph{Note}: slash denotes different antecedents; comma denotes different consequences.}}
\label{fig:sem_aim}
\end{figure*}

\subsection{\ACS Trustworthiness \& Acceptance}
\label{sec:aims_tam}
Our first line of inquiry is a sound model of \ACS trustworthiness and technology acceptance.
Here, we aim at fitting a theoretically founded model on the different relations impacting \ACS technology acceptance.

\begin{researchquestion}[\ACS Trustworthiness \& Acceptance]
\label{rq:sem}
We aim at establishing and confirming a robust structural latent variable model of the perceived trustworthiness and technology acceptance of \ACS.
\end{researchquestion}
We aim at a structural equation model based upon the Technology Acceptance Model (TAM 2.0)~\cite{venkatesh2000theoretical} and the empirical research by Benenson et al.~\cite{benenson2015user}. Figure~\ref{fig:sem_aim} illustrates the overall planned structure. The corresponding hypotheses are yielded by the underlying theory and not by establishing experimental cause-and-effect relations.

\subsubsection*{Hypotheses}
For brevity, we only name the alternative hypotheses and have it understood that the corresponding null hypotheses can be derived canonically.
\begin{compactdesc}
  \item[$H_1$:] Privacy Concerns yield a positive impact on the Intention to Use an \ACS.
  \item[$H_2$:] Faith in Technology yields a positive impact on the Intention to Use an \ACS.
  \item[$H_3, H_4$:] Privacy Concerns have a positive impact on Primary Usefulness and Privacy Usefulness, respectively.
  \item[$H_5, H_6$:] Faith in Technology has a positive impact on Primary Usefulness and Privacy Usefulness, respectively.
  \item[$H_7$:] Facilitating Conditions impact Perceived Trustworthiness positively.
  \item[$H_8$:] Facilitating Conditions impact Ease of Use positively.
  \item[$H_9, H_{10}$:] Results Demonstrability has a positive impact on Primary Usefulness and Privacy Usefulness, respectively.
  \item[$H_{11}$:] Results Demonstrability impacts Ease of Use positively.
  \item[$H_{12}, H_{13}$:] Perceived Trustworthiness has a positive impact on Primary Usefulness and Privacy Usefulness, respectively.
  \item[$H_{14}, H_{15}$:] Ease of Use has a positive impact on Primary Usefulness and Privacy Usefulness, respectively.
  \item[$H_{16}$ / $H_{17}$:] Primary Usefulness as well as Privacy Usefulness yield a positive impact on Intention to Use.
  \item[$H_{18}$:] Ease of Use impacts the Intention to Use positively.
  \item[$H_{19}$:] The Intention to Use impacts the Behavior to download positively.
\end{compactdesc}
We call the relations between Usefulness, Ease of Use, Intention to Use, governed by Hypotheses $H_{14}$--$H_{18}$, the \textit{Core Technology Acceptance Model}.

\subsection{Impact of \ACS Properties}
\label{sec:aims_causal}
Our second line of inquiry is the cause-effect impact of intrinsic and presentation properties of an \ACS on its perceived trustworthiness and technology acceptance. 
\begin{researchquestion}[Impact of \ACS Properties]
\label{rq:causal}
We investigate to what extent
\begin{compactenum}[(i)]
  \item intrinsic properties of a scheme and its provider, as well as
  \item the overall presentation and perception of the offering
\end{compactenum}
impact the perceived trustworthiness, overall technology acceptance and behavioral intention to follow through with an installation.
  \end{researchquestion}
The corresponding hypotheses model true cause-effect relations established by experimental manipulations.

\subsubsection*{Hypotheses}
Again, we only name the alternative hypotheses, the prefix $C$ indicating a cause-and-effect hypothesis.
All hypotheses operate on the perceived trustworthiness and the antecedents of the core technology acceptance model. Note that, therefore, each hypothesis is a compound hypothesis yielding multiple statistical hypotheses operating on low-level variables.

\paragraph{Intrinsic Properties}
\begin{compactdesc}
  \item[$H_{C, 1}$:] The provider of an \ACS impacts its perceived trustworthiness and acceptance.
  \item[$H_{C, 2}$:] The benefits of an \ACS yield an impact in that user benefits have a more positively effect than privacy benefits.
  \item[$H_{C, 3}$:] The usage of an \ACS makes a difference in that everyday use has a more positively effect that tech usage.
\end{compactdesc}  

\paragraph{Presentation Properties}
\begin{compactdesc}
  \item[$H_{C, 4}$:] Simplicity of the textual content has a positive impact on perceived trustworthiness and acceptance.
  \item[$H_{C, 5}$:] The presence of people has a positive impact on perceived trustworthiness and acceptance.
  \item[$H_{C, 6}$:] The available support makes a difference on perceived trustworthiness and acceptance in that more support options yield a more positive outcome.
\end{compactdesc}

%% file: method.tex
\section{Method}
This work was pre-registered in the Open Science Framework\footnote{\tiny\url{https://osf.io/w39bv/?view_only=199b1119d7134e6cb2f50109361a89bd}}.  
For reproducibility, results, graphs and tables were computed from the datasets with the \textsf{R} package \textsf{knitr}.
All test statistics are evaluated at a significance level of $\alpha = .05$. Statistical inferences are designed to be two-tailed and multiple-comparison corrected.

We hosted two experiments as part of this study, one on intrinsic properties, the other on presentation properties. The two experiments were created as a fractional factorial design, that is, each experiment is a factorial design of its own manipulations, and uses the default levels of the variables of its sister experiment as fixed values. The experiment on presentation properties of \ACSplname set the provider to \textsf{none}.

\subsection{Ethics}
The studies followed the ethical guidelines of the institution and received ethical approval.

Participants entered the study under informed consent and could leave the survey at any point.
Participants had the opportunity to contact the principal investigator to ask further questions.

Participants were paid at slightly greater rate than required by Prolific Academic for the expected completion time of the questionnaire: \pounds{4}. Participants could only enter the survey once.

\subsection{Sample}
\paragraph*{Sampling Process}
We set the target population as residents of the United Kingdom. 
As survey population, we chose the UK residents registered on the platform Prolific Academic, the sampling frame consisting of the corresponding user list as of August 2019. Both experiments were independently sampled, with participants registered for one experiment being excluded from the other, thereby maintaining independence of observations.
The sampling was conducted with a representative distribution by age, gender and ethnicity, with replacement in the case of participants not completing the full survey. Overall, the sampling process was a judgment sampling, in which participants matching the age/gender/ethnicity constraints self-selected to participate.

\paragraph*{Sample Size}
The sample size was determined with an \textit{a priori} power analysis. 
We planned for a range of analyses with the ultimate goal of covariance-based structural equation modelling (CB-SEM).
Using a comparable study design by Benenson et al.~\cite{benenson2015user} as guidance (with 8 variables), we computed an \textit{a priori} sample size of 553 cases for Structural Equation Modeling. For a greater model with $14$ latent variable and 53 measurement variables, we found \textit{a priori} that a sample of 840 would offer us a sensitivity of an effect size of $0.16$ at $80\%$ power. Consequently, designed the study to have an overall sample size of 420 and were to be merged into one sample with 840 cases for SEM.

\subsection{Assignment}
Both experiments were designed as random-controlled trials.
In both experiments, we used a \emph{simple random assignment} to allocate participants to their conditions.
The randomness was generated with a Javascript pseudo-random number generator.

During an experiment run, the assignment was \emph{blinded} to the experimenters as well as the participants.
We facilitated this blinding by segregating the experimental environments by conditions, that is, offering a separate survey as well as a separate Web page embedding the corresponding manipulations for each condition.
Experimenters were able to see the conditions assigned to participants in the final dataset.

\subsection{Operationalization}

\subsubsection{Measurement Instruments}

We adapted a range of standard questionnaires for the topic of \ACSplname as target technologies.
The overall purpose was to measure system trustworthiness and technology acceptance contextualized by privacy concerns and general faith in technology.
Specifically, we selected the following instruments for the following purposes:
\begin{compactdesc}
  \item[Privacy concerns (IUIPC-10])~\cite{malhotra2004internet} Elicits the long-term privacy concerns of users in the dimensions control (\textsf{ctrl}), awareness (\textsf{aware}) and collection (\textsf{collect}). We selected it as a covariate, informing technology adoption wrt. privacy usefulness.
  \item[Faith in Technology (FIT)~\cite{mcknight2011trust}] Elicits long-term attitudes on using technology and general trusting stance. We selected FIT as a covariate informing primary usefulness.
  \item[System Trustworthiness (ST)~\cite{mcknight2011trust}] Elicits perceived trustworthiness in a particular technology, with a range of sub-scales. \begin{inparaenum}[(i)]
  \item Social influence (\textsf{si}) elicits a subjective norm in the form perceived social pressure to engage or not to engage in a behavior. 
   \item Facilitating conditions (\textsf{fc}) models the available resources, knowledge and support engage with the technology to be trusted. 
   \item Performance Expectancy measures the expectation that the technology will be useful, which we used as primary usefulness (\textsf{primuse}) in the Technology Acceptance Model (TAM 2.0).
   \item Perceived Trustworthiness (\textsf{trust}) elicits to what extent participants are willing to trust the technology. 
   \item Behavioral Intention models is identical to the behavioral intention used in TAM 2.0.
\end{inparaenum}
\item[Technology Acceptance Model (TAM 2.0)~\cite{venkatesh2000theoretical}] Models to what extent users are willing to adopt a given technology.
\begin{inparaenum}[(i)]
   \item Results Demonstrability (\textsf{results}) models whether participants find the results apparent and easily communicable.
   \item Primary Usefulness (\textsf{primuse}) measures the expectation that the technology will be useful in general, adapted from ST Performance Expectancy.
   \item Privacy Usefulness (\textsf{privuse}) elicits whether participants find the technology useful to support their privacy.
   \item Behavioral Intention (\textsf{bi}) models the intent to use the technology, which is identical to the behavioral intention used in ST.
\end{inparaenum}
\end{compactdesc}
Table~\ref{tab:ops.m} contains an overview of the measurement variables with their operationalization in these instruments.

\paragraph*{Instrument Evaluation}
As we have adapted existing questionnaires in their wording to match them to the application area of \ACSname, we diligently evaluated the final instruments for their validity and reliability.
\processifversion{DocumentVersionAppendix}{Appendix~\ref{sec:instrument} contains the results of this systematic evaluation.}

Overall, the instruments used yielded considerable validity and reliability metrics.
For the TAM Results Demonstrability construct, we found that one item (\textsf{rTAR4}) was inconsistent with the rest of the construct and chose to eliminate this item, with no ill effect on the reliability of the measurement of the construct.
We removed the Social influence (\textsf{si}) subscale, because of excessively great indicator correlation.
With regards to behavioral intention, we removed \textsf{STBI3} due to great correlation with \textsf{STBI1}.

We noticed that two IUIPC sub-scales (\textsf{ctrl} and \textsf{aware}) yielded low internal consistency $\omega$ as discussed in earlier work. We chose to retain these constructs to keep privacy concerns in the model.

\begin{DocumentVersionTR}
\paragraph*{Further Measurement Instruments.} We note here that we administered more questionnaires than included in the final model (Table~\ref{tab:ops.m}), namely: 
\begin{inparaenum}[(i)]
  \item Perceived Provider Trustworthiness~\cite{buttner2008perceived}, considering perceived ability, benevolence, integrity and predictability;
  \item Perceived System Trustworthiness~\cite{corritore2005measuring}, based on honesty, reputation and risk,
  \item Perceived Reliability~\cite{mcknight2011trust}, considering system likelihood of failure.
\end{inparaenum}
We removed these constructs from the model, because the correlation with the included perceived trustworthiness construct was so great that we expected spurious results.\footnote{The full questionnaire is available at the OSF repository}.

\extend{Use auxiliary scales to validate used scales in terms of correlations. Concurrent reliability/measurement theory.}

\paragraph*{Inclusion Decision on Risk.} 
The construct perceived risk (\textsf{risk}) from Corritore's Perceived System Trustworthiness~\cite{corritore2005measuring} deserves special mention. While Benenson et al.~\cite{benenson2015user} modelled perceived risk as variable next to perceived trustworthiness, Harborth and Pape~\cite{harborth2018examining} did not. We found a statistically significant positive correlation between perceived risk and perceived trustworthiness at Pearson's $r = .63, p < .001$, that is, participants who reported greater perceived risks were also ready to trust more. Hence, we followed Harborth and Pape's approach to build the model without perceived risk as latent variable.
\end{DocumentVersionTR}

\begin{table*}[tb]
\centering
\footnotesize
\caption{Final Operationalization of Measurement Variables (MVs)}
\label{tab:ops.m}
\begin{tabular}{llllp{5cm}}
\toprule
                & Instrument & Sub-Scale & Variable & Items\\
\midrule
Privacy concerns & IUIPC-10~\cite{malhotra2004internet} & Control & \textsf{ctrl} & \textsf{ctrl1}, \textsf{ctrl2}, \textsf{ctrl3}, \\
			&	& Awareness & \textsf{aware} & \textsf{awa1}, \textsf{awa2}, \textsf{awa3} \\
			&		& Collection & \textsf{collect} & \textsf{coll1}, \textsf{coll2}, \textsf{coll3}, \textsf{coll4}\\
\addlinespace
Faith in Technology & McKnight~\cite{mcknight2011trust} & Faith & \textsf{fit} & \textsf{MKF1}, \textsf{MKF2}, \textsf{MKF3}, \textsf{MKF4}\\
				&	& Trusting Stance & \textsf{ts} & \textsf{MKTS1}, \textsf{MKTS2}, \textsf{MKTS3}\\
\addlinespace
System Trustworthiness (ST)~\cite{mcknight2011trust} & & \sout{Social Influence} & \sout{\textsf{si}} & \sout{\textsf{STSI1}}, \sout{\textsf{STSI2}}, \sout{\textsf{STSI3}} \\ 
				      & & Facilitating Conditions & \textsf{fc} & \textsf{STFC1}, \textsf{STFC2}, \textsf{STFC3}, \textsf{STFC4}\\
				      & & Perceived Trustworthiness & \textsf{trust} & \textsf{STT1}, \textsf{STT2}, \textsf{STT3}, \textsf{STT4}, \textsf{STT5}, \textsf{STT6}\\
\addlinespace
Technology Acceptance Model~\cite{venkatesh2000theoretical} & TAM 2.0 & Results Demonstrability & \textsf{results} & \textsf{TAR1}, \textsf{TAR2}, \textsf{TAR3}, \sout{\textsf{rTAR4}}\\
						& 		&Primary Usefulness  & \textsf{primuse} & \textsf{STPE1}, \textsf{STPE2}, \textsf{STPE3}, \textsf{STPE4} \\
						&               & := Performance Expectancy (ST) \\
						&		& Privacy Usefulness & \textsf{privuse} &  \textsf{TAU1}, \textsf{TAU2}, \textsf{TAU3}, \textsf{TAU4},  \\
						&		& Ease of Use & \textsf{ease} & \textsf{TAE1}, \textsf{TAE2}, \textsf{TAE3},  \textsf{TAE4}\\
						& 		& Behavioral Intention (TS) & \textsf{bi} & \textsf{STBI1}, \textsf{STBI2}, \sout{\textsf{STBI3}}, \textsf{STBI4}\\
\bottomrule
\end{tabular}
\\\emph{Note:} \textsf{rVAR} = reverse-coded variable \textsf{VAR}; \sout{\textsf{VAR}} = variable \textsf{VAR} removed after reliability analysis.
\end{table*}

\subsubsection{Manipulations}
We developed a Web site about a new \ACSlong, in which different pieces of content could be changed easily. In that, we enabled the manipulation of \ACS intrinsic properties---their provider, usage and claimed benefits---and \ACS presentation properties---simplicity of the content in terms of readability, presence of people and level of support. The intrinsic properties were changed in the core content, the presentation properties also in the Web design elements around the core content.
These manipulations were to some extent influenced by the work of Egger et al.~\cite{egger2001affective} on affective design of e-commerce user interfaces and design features that yield perceived trustworthiness.
Figure~\ref{fig:manipulations} illustrates the manipulated pieces of content in the example of a single Web page.
\processifversion{DocumentVersionAppendix}{Table~\ref{tab:manipulations_content} in the materials Appendix~\ref{sec:materials} specifies the exact content fragments used in different conditions.}
We controlled the possible confounder of readability with Flesch's Reading Ease, crafting the text fragments such that they are similar in readability even if they differed in content.

\definecolor{myblue}{RGB}{0, 0, 255}
\begin{figure*}[tbp!]
\centering
\includegraphics[keepaspectratio,width=.8\textwidth]{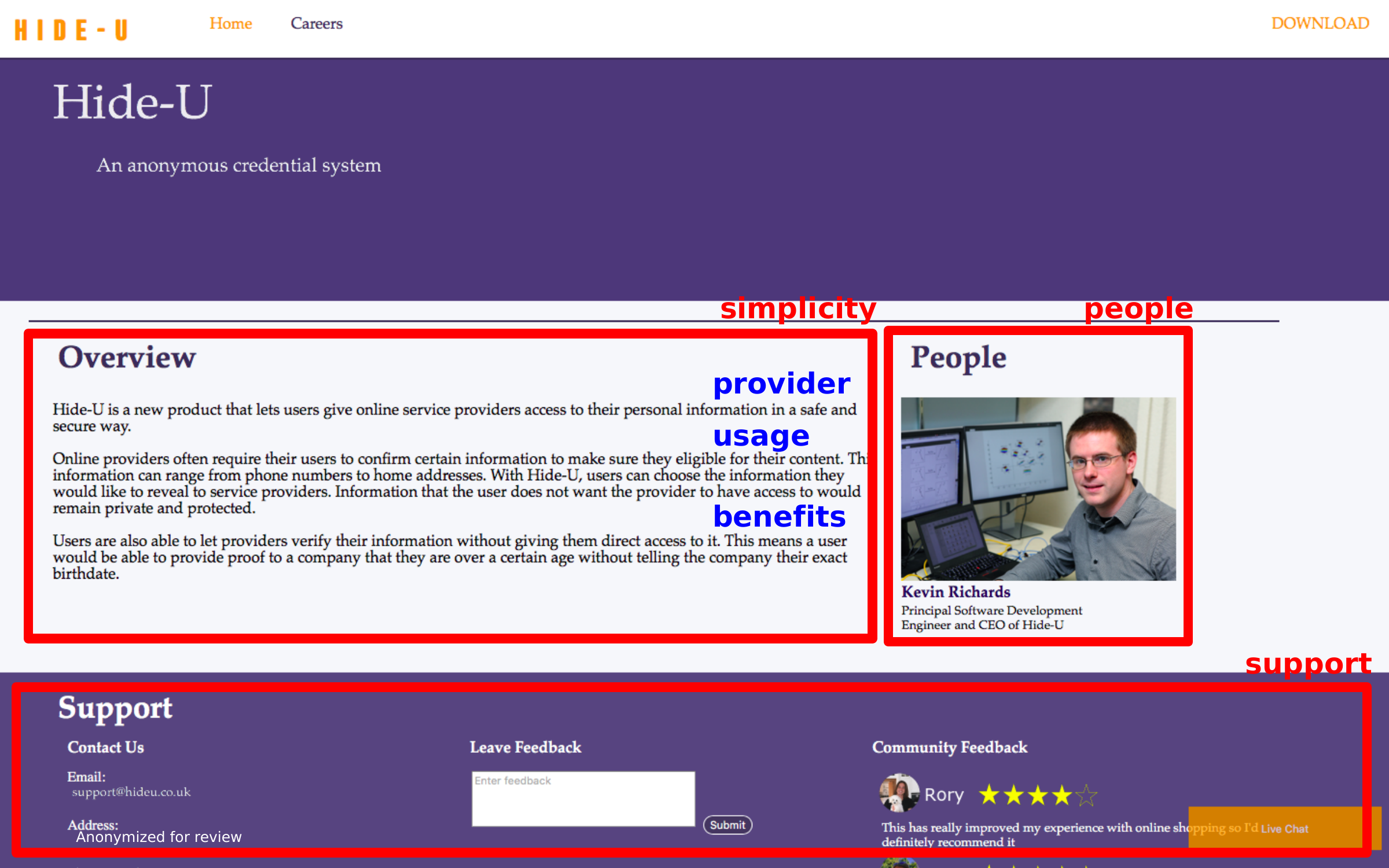}
\caption{Overview of the interventions placed on the \ACS Web site. \emph{Note:} Intrinsic variables are depicted in {\color{myblue}\textbf{\textsf{blue}}}; presentation variables in {\color{red}\textbf{\textsf{red}}}. The example site displays the condition: \textsf{provider}=\textsf{none}, \textsf{usage}=\textsf{everyday}, \textsf{benefits}=\textsf{user}, \textsf{simplicity}=\textsf{simple}, \textsf{people}=\textsf{photo}, \textsf{support}=\textsf{fullsupport} of the presentation study.}
\label{fig:manipulations}
\end{figure*}

\paragraph*{Intrinsic Properties}
\begin{compactenum}[(i)]
  \item \textsf{provider}: intrinsic properties of the provider, manipulated by which provider is being used, incl. a description of the organization in an about section. Four levels (with three binomial dummy variables):
    \begin{compactenum}[1.]
      \item \textsf{gov}: the UK government (\textsf{d\_gov} = 1),
      \item \textsf{company}: IBM (an internationally operating company which predominately developed \ACSplname in the past) (\textsf{d\_company} = 1),
      \item \textsf{uni}: a nationally known university (anonymized for submission) in the UK (\textsf{d\_uni} = 1).
      \item \textsf{none}: no provider is named, a condition reserved for the presentation trial (\textsf{d\_gov} = 0 \& \textsf{d\_company} = 0 \& \textsf{d\_uni} = 0).
    \end{compactenum}
  \item \textsf{benefits}: intrinsic benefits of an \ACSname, with two levels:
    \begin{compactenum}[1.]
      \item \textsf{privacy}: general benefits for privacy and data protection (\textsf{benefits} = 0)
      \item \textsf{user}: benefits specific for a user's life (\textsf{benefits} = 1),
    \end{compactenum}
  \item \textsf{usage}: description of how the system is used intrinsically, with two levels:
    \begin{compactenum}[1.]
      \item \textsf{tech}: Usage described in terms of the processes and procedures of the technology (\textsf{usage} = 0),
      \item \textsf{everyday}: Usage described in terms of every-day use of the system (\textsf{usage} = 1).
    \end{compactenum}
 \end{compactenum}
 We define as reference category: \textsf{university}-\textsf{privacy}-\textsf{tech}.

 \paragraph*{Presentation Properties}
\begin{compactenum}[(i)]
  \item \textsf{simplicity}: simplicity of the language used to convey the usage and benefits of the \ACS controlled by readability metrics. Two levels:
    \begin{compactenum}[1.]
      \item \textsf{complex}: employs language being complex, that is, a low readability in terms of Flesch Reading Ease (or correspondingly a high reading grade level) (\textsf{simplicity} =0),
      \item \textsf{simple}: employs simple language, that is, a high readability in terms of Flesch Reading Ease (and low reading grade level) (\textsf{simplicity} = 1). 
    \end{compactenum}
  \item \textsf{people}: presentation of the \ACS with a photo of a person to relate to or not, with two levels:
    \begin{compactenum}[1.]
      \item \textsf{nophoto}: a photo of a person to relate to is absent (\textsf{people} = 0).
      \item \textsf{photo}: a (stock) photo representing a leading developer is displayed next to the content (\textsf{people} = 1),
    \end{compactenum}
  \item \textsf{support}: presentation of different levels of support opportunities on the Web site. Three levels (with two binomial dummy variables):
    \begin{compactenum}[1.]
      \item \textsf{fullsupport}: support information containing contact information (e-mail and chat) as well as a user feedback system (\textsf{d\_fullsupport} = 1),
      \item \textsf{contact}: support information contains contact information only (e-mail and chat) (\textsf{d\_contact} = 1),
      \item \textsf{nosupport}: no support-related cues are given (\textsf{d\_fullsupport} = 0 \& \textsf{d\_contact} = 0).
    \end{compactenum}
 \end{compactenum}
 We define as \textit{reference categories}: \textsf{complex}-\textsf{nophoto}-\textsf{nosupport}.

\subsubsection{Manipulation and Attention Checks}
We integrated three kinds of manipulation checks in the survey:
\begin{compactenum}[1.]
  \item Factual manipulation checks, asking participants questions about facts of the inspected Web pages, such as, the color of the download button, or presence of manipulations, such as, the presence/absence of a photo as indicated by the condition.
  \item Impact manipulation checks, debriefing self-report of the participants how much they perceived a particular aspect of the Web page influenced them.
  \item Instructional Manipulation Checks (IMCs)~\cite{oppenheimer2009instructional}, instructions to select a particular option as confirmation of sustained attention.
\end{compactenum}
We pre-registered the plan to remove observations of participants who failed more than one attention check (IMC).

\subsection{Procedure}
As overall procedure, participants were asked to visit a Web site on \ACSplname and evaluate their properties.
The condition, that is, the combination of realizations of manipulated independent variables, was embedded in the Web site the participants were directed to.

\begin{figure*}
\centering
\includegraphics[keepaspectratio,width=.8\textwidth]{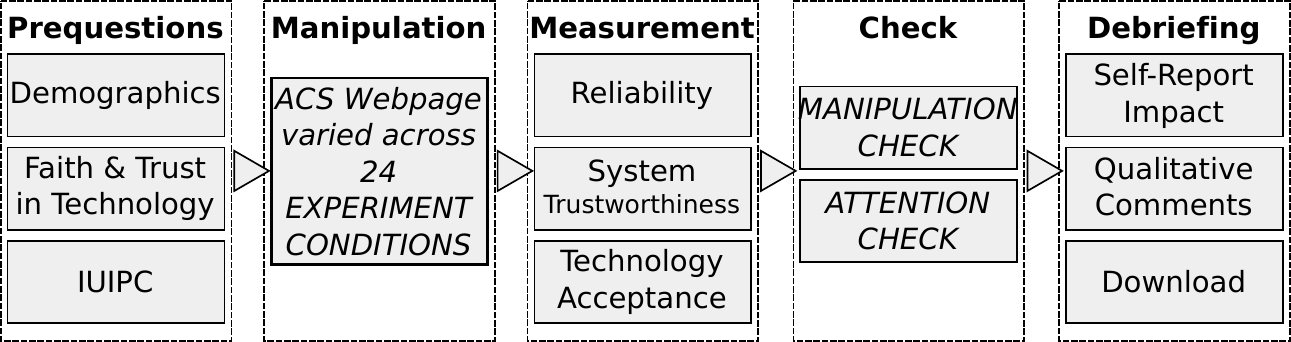}
\caption{Overview of the experimental procedure.}
\label{fig:procedure}
\end{figure*}

The participants proceeded as follows:
\begin{compactenum}
  \item First, the participants filled in questionnaires their demographics and on trait-like co-variates, such as IUIPC and general Faith and Trust in Technology.
  \item Second, participants were directed to a web site and asked to spend 10 minutes to read the Web site and form an opinion on the described \ACSlong.
  \item Third, the participants were asked factual questions on the Web site appearance to check that they indeed completed the task. (In Figure~\ref{fig:procedure} this step is shown with the other manipulation checks to save space)
  \item Fourth, the participants answered questionnaires on their impressions of the system, incl. on reliability, system trustworthiness, and technology acceptance, as well as their behavioral intention to use such a system.
  \item Throughout the questionnaire, the participants answered instructional manipulation checks (IMCs) that served as attention checks and re-confirmation to pay attention for the participants.
  \item Finally, in the debriefing, the participants reported how much they felt different aspects of the Web site contributed to their decisions, were offered an opportunity to give qualitative comments and finally reconfirm their download action with a code.
\end{compactenum}
Figure~\ref{fig:procedure} illustrates this procedure in a nutshell.

\subsection{Structural Equation Model}
The nomological network is theoretically founded on the Technology Acceptance Model (TAM 2.0)~\cite{venkatesh2000theoretical}.
It is empirically substantiated taking into account the correlation analysis of Benenson et al.~\cite{benenson2015user}.
We depicted the nomology designed for this study in Figure~\ref{fig:sem_aim}.

As summarized in Table~\ref{tab:steps}, we establish the structural equation model in three steps.
\begin{inparaenum}[1.]
\item The \textit{measurement model} is constructed by setting the measurement variables (MVs) defined in the operationalization of Table~\ref{tab:ops.m} as reflective measurements of the corresponding latent variables.
\item The \textit{correlational path model} is based on the theoretical nomology proposed in Section~\ref{sec:aims}, incorporating the variables primary usefulness (\textsf{primuse}), privacy usefulness (\textsf{privuse}), and ease of use (\textsf{ease}) forming the core of TAM. Results demonstrability (\textsf{results}) and facilitating conditions (\textsf{fc}) act as major antecedents.
\item The \textit{causal path model} is built by adding causal variables as single-item variables and encoded as binomial dummy variables for multinomial variables. They model the conditions of the experiments.
We incorporated them as regression antecedents for perceived trustworthiness (\textsf{trust}), primary usefulness (\textsf{primuse}), privacy usefulness (\textsf{privuse}), and ease of use (\textsf{ease}). This model also incorporates the download behavior as endogenous dichotomous variable and consequence of behavioral intention.
\end{inparaenum}

\subsubsection{Modelling Approach}
\begin{table}[tb]
\centering
\caption{SEM Modelling Phases}
\label{tab:steps}
\begin{tabular}{lp{.7\columnwidth}}
\toprule
Model & Description \\
\midrule
1. Measurement & Reflective measurement of latent variables (LVs) by the measurement variables (MVs, indicators); no regression equations.\\
2. Correlational & Path model incorporating the measurement model \emph{plus} regression equations according to statistical hypotheses from aims establishing ACS Trustworthiness and Acceptance (\S\ref{sec:aims_tam}; Figure~\ref{fig:sem_aim}).\\
3. Causal & Path model incorporating the correlational model \emph{plus} regression equations on dichotomous variables modelling the experiment conditions and, thereby, the causal impact of intrinsic and presentation properties (\S\ref{sec:aims_causal}).\\
\bottomrule
\end{tabular}
\end{table}

We followed a modelling approach that embeds the Two-Step Modelling approach proposed by Anderson and Gerling~\cite{anderson1988structural} and advocated by Kline~\cite{kline2015principles}. We base the SEMs on a Weighted Least Square estimation with mean and variance correction (WLSMV) to account for the ordinal, non-normal data~\cite{bovaird2012measurement}, evaluating the fit with the robust CFI and TLI, a robust RMSEA and SRMR.\footnote{We followed David A. Kenny's recommendation~\cite{Kenny2015fit} that for models with $N>400$ cases, the $\chi^2$ test will be almost always statistically significant, and de-emphasized this test. Further, we are aware that RMSEA may be less reliable than usually expected in WLSMV estimation and focus on the more robust SRMR~\cite{shi2020assessing}.} We compare nested models with the likelihood-ratio test (LRT) on the equal-fit hypothesis. For unnested models, we compare the $\mathsf{cn}_{05}$, that is, the critical $N$ for the $\chi^2$ test at .95 confidence~\cite{bollen1988some}, in addition to other fit indices.

In our variant, we proceed according to three steps outlined in Table~\ref{tab:steps}.
\begin{compactenum}
  \item We started with the theoretical model informed by general TAM 2.0~\cite{venkatesh2000theoretical} and the model proposed by Benenson et al.~\cite{benenson2015user}
  \item We computed univariate histograms and density diagrams for the inputs to diagnose distribution problems as well as covariance/correlation matrices to indicate variables with risk of substantial multicollinearity. In this process, we identified variables that need additional consideration in the final model.
  \item We evaluated the measurement model first (Table~\ref{tab:steps}, Step 1.), that is, computed a confirmatory factor analysis (CFA) model that only included the exogenous and endogenous latent variables (LVs) for the desired constructs and their indicator/measurement variables (MVs). In that, we established general fit, the significance of all reflective MV-LV relations, as well as factor loadings and the reliability in terms internal consistency.
  \item Based on this analysis, we diagnosed faults in the measurement model and underlying questionnaires, deriving a refined measurement model.
  \item We added the predicted regression equations to the measurement model to build the correlational model (Table~\ref{tab:steps}, Step 2.). We tested with a $\chi^2$ likelihood ratio test (LRT) that the improvement of the more complex correlational model over the simpler measurement model is statistically significant. We then expanded the correlational model with direct paths from the attitudes and subjective norms to the behavioral intention. We tested this expansion, in turn, with an LRT.
  \item Once convinced that the correlational model is a good fit, we expanded it by dichotomous variables designating the conditions of the experiments, forming the causal model (Table~\ref{tab:steps}, Step 3.). We tested the statistical significance of incorporating the direct paths on attitudes and subjective norms with an LRT.
\begin{DocumentVersionBootstrapping}   
   \item For the final selected SEM, we bootstrapped with 5,000 iterations, the \emph{Bollen-Stine} algorithm, and a fixed seed set in the \emph{L'Ecuyer-CMRG} pseudo-random number generator.
\end{DocumentVersionBootstrapping}   
\end{compactenum}

%% file: demoTableOne.tex
\begin{table}
\centering
\caption{Demographics of the final sample}
\label{tab:demo}
\begin{tabular}{ll}
\toprule
  & Overall\\
\midrule
$N$ & 812\\
Gender (\%) & \\
Male & 391 (48.2)\\
Female & 414 (51.0)\\
Rather not say & 7 ( 0.9)\\
\addlinespace
Age (\%) & \\
18-24 & 133 (16.4)\\
25-34 & 215 (26.5)\\
35-44 & 167 (20.6)\\
\addlinespace
45-54 & 115 (14.2)\\
55-64 & 141 (17.4)\\
65+ & 41 ( 5.0)\\
\bottomrule
\end{tabular}
\end{table}

%% file: privacy_comparison_total.tex
\begin{table*}[ht]
\centering
\caption{Privacy concern in indirect and direct modelling.} 
\label{tab:privacyComparison}
\begingroup\footnotesize
\begin{tabular}{crrrrrrrr}
  \toprule
                       & \multicolumn{4}{c}{indirect} & \multicolumn{4}{c}{direct} \\
                       \cmidrule(lr){2-5} \cmidrule(lr){6-9}
Relation & $B$ & $\vari{SE}$ & $p$ & $\beta$ & $B_d$ & $\vari{SE}_d$ & $p_d$ & $\beta_d$ \\ 
  \midrule
$\mathsf{primuse}\sim{}\mathsf{ctrl}$ & -0.314 & 0.069 & $<.001$ & -0.269 & -0.700 & 0.128 & $<.001$ & -0.585 \\ 
  $\mathsf{primuse}\sim{}\mathsf{aware}$ & 0.340 & 0.078 & $<.001$ & 0.303 & 0.819 & 0.149 & $<.001$ & 0.704 \\ 
  $\mathsf{primuse}\sim{}\mathsf{collect}$ & 0.136 & 0.044 & $\phantom{<}.002$ & 0.123 & -0.048 & 0.065 & $\phantom{<}.455$ & -0.044 \\ 
  $\mathsf{privuse}\sim{}\mathsf{ctrl}$ & -0.247 & 0.069 & $<.001$ & -0.208 & -0.783 & 0.159 & $<.001$ & -0.645 \\ 
  $\mathsf{privuse}\sim{}\mathsf{aware}$ & 0.390 & 0.078 & $<.001$ & 0.340 & 1.074 & 0.192 & $<.001$ & 0.910 \\ 
  $\mathsf{privuse}\sim{}\mathsf{collect}$ & -0.017 & 0.041 & $\phantom{<}.677$ & -0.015 & -0.258 & 0.084 & $\phantom{<}.002$ & -0.230 \\ 
  $\mathsf{bi}\sim{}\mathsf{ctrl}$ &  &  &  &  & 0.998 & 0.181 & $<.001$ & 0.797 \\ 
  $\mathsf{bi}\sim{}\mathsf{aware}$ &  &  &  &  & -1.223 & 0.213 & $<.001$ & -1.006 \\ 
  $\mathsf{bi}\sim{}\mathsf{collect}$ &  &  &  &  & 0.411 & 0.089 & $<.001$ & 0.355 \\ 
\midrule
Total Effect\\
\midrule
  $\mathsf{bi}\leftarrow{}\mathsf{ctrl}$ &  &  &  & -0.206 &  &  &  & 0.124 \\ 
  $\mathsf{bi}\leftarrow{}\mathsf{aware}$ &  &  &  & 0.256  &  &  &  & -0.139  \\
  $\mathsf{bi}\leftarrow{}\mathsf{collect}$ &  &  &  & 0.070 &  &  &  & 0.228 \\ 
   \bottomrule
\end{tabular}
\endgroup
\end{table*}

%% file: discussion.tex
\section{Discussion}

\subsection{Technology acceptance of \ACS}
Adoption of \ACSpllong is well modeled with the extended Technology Acceptance Model (TAM 2.0).
Taking into account the earlier work of Benenson et al.~\cite{benenson2015user}, we offer the first structural latent variable model from a large sample size with a sound WLSMV estimation for ordinal data.
Perceived trustworthiness (\textsf{trust}) is a major predictor of Privacy Usefulness (\textsf{privuse}) and Primary Usefulness (\textsf{primuse}) with medium standardized effects.

\subsection{Primary tasks trump privacy protection}
We found that the primary usefulness trumped the privacy usefulness. This is not surprising and is a common belief in the community, that is, that the user's focus is on the primary task and not on the secondary task (here, privacy protection), phenomenon also observed by Benenson et al.~\cite{benenson2015user}.
However, once direct and indirect effects of privacy concern are accounted for, the ratio of effect of primary usefulness and privacy usefulness is only 3:2.

\subsection{Demonstrable results and facilitating conditions are key drivers}
Our structural model design took into account antecedents from system trustworthiness as well as TAM 2.0: facilitating conditions and results demonstrability. Facilitating conditions strongly impacted trust and ease of use. They model whether a user has the right resources and knowledge, compatible systems and help available. It stands to reason that these questions indeed need to be answered well to convince a user to adopt an \ACS.

Similarly, TAM 2.0 results demonstrability, that is, whether the results are apparent to the user and whether the user is capable of explaining the results to others, yield considerable effects on ease of use as well as on privacy usefulness. Both results demonstrability and facilitating conditions had consistently positive effects as drivers of behavioral intention.

\subsection{Focus on simplicity}

Of the manipulated variables, \textsf{simplicity} has the most consistent positive effect:
We found that choosing simple language in terms of readability had considerable significant positive effects on primary usefulness and ease of use. Hence, simple language would make users more likely to adopt the technology. The question arises to what extent the \ACS can be presented in a simple fashion while still convincing users of their technical merits.

\subsection{Lessons learned for the privacy paradox}
Understanding how privacy concerns impact behavioral intention is crucial for the investigation of the privacy paradox. While earlier research already pointed out an attenuation effect in behavioral models due to middling reliability of prevalent privacy concern scales, we gain further insights from our model.

\subsubsection{Inconsistent impact of IUIPC factors}
We observed that the IUIPC factors control and awareness have contradicting effects on behavioral intention. While awareness increases the behavioral intention to use the \ACS, control diminishes it. Awareness emphasizes that the user be aware and knowledgable how information is used and that companies facilitate that by clearly disclosing how they handle information.
Control on the other hand emphasizes the right to control and autonomy over decisions over the user's information. 
One may hypothesize that the conviction that the user has the right to control already yields less behavioral intention to adopt further privacy preserving technologies to enforce that control.
With regards to the privacy paradox, the contradiction between impact of privacy antecedents on behavioral intention consequences offers an explanation why the privacy paradox occurs that has not been well researched previously.

\subsubsection{Inconsistent mediation}
Our analysis is the first to offer a mediation analysis of indirect and direct paths of privacy concern impacting behavioral intention.
Here, we observed an inconsistent mediation for all privacy concern factors, that is, that direct effects and indirect effects have different signs. For example, control yields a positive effect through a direct path to behavioral intention, but negative indirect paths through usefulness.

%% file: limitations.tex
\subsection{Limitations}

\subsubsection{Ecological Validity}
There are limitations to the ecological validity as the \ACS Web pages used as stimuli were mock-ups.
With respect to manipulating the provider, the Web sites did not use the provider's logos or domains due to licensing and legal restrictions. Otherwise, they mimicked the look-and-feel of existing sites of providers on \ACSplname.

In order to gain a large representative sample and to be able to manipulate aspects of the \ACS Web site, we paid a price in terms of ecological validity. The study of Benenson et al.~\cite{benenson2015user} was more hands-on in offering their students physical artifacts (smart cards, etc.), thereby yielding stronger ecological validity.

\subsubsection{Generalizability}
The large sample was representative for the UK, in terms of age, gender and ethnicity. Due how Prolific operates, the sampling process is judgment sampling with self-selection, rather than a random sampling process on UK residents.
However, in contrast to prior work, this study was neither limited to university students~\cite{benenson2015user} nor constrained to anonymous users without reported demographics~\cite{harborth2018examining}, affording us greater generalizability than in any previous study.

\subsubsection{Measurement of Privacy Concerns}
As indicated in the background section and observed in earlier studies~\cite{Gross2020IUIPCPETS,Gross2023IUIPC-8}, the measurement of privacy concern with IUIPC-10 only offered middling reliability. As a consequence, we face a low signal-to-noise ratio on the privacy concern measurement and an attenuation of its effects on other factors. While most of the relations could still be estimated with considerable certainty, we incurred lower certainty for smaller regression coefficients.

%% file: conclusion.tex
\section{Conclusion}
In having established the first comprehensive structural latent variable model for the trustworthiness and acceptance of \ACSlong, we offer a sound foundation for research in user decision making on \ACS. Our latent variable model is based on a large sample representative of the UK population in gender, age and ethnicity.

While our model confirms and expands on existing research on \ACS trustworthiness and acceptance as well as re-confirms conventional wisdom that perceived primary-task usability typically trumps perceived privacy usability, we show new insights which factors drive the users decisions. 
In that, we are the first to include results demonstrability and facilitating conditions as key drivers and show that they yield a consistent, medium-large positive effect on the ultimate technology acceptance of \ACS.
Overall, our model constitutes a comprehensive nomological network that can be used for other privacy-enhancing technologies beyond the given use case of \ACS.

Furthermore, having established the first truly causal-effect analysis of the impact of intrinsic and presentation properties, we can show that simplicity of content yields a consistently positive impact.
While a support framework (e.g., support contact or forum) also yielded a positive impact: it benefited the perceived trustworthiness.

We can show that using simple language to instruct the user on facilitating conditions, that is, what knowledge and resources are needed to run the \ACS successfully, and on demonstrable results, that is, what the \ACS does in day-to-day and how to convey that to others, will yield a considerable boost in technology acceptance.

Our work, finally, offers new insights into the study of the privacy paradox, the attitude-behavior dichotomy of privacy: we observed inconsistent effects of different privacy concern factors and an inconsistent mediation.

%% file: acknowledgments.tex
\section*{Acknowledgments}
This work was supported by \CASCAde.